\documentclass[rnote]{aa}
\usepackage{natbib,graphicx,txfonts}
\usepackage{colortbl}

\newcommand{\Hi}{\textup{H\,{\mdseries\textsc{i}}}}
\newcommand{\HI}{\textup{H\,{\mdseries\textsc{i}}}~}

\newcommand{\kms}{{km~s$^{-1}$}}

\newcommand{\gmr}{($g$-$r$)}
\def\kms{{km~s$^{-1}$}}

\def\arcmin{\hbox{$^\prime$}}
\def\arcsec{\hbox{$^{\prime\prime}$}}

\def\farcmin{\hbox{$.\mkern-4mu^\prime$}}
\def\farcsec{\hbox{$.\!\!^{\prime\prime}$}}

\usepackage{amstext}

\begin{document}

\title{Morphology and surface photometry of a sample of isolated early-type galaxies from deep imaging}
%with {\tt 4KCCD}@VATT}
\author{ R. Rampazzo\inst{1}\and A. Omizzolo\inst{2,1}\and M. Uslenghi\inst{3}\and J. Rom\'an\inst{4}\and 
 P. Mazzei\inst{5}\and L. Verdes-Montenegro\inst{4}\and A. Marino\inst{5}\and M. G. Jones\inst{4}
}

\institute{
INAF Osservatorio Astronomico di Padova,  
Via dell'Osservatorio 8, 36012 Asiago, Italy
\email{roberto.rampazzo@inaf.it}
\and
Vatican Observatory, Vatican City 
\email{aomizzolo@specola.va}
\and
INAF- IASF, Via A. Curti 12, 20133 Milano, Italy
\email{michela.uslenghi@inaf.it}
 \and
Dep.to Astronomia Extragal\'actica Istituto Astrofisica de Andaluc\'ia, 
Glorieta de la Astronomia s/n 18008 Granada, Spain
\and
INAF Osservatorio Astronomico di Padova,  
Vicolo dell'Osservatorio 5, 35122 Padova, Italy
}
  \authorrunning{Rampazzo et al.}
  \titlerunning{Morphology and deep surface photometry of iETGs}
  \date{Received; accepted}
  
\abstract
{Isolated early-type galaxies  are evolving in unusually poor environments 
for this morphological family, which is typical of cluster inhabitants. We 
investigate the mechanisms driving the evolution of these galaxies.} 
{Several studies indicate that interactions, accretions, and merging episodes 
leave their signature on the galaxy structure, from the nucleus down to the 
faint outskirts. We focus on revealing such signatures, if any, in a sample of
isolated early-type galaxies, and we quantitatively revise their galaxy classification. }
{We observed  20 (out of 104) isolated early-type galaxies, selected
from the AMIGA catalog, with the {\tt 4KCCD} camera at the Vatican 
Advanced Technology Telescope (VATT) in the Sloan Digital Sky Survey
(SDSS) $g$ and $r$ bands. These are
the deepest observations of a sample of isolated early-type galaxies so far: on average,
the light profiles reach $\mu_g\approx28.11\pm0.70$ mag arcsec$^{-2}$  
and $\mu_r\approx27.36\pm0.68$ mag arcsec$^{-2}$. The analysis
was performed using the {\tt AIDA} package, providing point spread function-corrected 
2D surface photometry  up to the galaxy outskirts. The package provides a model 
of the 2D galaxy light distribution, which after model subtraction enhances the
fine and peculiar structures in the residual image of the galaxies.} 
{Our re-classification suggests that the sample is composed of {\it \textup{bona fide}} early-type 
galaxies spanning from ellipticals to late-S0s galaxies.
Most of the surface brightness profiles are best fitted with a bulge plus
disc
%\LEt{you opted for British English spelling conventions. Please change all "disk" to "disc", thank you} 
model, suggesting the presence of an underlying disc
structure. The residuals obtained after the model subtraction 
show the nearly ubiquitous presence of fine structures, such as shells, stellar fans, rings, and tails. 
Shell systems are revealed in  about 60\% of these galaxies.} 
{Because interaction, accretion, and  merging events are widely interpreted 
as the origin  of the fans, ripples, shells and tails
%\LEt{the slash is commonly used to signify "ratio", as in signal-to-noise ratio, "S/N". You clearly misuse this here for commas in a list. Please check this throughout and use commas or "and" for two-item lists. \ In longer lists, the last item is set off with a comma and "and§", as in "interaction, accretion, and merging" above. I have counted 136 instances with various uses of the slash. They include some examples where you may mean a range, as in "B/T", which then should be written "B-T", and in some instances it it okay, as in "and/or" or "0376/px". To avoid me making incorrect changes, I leave this to you, but please make sure that you do this throughout} 
in galaxies, we suggest that
 most of these isolated early-type galaxies have experienced such events.   
 Because they are isolated (after 2-3 Gyr), these galaxies are the {\it \textup{cleanest 
 environment}} 
 %\LEt{please change all italics for emphasis and for Latin expressions to upright, thank you}
 in which to study phenomena connected with events like these.}


\maketitle

\section{Introduction}

Since the discovery of the morphology-density relation,  early-type
galaxies (Es+S0s=ETGs hereafter) have been known to be typical inhabitants
of dense clusters \citep{Dressler1980,Houghton2015}. Although with the large uncertainty 
introduced by automatic galaxy classifications, studies based on the Sloan Digital Sky Survey 
(SDDS hereafter) \citep[see e.g.][]{York2000,Abazajian2009} have shown that
ETGs represent a minority ($\approx$10\% E and $\approx$20\% S0s) 
of  the galaxy population in the lowest density fields 
\citep[see e.g.][their Figure 12]{Goto2003}. One of the pioneering attempts to
characterise the population of galaxies in relatively high isolation
produced the {\it Catalog of Isolated Galaxies in the Northern Hemisphere} by
\cite{Karachentseva1973}, which includes a small set of isolated ETGs (iETGs hereafter).
 Several catalogues of isolated galaxies have been
produced since then \citep[e.g.][]{Verdes-Montenegro2005,Hernandez2010,Argudo2013,Argudo2015}.
Historically, these efforts are intended to provide the best sample of
{\it \textup{unperturbed}} galaxies  to be used as a baseline for comparison of
galaxy properties with interacting or cluster samples, for example \citep[see
e.g.][]{Rampazzo2016}.

%------------------------- begin Table 1 -----------------------------------
\begin{sidewaystable*}
%\begin{table*}
\caption{The sample: KIGs morphology, heliocentric velocity, and adopted distance}
\centering
%% \tablesize{} %% You can specify the fontsize here, e.g.  \tablesize{\footnotesize}. If commented out \small will be used.
\begin{tabular}{lcclllcc}
\hline
\textbf{KIG} & RA & Dec.        & & \textbf{Morphology (Type)} &        & \textbf{V$_{hel}$} & \textbf{D}\\
\textbf{}       & J2000& J2000 & Buta & HyperLeda & F-L+        & [{km~s$^{-1}$}] & [Mpc] \\
\hline
 \rowcolor[gray]{0.8} 264       & 08 36 01.5 & +30 15 59        & SB0$^-$(-3.0)& SB0-a (-0.7$\pm$1.6) & E/S0 (-3.0$\pm$1.5) & 7715$\pm$26 & 113.9$\pm$4.4  \\
378  & 09 51 28.1  & +10 55 12 & E2 (-5.0) & E (-5.0$\pm$2.0) & E/S0 (-3.0$\pm$1.5) & 10385$\pm$54 & 153.8$\pm$5.4   \\
 \rowcolor[gray]{0.8} 412  & 10 24 46.4 & +46 27 22  & E1 (-5.0)  & E (-5.0$\pm$2.0) & E (-4.0$\pm$1.5) & 12789$\pm$120 & 187.1$\pm$6.1   \\
481  & 11 27 41.2 & +66 35 23  & SA(l,rs)a pec/E-S0 pec (1.0) & S0-a (-0.1$\pm$0.5) & S0 (-2.0$\pm$1.5) & 1608$\pm$16 &28.0$\pm$3.0  \\
 \rowcolor[gray]{0.8}490  & 11 36 39.6 & +06 17 31  & SA(r$\underbar{s}$:)0/a pec (0.0) & S0(r) (-1.1$\pm$0.9) & S0 (-2.0$\pm$1.5) & 5700$\pm$36 & 89.4$\pm$4.0  \\
 517    & 12 02 52.1 & +26 15 10        & E3 (-5.0)     & S0  (-2.2$\pm$1.3) & S0 (-2.0$\pm$1.5)& 9633$\pm$30 & 144.0$\pm$5.1  \\
 \rowcolor[gray]{0.8}578  & 13 16 15.4 & +20 02 52 & E1 (-5.0)  & E  (-5.0$\pm$2.0) & E  (-4.0$\pm$1.5) & 9191$\pm$18 & 139.5$\pm$5.0   \\
595  & 13 39 11.6  & +61 30 22 & E4-5 (-5.0)    & E (-4.8$\pm$0.6)  & S0 (-2.0$\pm$1.5) & 9406$\pm$53 & 139.6$\pm$5.0  \\
 \rowcolor[gray]{0.8} 599  & 13 48 34.6 & +37 06 48 & SA0$^-$(-3.0)& (R)S0  (-2.3$\pm$1.7)  & S0/a (0.0$\pm$1.5) & 10248$\pm$17 & 153.6$\pm$5.4  \\
620  & 14 13 49.2 & +37 16 09 & SA($\underbar{r}$l)0/$\underbar{a}$:/E(d)2 (0.5) & SBa (1.0$\pm$1.6) & S0 (-2.0$\pm$1.5) & 6621$\pm$29 & 102.2$\pm$4.1  \\
 \rowcolor[gray]{0.8} 636  & 14 33 31.4 & +57 42 42 & SAB(l:)0$^{-0}$ (-2.5) & SB0 (-2.0$\pm$2.0) & E/S0 (-3.0$\pm$1.5) & 11238$\pm$58 & 165.8$\pm$5.6  \\
637  & 14 34 52.4 & +54 28 33 & E4 (-5.0) & E-S0 (-3.2$\pm$1.0) & E (-4.0$\pm$1.5) & 2119$\pm$59 & 37.4$\pm$3.1   \\
 \rowcolor[gray]{0.8} 644  &  14 43 54.7 & +43 34 50 & SAB$\underbar{x}$a(r)0$^+$ (-1.0) & (R)Sab (2.2$\pm$1.8)  & S0/a (0.0$\pm$1.5)  & 8117$\pm$30 & 122.9$\pm$4.6   \\
670  & 15 19 30.2 & +67 30 17  & E3-4 (-5.0) & E/S0 (-3.0$\pm$1.7) & S0 (-2.0$\pm$1.5) & 12477$\pm$35 & 183.2$\pm$6.1  \\
 \rowcolor[gray]{0.8} 685  & 15 30 15.2 & +56 49 56  & E$^+$0: pec (-4.0) & E (-3.9$\pm$2.4) & E/S0 (-3.0$\pm$1.5) & 15383$\pm$150 & 224.5$\pm$7.1  \\
705  & 15 47 44.4  & +37 12 18 & E$^+$0 (-4.0) & E-S0 (-3.5$\pm$2.1)   & E (-4.0$\pm$1.5) & 11947$\pm$72  & 172.2$\pm$5.9  \\
 \rowcolor[gray]{0.8} 722  & 16 08 32.7  & +09 36 24 & E/E$^+$1 (-4.5) & E (-4.0$\pm$1.8)    & E (-4.0$\pm$1.5) & 10238$\pm$31 & 154.1$\pm$5.3   \\
 732  & 16 16 52.   & +53 00 22 & E/E$^+$1 (-4.5) & E (-4.9$\pm$0.6)  & E (-4.0$\pm$1.5) & 5615$\pm$31 &  86.3$\pm$3.8 \\
 \rowcolor[gray]{0.8}733  & 16 17 56.9 & +22 56 44 & (R1P)SAB(s)0/a pec (0.0) & SABa (1.5$\pm$2.1)  & S0/a (0.0$\pm$1.5) & 4512$\pm$32 & 72.2$\pm$3.6\\
841  & 17 59 14.7  & +45 53 13 &SA(rpl)0/s pec (0.0) & E-S0 (-3.0$\pm$0.5) & S0/a (0.0$\pm$1.5) & 5763$\pm$55 & 87.6$\pm$3.8  \\
\hline
\label{tab-1}
\end{tabular}
\tablefoot{Classifications are from \citet{Buta2019} (col.4), 
{\tt HyperLeda} (col. 5), and \citet{Fernandez2012} (F-L+ col. 6). 
The heliocentric velocity (col. 7) is from {\tt NED} and the distance (col. 8) 
is taken from
%\LEt{to properly include this reference into the footnote, please move the opening parenthesis to before the year. This is a LaTeX command error that I cannot fix for you with the program I work with (citet and citep)}  
\citet{Jones2018} (see \S~\ref{Sample}). }
%\end{table*}
\end{sidewaystable*}
%------------------------- end Table 1 -----------------------------------

%------------------------- begin Table 2 -----------------------------------
%\begin{sidewaystable*}
\begin{table*}
\caption{KIG relevant geometric and photometric properties}
\centering
%% \tablesize{} %% You can specify the fontsize here, e.g.  \tablesize{\footnotesize}. If commented out \small will be used.
\begin{tabular}{lccccccccc}
\hline
\textbf{KIG} & \textbf{Other ID} & \textbf{$\epsilon$} & \textbf{PA}  & d$_{25}$  & \textbf{$g_T$}   & \textbf{$r_T$}  & \textbf{$NUV_T$} & \textbf{E(B-V)}  & \textbf{A$_r$(SDSS)} \\
\textbf{}          &                              &                                & [deg]            & [arcsec]   &[ABmag]   & [ABmag]           & [ABmag]                 & [ABmag]            & [ABmag]\\
\hline
 \rowcolor[gray]{0.8}  264 &           &0.26$\pm$0.08 & 95.6   & 38.83 & 14.81& 14.09 & 19.30$\pm$ 0.03 & 0.033 & 0.077\\
                                 378  & IC 569&0.28$\pm$0.13 & 164.5& 47.66  &14.99 & 14.13 & 20.67$\pm$0.16  & 0.022 & 0.052 \\
 \rowcolor[gray]{0.8} 412 &            &0.03$\pm$0.06 & \dots & 25.01  &15.55 & 14.72 & 20.21$\pm$0.12  & 0.016 & 0.037 \\
                                 481 &NGC 3682&0.37$\pm$0.12& 93.3 & 134.01 & 12.66 & 11.94&  15.77$\pm$ 0.02  & 0.080 & 0.020 \\
 \rowcolor[gray]{0.8} 490  &           &0.29$\pm$0.09 & 36.3  & 43.47 &14.60   & 13.83 & 19.52$\pm$0.15  & 0.024 & 0.056 \\
                                517     &           & 0.34$\pm$0.10  & 65.9   & 37.68 & 14.98 &14.13 & 20.17$\pm$0.21 & 0.018& 0.042\\
 \rowcolor[gray]{0.8} 578 & IC 862  & 0.00$\pm$0.09 & \dots & 23.29  & 15.13 & 14.26  & 19.78$\pm$0.15  & 0.019 & 0.051 \\
                               595 & UGC 8649 & 0.44$\pm$0.11&50.8 & 93.57 &14.29  & 13.45  & 19.41$\pm$0.11  & 0.013& 0.034\\
 \rowcolor[gray]{0.8} 599  &          & 0.13$\pm$0.07 &94.6 & 64.14 & 14.14 & 13.37  & 19.31$\pm$0.13 & 0.010& 0.027\\
                                620 &           & 0.48$\pm$0.19 & 4.4 & 35.41 & 14.91   & 14.15   &  18.14$\pm$0.04 & 0.006 & 0.015\\
 \rowcolor[gray]{0.8} 636  &        & 0.17$\pm$0.07 & 116.8 & 35.25 & 15.03  & 14.19 & 20.14$\pm$0.06 & 0.08& 0.020\\
                                637 & NGC 5687 & 0.43$\pm$0.19 & 103.3  & 144.59 & 12.98     & 12.19  & 17.20$\pm$0.01 & 0.010& 0.026 \\
 \rowcolor[gray]{0.8} 644  &             & 0.59$\pm$0.25  & 85.9  & 55.23 & 14.77    & 13.95    & 19.74$\pm$0.10 & 0.017& 0.044\\
                                670 &              & 0.41$\pm$0.14 & 148.0  & 62.83 & 14.59    & 13.78  & 20.57$\pm$0.21 & 0.020& 0.051\\
 \rowcolor[gray]{0.8} 685  &            & 0.14$\pm$0.08 & 110.6 & 39.28 & 15.41 & 14.56  & 20.87$\pm$0.18 & 0.009& 0.024\\
                                705 & I Zw 126 & 0.03$\pm$0.05 & \dots & 36.66 & 14.57   & 13.81  & 18.38$\pm$0.06 & 0.017& 0.044\\
 \rowcolor[gray]{0.8} 722  &             & 0.27$\pm$0.14 & 94.2  & 63.26  & 14.15   & 13.29   & \dots & \dots& \dots\\
                                732  & IC 1211& 0.07$\pm$0.05 & \dots  &80.38 &13.27  &  12.42  & 18.64$\pm$0.08 & 0.020 & 0.053\\
 \rowcolor[gray]{0.8}733  &             & 0.50$\pm$0.18& 165.2  & 53.11 &  14.49  & 13.65  & 19.10$\pm$0.11 &0.086 & 0.223\\
                               841 & NGC 6524  & 0.39$\pm$0.09 & 156.0 & 97.53 & 13.36   & 12.52   & 16.78$\pm$0.02 & 0.035 & 0.090\\
\hline
\label{tab-2}
\end{tabular}
\tablefoot{Ellipticity, position angle (measured NE), and the isophotal diameter at $\mu_B=25$ mag~arcsec$^{-2}$ 
are from \citet{Jones2018} (col. 3, 4 and 5). Total $g$, $r$   and 
%{\it GALEX} near-UV ($NUV$ hereafter)
 %\LEt{the notes to a table are not the correct place to introduce an abbreviation. Please move this introduction to the first mention of NUV in the main text and remove it here} 
 NUV integrated magnitudes are reported from {\tt NED}.  
 The errors reported in {\tt NED} for SDSS magnitudes  (CModel) are of about 0.002-0.003 mag. }
\end{table*}
%\end{sidewaystable*}
%------------------------- end Table 2 -----------------------------------

The study of the evolutionary scenario that gives rise to iETGs 
is of particular interest because their environment is so unusual for this class of galaxy.
ETGs are thought to be the final product of the galaxy evolution, where different
phenomena, from secular to accretion driven, can play a role
\citep{Cappellari2016}. Simulations suggest that
interaction, accretion and merging episodes \citep[see  ][and
references therein]{Eliche2018,Mazzei2019} leave their signatures on
galaxies, from the nuclear cuspy versus core shape
\citep[][]{Lauer1991, Lauer1992,Lauer2012,Cote2006,Turner2012} to their
outskirts, where tails, streams, fans, and shells may be found
\citep[][]{Arp1966,Malin1983,Prieur1988,Wilkinson2000,Duc2015}. Shells 
are often found in ETGs. They are described as 
 interleaved stellar tidal debris with large 
opening angles and low surface brightness that are often situated 
on either side of the galaxy centre and have
regular as well as irregular shapes \citep{Dupraz1987,Weil1993,Pop2018,Mancillas2019}.
These signatures have different lifetimes and originate from different mechanisms.   
Recently, \citet[][]{ Mancillas2019} pointed out that tails emerge from the
primary galaxy and streams are generated from the secondary galaxy. Tails
have shorter lifetimes (2 Gyr) than shells. Shells  are generated  by
minor and major either wet or dry mergers and are long lasting (3 Gyr)\citep[see
also][]{Dupraz1987,Weil1993,Longhetti1999}. Streams remains visible 
in all phases of galaxy evolution.

For iETGs, the detection of these fine structures may help to understand  
the time that elapsed from their last interaction, accretion, and merging episode. In
general, ETGs in low-density environments are found to be 
rejuvenated  in their centre as well as in their outskirts, as shown by the
{\it Galaxy Evolution Explorer (GALEX)} \citep{Martin2005}, 
suggesting that some form of activity might be induced by interaction, accretion, 
and merging episodes  \citep[][and references
therein]{Clemens2006,Clemens2009,Shawinski2007,Rampazzo2007,
Marino2011a,Marino2011b,Mazzei14a,Mazzei14b,Mapelli2015,Hagen2016,Mazzei2019}. 

The SDSS  has widely contributed  to the
investigation of iETGs. Many studies have been dedicated to a visual
re-classification of the galaxies in the \citet{Karachentseva1973}
catalog and/or revisions of it
\citep{Sulentic2006,Fernandez2012,Buta2019}. \citet{Hernandez2008} (H-T
hereafter) performed a study (using DR6) of the 579 galaxies in the
\citet{Karachentseva1973} sample, including iETGs (58 E, 14 E/S0, 67 S0,
and 19 S0/a). They revised the  ETG classification on the basis of the
geometric profiles obtained by $r$ -band images using {\tt IRAF}
\citep{Jedrzejewski1987}. This process classified only 18 galaxies 
(3.5\% of the entire sample)  as bona fide E. They are  listed in their
Table~6. The authors report a 50\% incidence of ripples
and shells in these Es (9 out of 18),
%\LEt{this should be "9 out of 18", please change to that also for the following instances}, 
see Table~6 of the paper, although 4 are
uncertain detections) as well as a high fraction of Es with diffuse halo
(7 out of 18).  Recently, \citet[][]{Rampazzo2019} found interaction and merging
signatures in a high spatial resolution study in K band, performed with
{\tt ARGOS+LUCI} at the Large Binocular Telescope (LBT) \citep{Rabien2019}, of KIG 685 and KIG 895. 
This latter has been found to be a misclassified spiral with a tail. 
KIG 685 is a bona fide ETG with shells that would suggest
an eventful life like that of ETGs in loose galaxy associations, 
where shells, for instance, are found with a higher frequency than 
in clusters \citep[see e.g.][]{Malin1983,Reduzzi1996}. 

The above studies emphasised two aspects. The first aspect is the lack of a
quantitative classification of the morphology of isolated galaxies, in 
particular, of iETGs whose classification is still very uncertain (cf.
\S~\ref{Sample}). For example, H-T found that E+S0 make up 
8.5\% (3.5\%+5\%)   of E+S0+E/S0 (5.7\%+6.6\%+1.4\%), 
while \citet{Sulentic2006} found them to make up 13.7\%.
The other aspect is that we need surface photometric studies
that are deep enough to  reveal the fine structure of iETGs, if any. These
considerations motivated the present study. Using $g$ and $r$ deep
surface photometry, we search for such faint
features in iETGs and refine the previously proposed iETG galaxy morphological 
classification \citep{Sulentic2006,Fernandez2012,Buta2019}
by substantiating the view of H-T  about the rarity of Es versus S0s
in samples of isolated galaxies. This will help understand their evolutionary
mechanisms and the meaning of their current relative isolation. 

The paper is structured as follows. The observed sample is presented in
\S~\ref{Sample}. In \S~\ref{Observations} we present the observations
performed at the Vatican Advanced Technology Telescope (VATT) with the 4K CCD camera. This section also describes the
reduction method we implement that fully exploits the quality of the
observations. Results are presented in
\S~\ref{Results}. In \S~\ref{Discussion} we discuss our results  to
understand the  nature and evolutionary paths of our iETGs.

%-------------------------------- Figure 1 ------------------------
\begin{figure}
\centering
\includegraphics[width=8.4cm]{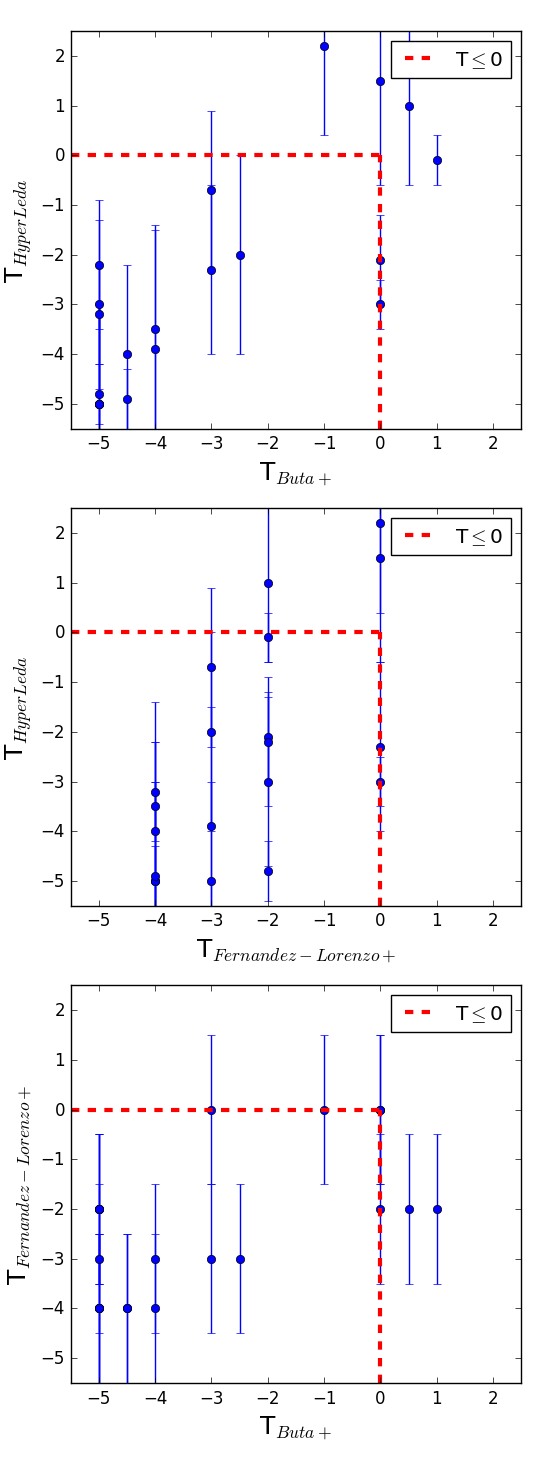}
\caption{Comparison between the three classifications into morphological type provided
in Table~\ref{tab-1}. The dashed red rectangle encloses iETGs
(T$\leq$0) in the RC3 classification. Comparisons  are made between {\tt
HyperLeda} vs. \citet{Buta2019}({\it top panel}),  {\tt HyperLeda} vs.
\citet{Fernandez2012} ({\it middle panel}), and \citet{Buta2019} vs.
\citet{Fernandez2012} ({\it bottom panel}).
%\LEt{the figure caption should provide only the information necessary to understand the figure itself. Any interpretation should be moved to the main text. To my understanding, the following sentences do not belong in the caption. Please remove} 
%On average, \citet{Buta2019} provided an {\it earlier} type ($\Delta$ T=-0.5) for galaxies
%than  {\tt HyperLeda} and \citet{Fernandez2012}. {\tt
%HyperLeda} and \citet{Fernandez2012} differ by $\Delta$T=-0.05 on average.
%Single cases may differ significantly in the assigned type, as can be
%deduced from the panels and Table~\ref{tab-1}. The \citet{Buta2019}
%classification also provides a description of the galaxy substructures
%that are only partially provided by the {\tt HyperLeda} classification.
}
\label{figure-1}
 \end{figure}
%--------------------------end figure 1 -----------------------

%-------------------------------- Figure 2 ------------------------
\begin{figure}
\center
\includegraphics[width=8.9cm]{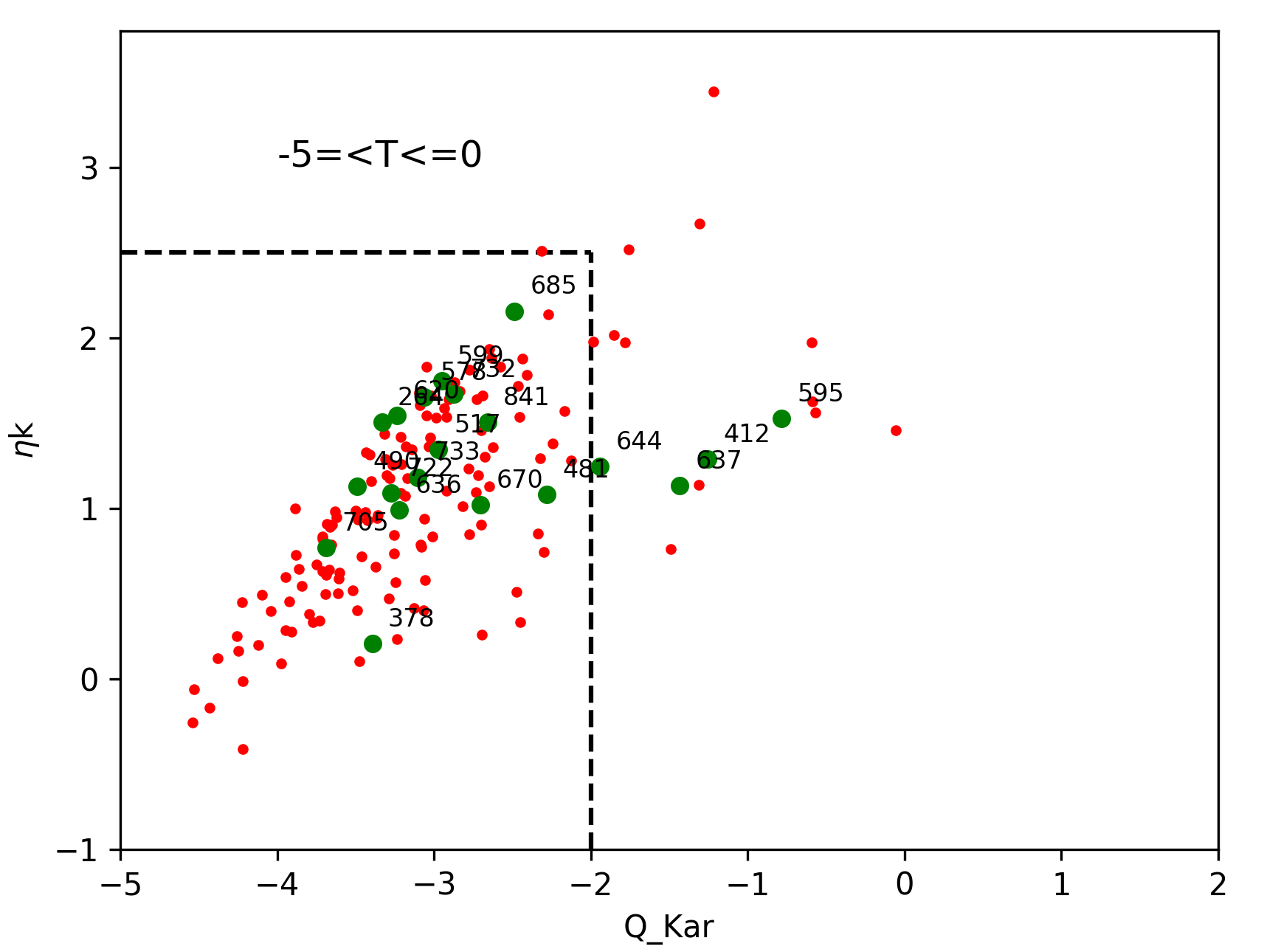}
\caption{Degree of isolation of the observed sample (green circles)
over-plotted on all iETGs  (red circles) in the  AMIGA sample.
According to \citet {Verley2007b}, the horizontal and vertical 
dashed lines enclose the fiducial range of isolated galaxies in the AMIGA 
sample.}
\label{figure-2}
 \end{figure}
%--------------------------end figure 2 -----------------------

\section{The sample}
\label{Sample}

The sample set is composed of 20 iETGs  that are included in the 
{\it Analysis of the interstellar Medium of Isolated Galaxies} sample
%\LEt{please change this so that the spelled-out version is part of the main text and the abbreviation is tinroduced in parentheses}
(AMIGA, \citet{Verdes-Montenegro2005}), which is a revision of the 1973 catalogue of
\citet{Karachentseva1973}. The analysis of the AMIGA sample revealed a
set of galaxies that should not have interacted for at least 3 Gyr
\citep{Verdes-Montenegro2005,Verley2007a,Verley2007b}.

In addition to the morphological classification, we selected the sample 
by considering the galaxy observability (see
\S~\ref{Observations}). We selected galaxies using the classification of
\citet{Fernandez2012}, which is an updated version  of the \citet{Sulentic2006}
classification. For comparison, we also considered the classification
provided by {\tt HyperLeda} and the recent morphological classification
by \citet{Buta2019}. We collect in Table~\ref{tab-1} and
Table~\ref{tab-2} the classifications and the main galaxy
characteristics.  In the results section (\S~\ref{Results}) we also consider
the classification of H-T, which is based on the geometrical profiles of the galaxies.

The differences among the classifications are summarised in
Figure~\ref{figure-1}. The morphological type distribution, provided by
\citet{Fernandez2012}, ranges from -5$\leq T \leq$ 0, that is, it covers the
entire range of iETGs, and contains a large portion of galaxies
with a disc component (-3 $< T < $0). However, we note that the galaxy
morphological classification is often (largely) discordant among these
authors,  as shown in Table~\ref{tab-1} and Figure~\ref{figure-1}. 
Some galaxies, that is, KIG 481 and KIG 620, are late type (T$>$0) according to
\citet{Buta2019}, and  KIG 620, KIG 644, and KIG 733, which are all seen edge-on, are
classified as late type by {\tt HyperLeda}.

\citet{Buta2019}, who most recently classified our iETGs visually,
on average tend to provide earlier $\Delta$T=-0.5 morphological type than
both {\tt HyperLeda} and \citet{Fernandez2012}. These latter
classifications differ by $\Delta$T=0.05 on average, but single cases may
differ more significantly, as shown in the middle panel of Figure~\ref{figure-1}.
We here  quantitatively verify the classification using surface photometry, 
for example, by determining the presence of a disc. 
Out of 20 ETGs, 5 show peculiarities ({pec}) according to the
\citet{Buta2019} classification: KIG 481, KIG 490, KIG 685, KIG
733, and KIG 841.

The heliocentric velocity of only two objects is lower than 3000
km~s$^{-1}$. These are KIG 481 and KIG 637. The average heliocentric velocity is
V$_{hel}$=8535 km~s$^{-1}$ (Table~\ref{tab-1}). 
Distances are taken from \citet{Jones2018}, who analysed\ the
{\it AMIGA} sample  in \Hi. They used the \citet{Mould2000} model
for distances. This model corrects for Local Group motion and uses separate attractor 
velocity fields for the Virgo cluster, the Shapley supercluster, and the Great Attractor.
The distance is finally obtained by adopting H$_0$=70 km s$^{-1}$ Mpc$^{-1}$.
The relevant photometric properties of iETGs are collected in
Table~\ref{tab-2}. The  $g$ and $r$  photometric data sets are taken from 
the SLOAN (CModel)  values reported in {\tt NED}. 
The {\it GALEX} near-UV ($NUV$ hereafter) integrated magnitudes are also 
taken from {\tt NED}.

\subsection{Degree of isolation in the sample}
\label{isolation} 
 
The environment of the AMIGA galaxies was investigated to
avoid similar size companions. Different criteria were developed
 to define  the isolation parameters 
 \citep{Verley2007b,Argudo2013,Argudo2015}.
 Figure~\ref{figure-2} shows the isolation  parameters of the 114 iETGs 
 in the  AMIGA sample classified  by \citet{Fernandez2012},
 devised by \citet{Verley2007b}. \citet{Verley2007b} calculated 
 %\LEt{please repeat for clarity whom you refer to here with "they": Frenandez-Lorenzo et al. or Verley et al.?} 
the $\eta_K$ and $Q_{Kar}$ parameters by defining the local galaxy
number density and the distribution of the tidal strength, respectively.
According to the \citet{Verley2007b} parameters, the four galaxies KIG 412, KIG 595, KIG 644, and KIG 637
 are located outside the fiducial range in the $Q$ versus $\eta_k$ plane for
isolated galaxies (dashed horizontal and vertical lines in Figure~\ref{figure-2}).  

More recently, \citet{Argudo2013} further revised the isolation criteria of the 
AMIGA galaxies using both photometric and spectroscopic data from SDSS 
to refine the $\eta_K$ and $Q_{Kar}$ parameters. Of the iETGs in this work, 
\citet{Argudo2013} found that only KIG 599 is considered isolated when 
the original isolation criteria of \citet{Karachentseva1973} are applied to 
SDSS images. However, most of the iETGs pass the criteria for the 
(photometric) tidal interaction and neighbour density parameters 
recommended by \citet{Argudo2013}, $Q_{Kar} < -2$  and $\eta_K < 2.7$, 
meaning that they are assumed to be minimally affected by any neighbours. 
The exceptions are KIG 517, KIG 644, KIG 722, and KIG 733, all of which 
violate the neighbour density criterion, but not the tidal interaction criterion. 
In their spectroscopic analysis, \citet{Argudo2013} found that about half of 
the iETG sample do fulfil the Karachentseva isolation criteria when neighbours 
separated by more than 500 km~sec$^{-1}$ from the target (in redshift) are removed. 
This includes KIG 644 and KIG 733, which did not pass the photometric criteria. 
The galaxies that fail are KIG 264, KIG 517, KIG 595, KIG 620, KIG 637, and KIG 722. 
There are no measurements (photometric or spectroscopic) for KIG 481, KIG 670, 
or KIG 732, and KIG 841 lacks spectroscopic measurements. In the cases of 
KIG 412 and KIG 644, \citet{Argudo2013} is at odds with \citet{Verley2007b} 
because they did not meet the previous criteria.
%We conclude that half of our iETGs are {\it strictly} isolated according to the 
%\citet{Argudo2013} spectroscopic criterium, while \citet{Verley2007b}
%16 iETGs in our sample are isolated. Although located in low density environment,
%companions, most of them dwarfs, {\it contaminate} the environment of the remaining
%galaxies, at different level.  
When they used the spectroscopic data, \citet{Argudo2013} found that approximately 
70\% of the AMIGA galaxies meet the Karachentseva criteria for isolation. For 
these iETGs, this number is 63$\pm$10\% (the error is due to galaxies without data). 
About half  of the iETGs do not pass the strictest test, but are still expected to be only 
minimally affected by any neighbours (except possibly KIG 595, those with missing data, 
and the two conflicts with \citet{Verley2007b}).

We conclude that the our iETG sample is consistent with the level of
 isolation in the KIG population as a whole, and they mostly fulfil the $Q$ 
 and $\eta$ criteria for isolation. We consider the galaxy sets of \citet{Verley2007b} and  
 \citet{Argudo2013}, which differ in their isolation criteria, in the discussion.

\section{Observations, data reduction, and analysis}
\label{Observations}

\subsection{Observations}
Observations have been performed during  a single run from April 9 to April 15, 2018, 
at the 1.8m VATT. Photometric
conditions varied greatly: the night of 12 April was lost, as were parts of 
some other nights. We used the back-illuminated 4K CCD (STA0500A), 
which was installed in March 2017  and was updated on October 8, 2017. 
The full well is about 117000 e$^{-}$, limited by the ADC (16 bit, 65536 DN). The
readout noise is 3.9 e$^-$ rms and the gain is 1.8 e$^{-}$ADU$^{-1}$.
The number of pixels is 4096$\times$4096 (15$\times$15 microns) for a
total field of view (FOV) of 12\farcmin5 arcmin square with a pixel scale of
0\farcs188/px. 
%The wavelength range is 300 - 1000 nm, peak 96\% at 450 nm. 
The galaxies were observed in the $g$ and $r$ SDSS filters
with a binning of 2$\times$2 (0\farcs376/px) of the images. The observation log 
is reported in Table~\ref{tab-3}.

%------------------------- begin Table 3 -----------------------------------
\begin{table}
\caption{Observation log}
\centering
%% \tablesize{} %% You can specify the fontsize here, e.g.  \tablesize{\footnotesize}. 
%If commented out \small will be used.
\begin{tabular}{cccccc}
\hline
\textbf{KIG}    & \textbf{Exp. time}    & \textbf{Date} & \textbf{Filter} & \textbf{FWHM} & \textbf{ZP}\\
%                            & [s]      &  [2018, 04]  &  SDSS        & [arcsec]   & mag\\
                             & [s]      &    &  SDSS        & [arcsec]   & mag\\
\hline
 \rowcolor[gray]{0.8}264        & 1800  & 9 & $g$  & 1\farcs2 & 22.83$\pm$0.03\\
 \rowcolor[gray]{0.8}        &  660  & 10 & $r$ &   1\farcs1 &23.18$\pm$0.05 \\
378     & 600   & 15 & $g$ &  1\farcs3 & 21.81$\pm$0.02\\
        & 540   & 15 & $r$ &   1\farcs0 &21.94$\pm$0.02\\
 \rowcolor[gray]{0.8}412        & 1500  & 11  & $g$ & 1\farcs3 & 23.01$\pm$0.03\\
 \rowcolor[gray]{0.8}        & 2340     & 13 & $r$ & 1\farcs0 & 21.64$\pm$0.04\\
481     & 2100  & 13& $g$      &  2\farcs0 &22.62$\pm$0.04\\
        & 1080          & 13  & $r$ &  2\farcs0 &22.44$\pm$0.02\\
 \rowcolor[gray]{0.8}490        & 1500  & 14 & $g$      & 1\farcs3 & 22.98$\pm$0.05\\
 \rowcolor[gray]{0.8}        & 1260     & 14 & $r$ & 1\farcs0  & 22.27$\pm$0.01\\
517     & 2700  & 9, 15& $g$      &  1\farcs2 &22.27$\pm$0.12\\
        & 2340  & 9, 15 & $r$ &  1\farcs0 & 21.64$\pm$0.09\\
 \rowcolor[gray]{0.8}578        & 2400  & 15  &$g$      & 1\farcs2 & 22.56$\pm$0.11\\
 \rowcolor[gray]{0.8}        & 1440     & 15 & $r$ &   1\farcs2 & 22.15$\pm$0.01\\
595     & 3000  & 11 & $g$  &  1\farcs0 & 22.65$\pm$0.05\\
        & 1800  & 11 & $r$ &       1\farcs1 &22.46$\pm$0.06\\
 \rowcolor[gray]{0.8}599        & 1800  &9   & $g$          & 1\farcs6 &22.69$\pm$0.06\\
 \rowcolor[gray]{0.8}        & 1650     & 9  & $r$ &  1\farcs1 &22.46$\pm$0.06\\
620     & 2400  & 13 & $g$   & 2\farcs0 & 22.44$\pm$0.02\\
        & 1440  & 13  & $r$ &    1\farcs8 &22.09$\pm$0.01\\
 \rowcolor[gray]{0.8}636            & 2400      & 15 & $g$      & 1\farcs1 & 22.50$\pm$0.04\\
 \rowcolor[gray]{0.8}        & 1440  & 15 & $r$ & 1\farcs4 &22.17$\pm$0.02\\
637     & 1800  & 14 & $g$      & 1\farcs2 & 22.78$\pm$0.01\\
        & 1440  & 14 & $r$ &   1\farcs4 &22.11$\pm$0.01\\
 \rowcolor[gray]{0.8}644        & 1800  & 10 & $g$      &  1\farcs3 &22.85$\pm$0.03\\
 \rowcolor[gray]{0.8}        & 1800     & 10 & $r$ &   1\farcs3 &21.92$\pm$0.02\\
670     & 2100  & 13 & $g$   & 1\farcs8 &22.62$\pm$0.04\\
        & 1800  & 13 & $r$ & 1\farcs6 &22.45$\pm$0.01\\
 \rowcolor[gray]{0.8}685        & 2400  & 14 & $g$      & 0\farcs9 & 22.27$\pm$0.03\\
 \rowcolor[gray]{0.8}        & 1440     & 14 & $r$ &   1\farcs2 &22.12$\pm$0.04\\
705     & 1510  & 9 & $g$   &   1\farcs3 & 22.50$\pm$0.05\\
        & 1200  & 9 & $r$ &  1\farcs0 & 21.87$\pm$0.02\\
 \rowcolor[gray]{0.8}722        & 1800     & 13 & $g$  &   2\farcs6  &22.74$\pm$0.07\\
 \rowcolor[gray]{0.8}        & 900      & 14 & $r$ &   0\farcs8 &22.65$\pm$0.04\\
732     & 3150  & 11 & $g$      & 1\farcs5 &22.41$\pm$0.02\\
        & 1800  & 11 & $r$ &   1\farcs3 &21.80$\pm$0.01\\
 \rowcolor[gray]{0.8}733        & 3000  & 15 & $g$      &  1\farcs2 & 22.24$\pm$0.06\\
 \rowcolor[gray]{0.8}        & 1800     & 15 & $r$ &   1\farcs1 &21.68$\pm$0.04\\
841     & 1500  & 10& $g$  & 1\farcsec0 & 23.04$\pm$0.01\\
        & 1800          & 10& $r$ &   1\farcs1 &21.92$\pm$0.03\\
\hline
\end{tabular}
\label{tab-3}
\tablefoot{We report the total exposure time in Col.2. Galaxies were
observed in April 2018: the observing date is given in column 3. The
filter and average Gaussian FWHM of stars nearby the galaxy, measured on the co-added images
using the {\tt IRAF} task {\tt IMEXAMINE}, are reported in columns 4 and 5
respectively (see \S~\ref{Observations}). The adopted ZP
%\LEt{again, please move this introduction of ZP to the first time it appears in the main text} 
are provided in col. 6. 
%The run average zero points are
%ZP$_g$=22.60$\pm$0.30 and ZP$_r$=22.18$\pm$0.39. 
}
\end{table}
%--------------------------------- end Table 3 -------------------------------

In each band, short multiple exposures of single objects
were obtained in order to avoid the saturation 
of the galaxy centre and strong stellar ghosts in the field, and to properly remove cosmic rays. 
Column~2 reports the total integration time.
A pipeline was developed for bias and dark subtraction to
create a master flat in each band for image correction 
before the final images were registered and co-added.  While the
bias and dark corrections are standard, the flat-fielding was performed 
by preparing an {\tt autoflat} procedure. 
It consists of an accurate masking of all sources in the images, which were stacked after normalisation. For the masking we used \texttt{Noisechisel}
\citep{Akhlaghi2015}. This final co-added image is the master flat 
that we applied to single images in each band.  
After bias, dark, and flat-field corrections, the images were finally registered 
for astrometry and co-added  using {\tt SWarp} \citep{Bertin2002}.
Unfortunately, the flat fielding is not 
able to completely eliminate certain artefacts, such as dust spots 
on the filters, because the filter wheel is instable and the filters 
were frequently changed between exposures. This is identified and 
masked in the image analysis.

\begin{figure}
\center
\includegraphics[width=7.7cm]{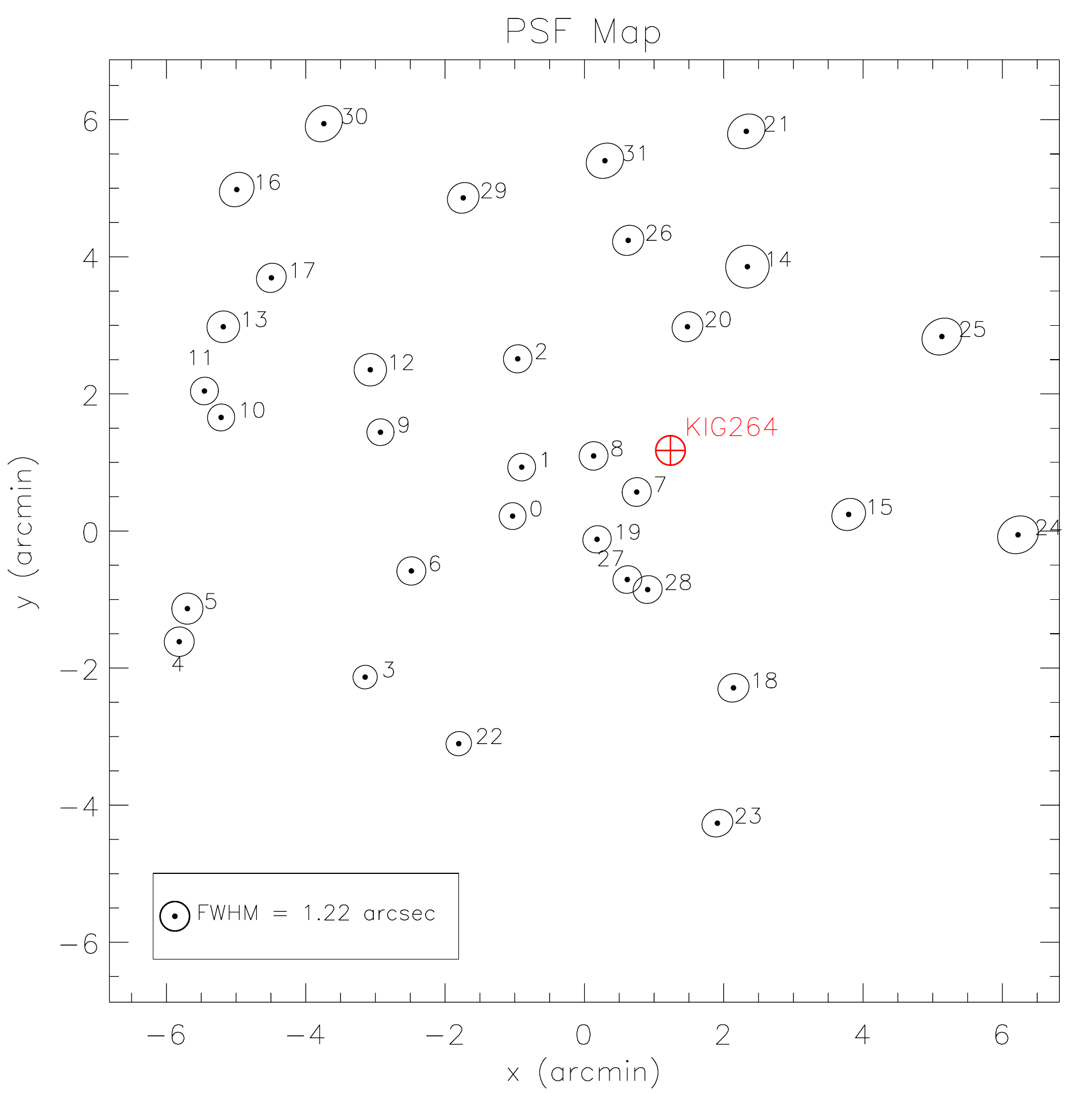}
\includegraphics[width=7.7cm]{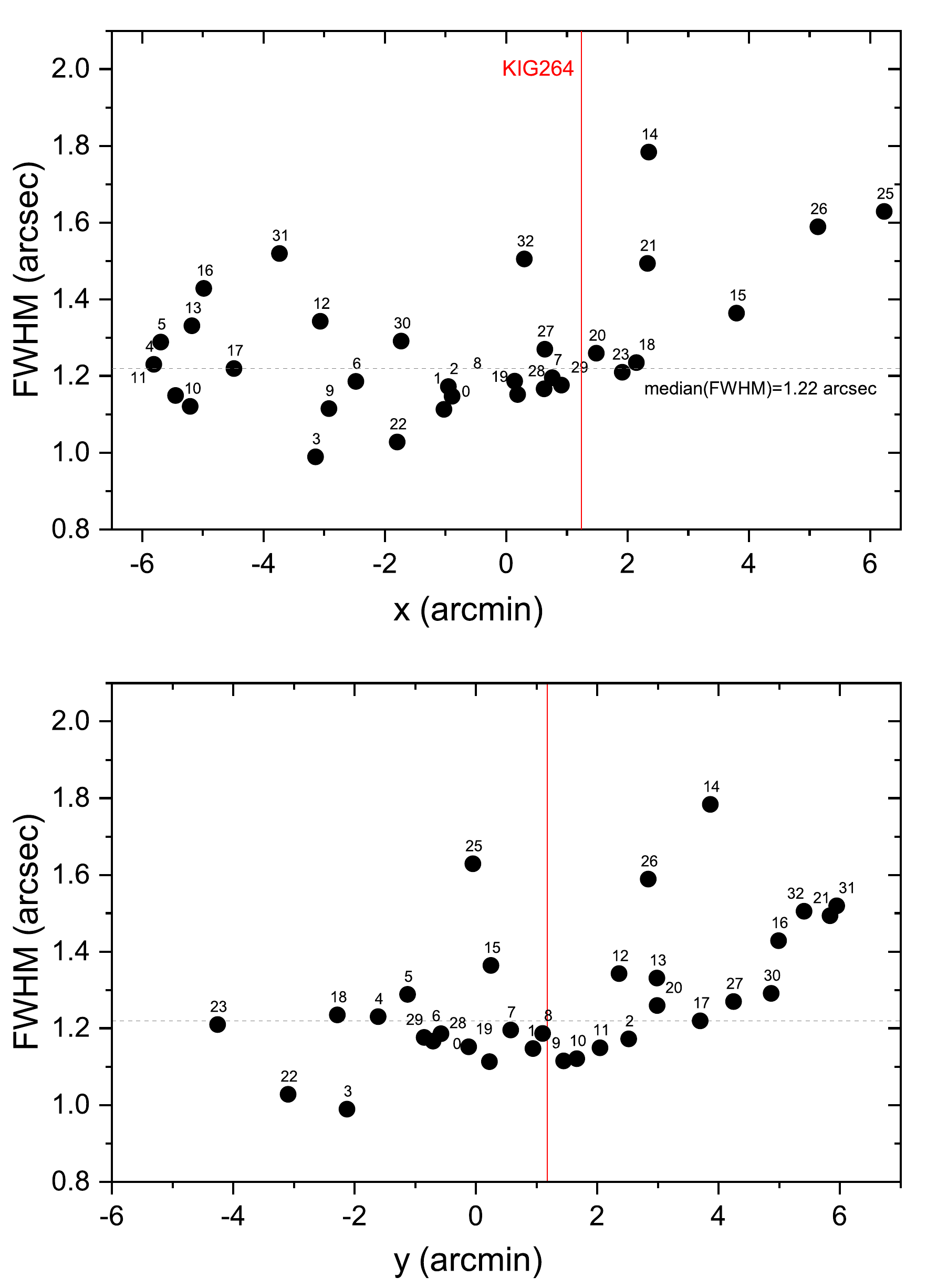}
\caption{{\it Top panel.} 2D map of the PSF FWHM variation across the $g$ 
-band FOV of KIG 264. 
%The semi-major axis and the position angle 
%of the plotted ellipses are proportional to the FWHM and provide the 
%direction of its elongation. 
{\it Middle panel.} The FWHM measured
along the x-axis (R.A.) and along the y-axis (Decl., bottom panel) of this
frame. The vertical red line marks the position of KIG 264.}
\label{figure-3}
 \end{figure}

For each image, column 5 in Table~\ref{tab-3} lists  the  average
Gaussian stellar full width at half-maximum (FWHM) 
measured with the {\tt IRAF} task {\tt IMEXAMINE}.
The measure was performed on stars nearby the galaxy. Figure~\ref{figure-3} 
shows the 2D map of the FWHM variation across the FOV
%\LEt{you introduced an abbreviation for this a few lines above. ÜPPlease decide whether you wish to use this or not and then stick to it throughout for consistency. This includes figure captions} 
of KIG 264. There is optical distortion towards the outskirts of the CCD fields.
However, the plot shows no significant geometric distortions where 
the galaxy is located. One direct comparison can be made using KIG 685, which was observed
by \citet[][see their Table~3]{Rampazzo2019} with adaptive optics at the LBT.
The ellipticity measured by these authors in K band is 0.16$\pm$0.05, which is well
comparable within the errors with the values of 0.18$\pm$0.05 for the bulge that
dominates the light profile (B/T=0.74) in both bands.
This observing position on the CCD was maintained for all
galaxies during the run. A few nearby very extended galaxies fell into the  gap
between the two CCDs.

%-------------------------------- Figure 3 ---------------------
%--------------------------end figure 3 -----------------------
%-------------------------------- Figure 4 ---------------------
\begin{figure*}
\center
\includegraphics[width=5.4cm]{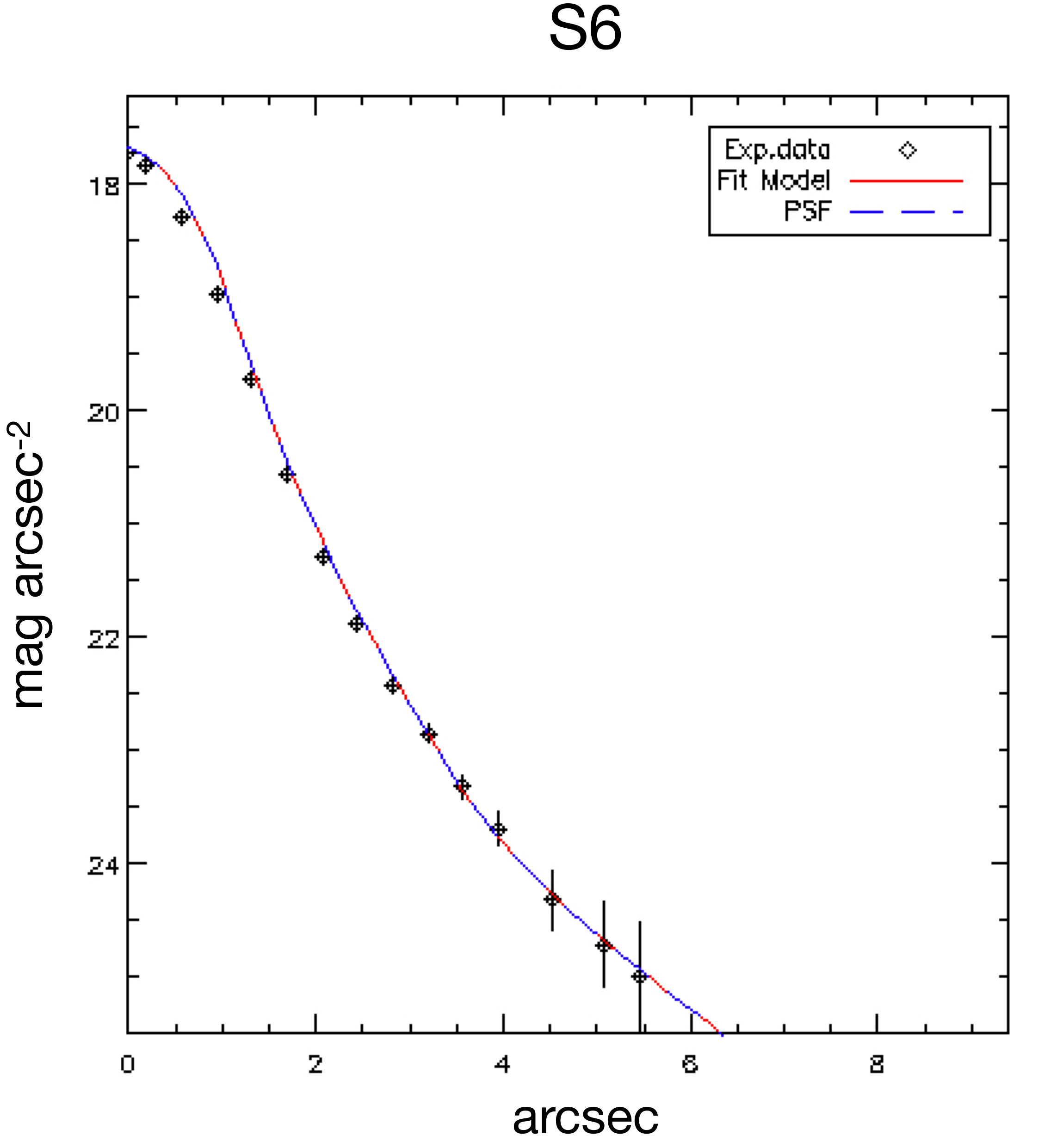}
\includegraphics[width=5.7cm]{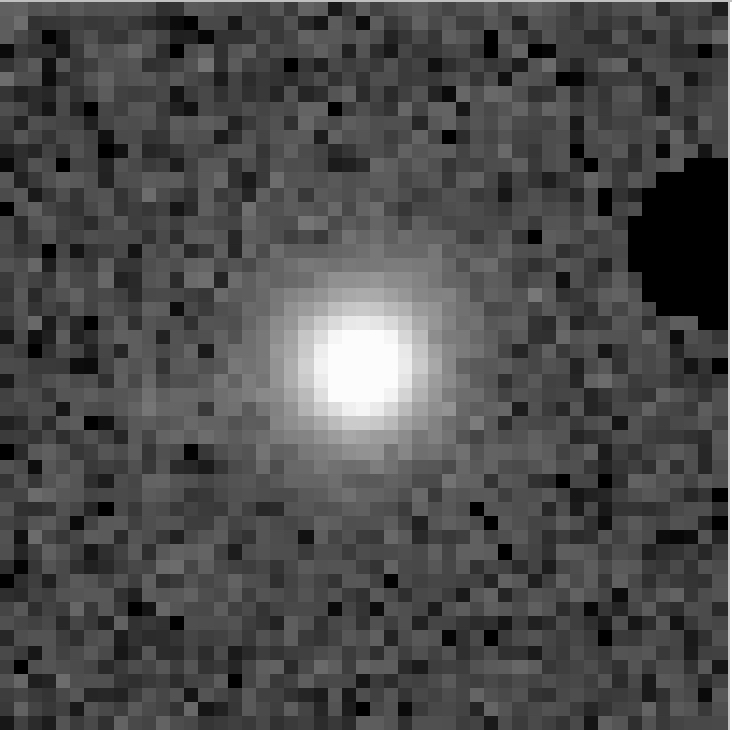}
\includegraphics[width=5.7cm]{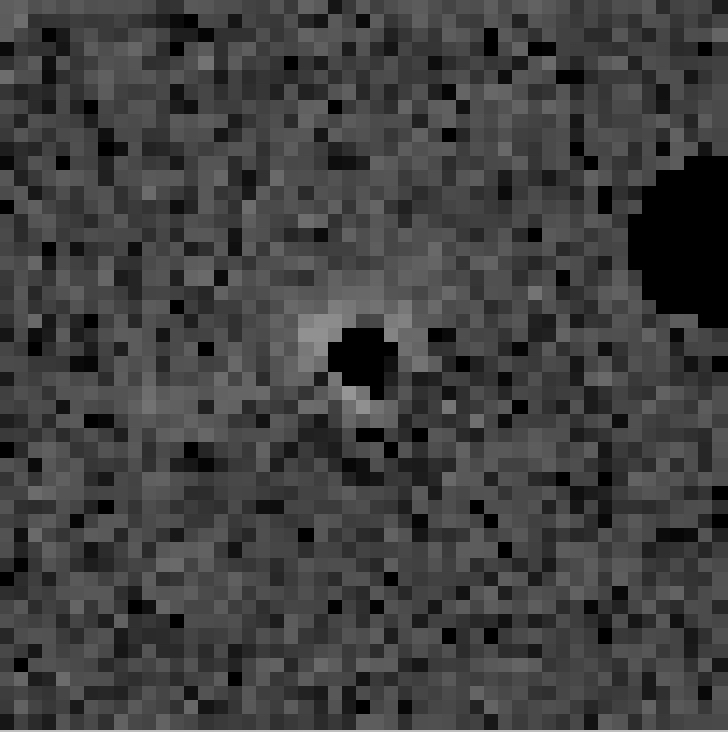}
\caption{PSF in KIG 264 in the $g$ band. 
The 2D PSF is best fitted with a composite model (solid line) combining three Gaussians
 plus an exponential. The adopted PSF has been generated from a set of stars 
 (star 6 is shown) close to the galaxy. The seeing 
 correction is extrapolated up to the galaxy outskirts 
 \citep[see the discussion in][and references therein]{Rampazzo2019}. A masked star
 is visible on the NW side of the field.}
\label{figure-4}
 \end{figure*}
%--------------------------end figure 4 -----------------------

\subsection{Data reduction}

The photometric calibration, the point spread function (PSF) study 
and the surface photometric analysis 
were performed using {\it Astronomical Image Decomposition
and Analysis}
%\LEt{again, please switch this } 
 ({\tt AIDA}) by \cite{Uslenghi2011}. This software package, written in {\tt IDL} 
(Image Display Language), was originally designed to analyse images of galaxies 
with a bright nucleus and to decompose them into the nuclear 
and galaxy components. {\tt AIDA} includes tools for PSF characterisation.

Because of the photometric instability and owing to the full-sky coverage of the 
SDSS survey \citep{York2000}, we corrected the calibration 
of our $g-$ and $r$ -band images by comparing the photometry of stars in each 
field with their SDSS magnitudes.
Using the SDSS navigator tool, we inspected each target field and identified 
a set of stars nearby the galaxy. We bootstrapped our instrumental magnitude of these isolated
unsaturated stars to their corresponding $g$ and
$r$ SDSS magnitude in the star catalogue. The instrumental stellar magnitude
in {\tt AIDA} was calculated within different apertures 
in order to include the entire  star, while the nearby residual sky was calculated
in a ring cleaned from possible faint nearby stars and objects. 
The adopted zero-points (ZPs hereafter) and their errors, reported in column 6 of  
Table~\ref{tab-3}, are the average values and the standard deviation 
of the scatter in magnitude of the set of stars used.

The PSF shape was obtained from the same set of stars for each $g$ and
$r$ field.  {\tt AIDA} characterises the PSF using 2D models (both analytical and
empirical, or a combination of them), even when they are variable in the FOV. 
PSF models can be provided by the user or
modelled by {\tt AIDA} itself using reference stars in the images.
Figure~\ref{figure-4} shows one of the stars we used to obtain 
the PSF  in the analysis of KIG 264. 
Although characterisations of PSF models exist 
(three Gaussians plus an exponential) up to 
large radii \citep[several arcminutes;][]{Slater2009,Sandin2014,Trujillo2016,
Karabal2017,Infante2020} , the relatively small extent of our targets means that 
the analytical PSF model that is adapted is indeed extended to the galaxy outskirts 
to obtain seeing-corrected parameters \citep{Sandin2015}.

For target galaxies, as in the case of stars, the residual sky background was determined
within a ring of variable radius, well outside the galaxy and its peculiar outskirts,
masking sources within it. Only a few very extended galaxies
are crossed by the CCD gap. This latter was masked during the reduction
and analysis procedure. The sky background value was obtained as the
average of sectors of the ring, and the errors are the standard deviation
of averages. 
%\LEt{please rephrase this next sentence for grammatical reasons. 
%I don't quite know how to change it for you so that it is correct}
%Computed the average value of the sky background
%{\tt AIDA} provides the radial light profile and measurement errors
%as the average and the standard deviation obtained from integrating
%the light within concentric rings centred on the galaxy. The
%uncertainty on the sky background determination was added in
%quadrature to the above errors.
{\tt AIDA} also provides the radial light profile, with associated measurement 
errors, of the objects. The radial profile is computed by averaging
the signal within concentric rings centred on the galaxy. The standard deviation
 in the rings is then used to evaluate the error bars, after adding in quadrature 
 the uncertainty on the sky background.

In the 2D approach (see also {\tt GALFIT} by \citet{Peng2010} or {\tt
IMFIT} by \citet{Erwing2015}), all pixels, except for the masked pixels,
contribute to the fitting process of the galaxy luminosity distribution,
but suffer from the fact that the components have single fixed values for
the ellipticity, position angle, and Fourier moments. The information
about the variation with radius of  the ellipticity, position angle,
and isophotal shape parameters obtained from the analysis of the
galaxy azimuthal luminosity profile \citep[see e.g.][]{Jedrzejewski1987}
is lost. In modelling the galaxy light profile, {\tt AIDA} adopts two values
of the ellipticity and position angle in the case of a bulge plus disc
(B+D) decomposition. Before the analysis of our target galaxies, the masking 
of the foreground and background objects superposed on the target 
galaxy was also performed manually, that is, masks were tailored to 
the extension of the objects that were to be removed. 
The fitting algorithms used by {\tt AIDA} are {\tt MPFIT} \citep{Markwardt2008} and
a modified version of the IDL standard library function {\tt CURVEFIT},
which uses a gradient-expansion algorithm (based on CURFIT
\citep{Bevington1994}).  The original algorithm was modified in
order to allow boundary constraints on the parameters. 

\bigskip
\subsection{Data analysis}

Our scientific aim is twofold: to unveil fine structures, and to 
quantitatively refine the galaxy morphology given in Table~\ref{tab-1}.  The
sample is composed of Es and S0s to about the same portions. However, the
three classifications considered in Table~\ref{tab-1} often differ
significantly, as shown in Figure~\ref{figure-1}. To ascertain the
presence of a disc is very valuable information about the role  of
{\it dissipative processes} in the ETGs evolution. We are aware that
the photometric evidence is, however, a necessary but not sufficient condition
\citep[see e.g.][for an ample discussion]{Meert2015,Costantin2018} 
to determine the presence of a kinematical disc.

We remark that our purpose does not consist of introducing additional
functions to possibly best fit {\it regular} features (e.g. analytic
functions to mimic a bar, rings etc., as e.g. in {\tt GALFIT} or {\tt
IMFIT} approaches). On the one hand, the fine structure we aim to reveal,
such as shells, ripples, and tails, are often if not always irregular. Fine
structures should be visible in the image itself. Fitted models are
intended to enhance fine structures when they are subtracted from the original image.
On the other hand, the visual classification in Table~\ref{tab-1}
reports only few iETGs with a ring or bar, or mixed AB types in the \citet{Buta2019}
classification. In addition, the galaxies that are acknowledged to possess these features
are not necessarily the same in the three classifications presented 
in Table~\ref{tab-1}. Only KIG 264 is considered barred by
\citet{Buta2019}, while KIG 264, KIG 620, and KIG 636 for {\tt HyperLeda}
are  barred S0s. \citet{Buta2019} considered
KIG 620, KIG 636 and KIG 733 as mixed AB type. Some of these galaxies have an inner ring. 
Only KIG 733 for \citet{Buta2019} 
and KIG 599 and KIG 644 for {\tt HyperLeda} possess outer rings. 

To conclude, we are aware that bars and inner rings introduce an erroneous
evaluation of the bulge parameters if they are not considered in the fit.
\citet[see e.g.][]{Meert2015} remarked that the presence of a bar  affects
%\LEt{please check the next part of the sentence, I don't know how to fix this for you. Do you mean "... affects the ftting by changing the ellipticity?" or "... affects the fitting, which changes..."?}
the fitting by changing the ellipticity and S\'ersic index of the bulge
component in their two-component models.  However,  the classifications
severely differ in our case, so that the  presence of bar and rings needs 
to be carefully determined
%\LEt{both? either? which? or do you mean "determined"?} 
by our surface photometry.  Our results are described in
detail in Section~\ref{Individual_notes}.

We used  two models to best fit the galaxy light profile. We best fit a
single S\'ersic law \citep{Sersic1968}, which is intended to represent  
Es \citep[see e.g.][and reference therein]{Ho11,Li2011} and   a classic bulge
\citep{deVauc1953} plus an exponential disc \citep{Freeman1970}, 
labelled B+D hereafter, for disc galaxies. The S\'ersic law exponent 
 was left free to vary up to $n\leq10$ in the single S\'ersic fit 
 (the values $n=1$ and $n=4$, the Freeman and de Vaucouleurs laws, 
 respectively, are special cases).  The selection of the B+D model is justified
by the following consideration that takes  into account that our range 
of morphological type is $-5\leq T \leq 0$ according to
\citet{Fernandez2012}, which  we used for sample selection.
Large surveys based on SDSS, which used automated decomposition algorithms, 
widely debated the biases introduced in the galaxy final 
structural parameters by fitting the light profiles using different models,
mostly in connection with image resolution and galaxies with B/T$\leq$0.5
 \citep[see e.g.][]{Gadotti2009,Simard2011,Meert2015}. 
Our targets are nearby (see Table~\ref{tab-1}), extended, and relatively bright galaxies
 (see Table~\ref{tab-2}). The galaxy light is dominated by the bulge 
 (the average is $\langle B/T \rangle=0.67$),  as discussed in \S~\ref{Results}. 
\citet{Meert2015} showed that  B/T, bulge radius, and bulge S\'ersic index 
all decrease with increasing T-type. The median bulge S\'ersic index in their  fit of a
S\'ersic plus exponential disc for the earliest T-types ($-5\leq T \leq0$)  
is approximately 5.0$\pm$1.5 (see their Figure~23), which is 
fully consistent with our use of a de Vaucouleurs law.  
Their median B/T decreases from 0.8 to 0.5 over the same 
range as in our case (see our \S~\ref{Results}).

The fit quality, as described in \citet{Uslenghi2011},  was estimated by minimising 
the $\chi^2$. In {\tt AIDA,}  the minimised $\chi^2$ is defined as

\begin{equation}
$$ \chi^2=\sum_y\sum_x\frac{(flux_{x,y}-Model)^2}{\sigma_{x,y}^2}Mask_{x.y}.$$
\end{equation}

The masked pixels are excluded from the fit. The weighted model
is computed defining $\sigma_{x,y}$ as the sum in quadrature of three
distinct components with different dependence on the signal level,

\begin{equation}
$$\sigma_{x,y}=\sqrt[2]{C^2+{SY\times(\sqrt[2]{Flux_{x,y}})^2} + (\alpha \times Flux_{x,y})^2}.$$
\end{equation}

$C$, $SY,$ and $\alpha$ can be provided either by the user or 
%\LEt{please check if this is correct or change} 
%C and SY, or 
they can be computed by AIDA based on readout noise and gain. If $SY$= $\alpha$=0 ,
the fit is unweighted. In general, $C$ describes the constant component of the noise, 
which is independent of the signal (in the ideal case, it coincides with the readout noise,
without sky background), $SY$ describes the component proportional to the 
square root of the signal  (shot noise, in this case, $SY$ is related to the 
conversion factor Ke-/ADU), and $\alpha$ represents the fixed pattern noise.
%\LEt{it is unusual to have discursive footnotes in a paper (urls and such are okay). If this information is too extraneous to move it to the main text, please consider moving it to a separate additional appendix. You need to number the equations for reference, too, and please add a period after the equations for grammatical reasons}
We selected the weighted fit in both S\'ersic and B+D models, tailored 
to the 4KCCD.

\citet{Meert2015} noted that when multiple components are fitted, a
significant second component may only indicate substantial departure
from a single-component profile rather than the presence of a physically
meaningful second component. They quoted as examples
\citet{Gonzalez2005}, \citet{Donzelli2011}, and \citet{Huang2013}, who 
fitted multiple components to ellipticals (E) and bright cluster galaxies
without necessarily claiming the existence of additional physically
distinct components. Our data set is deeper than the SDSS data set,
and this significantly affects the model selection. The single
S\'ersic fit often tends to overestimate the galaxy luminosity  at low surface
brightness levels, for instance, also when fine structures are present. The effect on the
magnitude estimate is weak, but the B+D model is statistically 
better by far (even by eye) in describing the galaxy light profile for most
of our iETGs.
The images in $g$ and $r$ bands of the iETGs in our sample and a summary
of the analysis performed are collected in Figure~\ref{figure-5} and 
in Appendix~\ref{AppendixA}. 

%-------------------------------- Figure 5 ---------------------
\begin{figure*}
\center
\includegraphics[width=18.3cm]{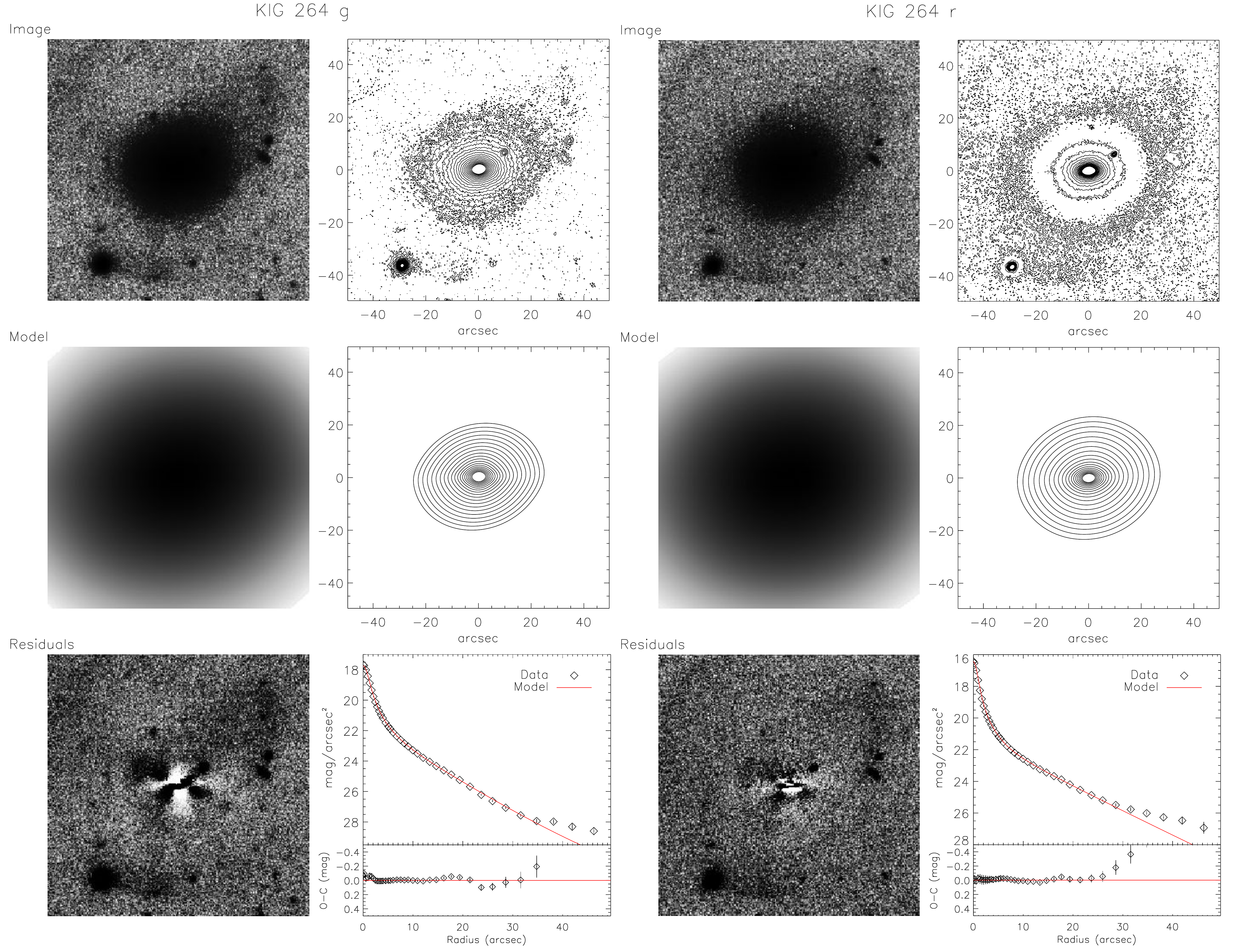}
\caption{Summary of the $g-$ ({\it left panels}) and $r-$ ({\it right panels})
band surface photometry of KIG 264. The adopted masks of the foreground
and background objects are not shown. North is up, and east is to the left. The right panels provide from
top to bottom ({\it left}) the original image, the best-fit  model
image, and the image of residuals after model subtraction for the g band. ({\it right})
Isophotal contours of the image, of the model, and of the azimuthal light
profile. For clarity, only the model (red line) is over-plotted on the
azimuthal light profile, and the (O--C) residuals are shown.
Table~\ref{tab-4} reports the model parameters. We show 20 isophote  levels,
between 500 and 2 $\sigma$ of the sky level ($\mu_g$=21.6$\pm$0.02 and 26.9$\pm$0.34
mag arcsec$^{-2}$ and  $\mu_r$=21.0$\pm$0.02 and 26.6$\pm$0.29 mag arcsec$^{-2}$), 
 for the original and model images. The same panels are shown for the $r$  -band
photometry. KIG 264 reveals ripples and tail at the NW side of the galaxy
body and a shell system in the outskirts. These features are revealed in
both the $g$ and $r$ images above 2$\sigma$ of the sky level and in the
light profiles, starting at $\approx$30\arcsec.}
\label{figure-5}
 \end{figure*}
%--------------------------end figure 5 -----------------------

%-------------------------------- Figure  6 ---------------------
\begin{figure}
\center
\includegraphics[width=8.7cm]{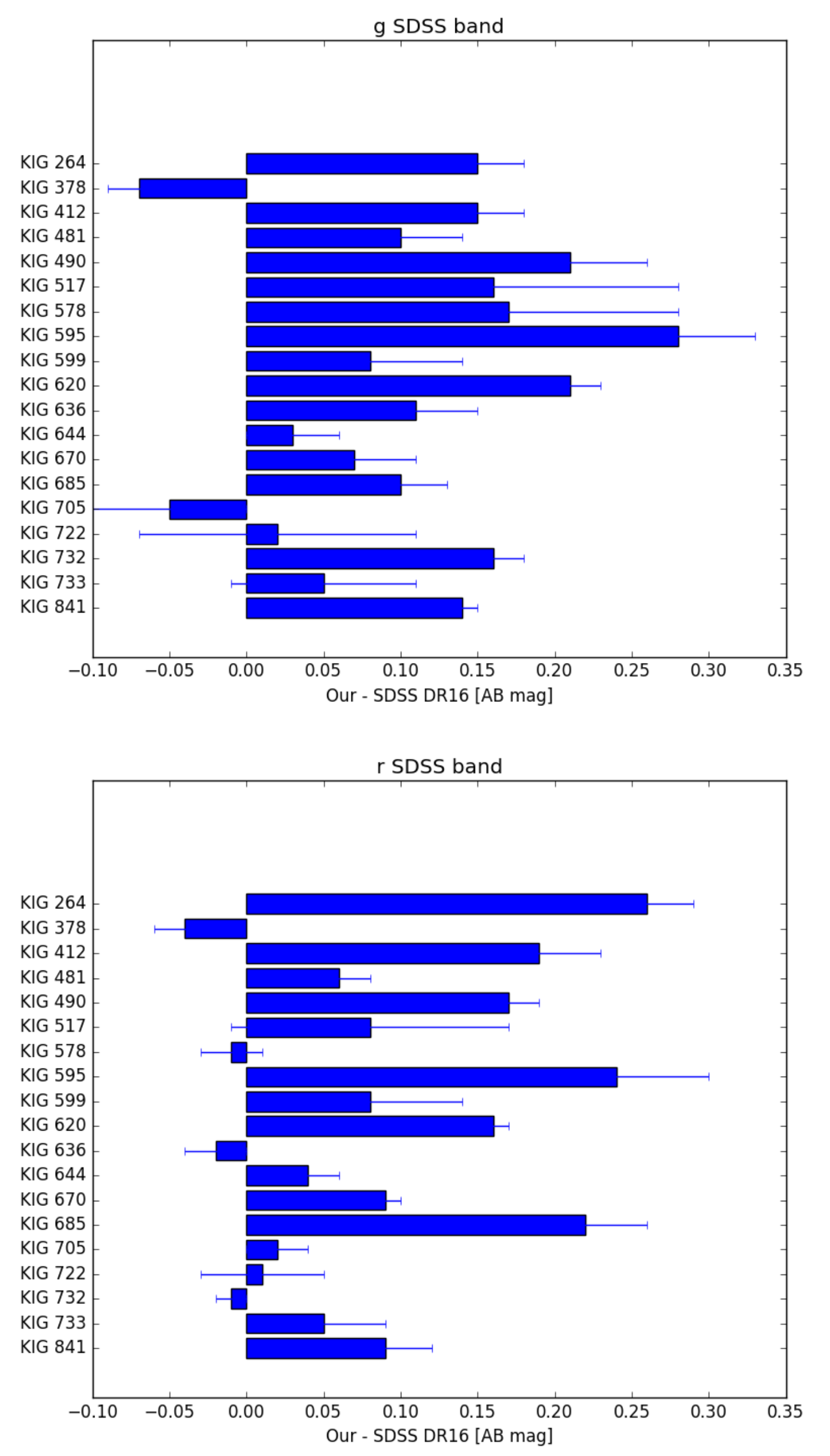}
\caption{Comparison between our integrated magnitudes in $g$ and $r$
bands with the SDSS DR16 values reported in Table~\ref{tab-2}. 
%Ourintegrated magnitudes are brighter than SDSS, on the average,
%by 0.10 mag and 0.09 mag in $g$ and $r$ bands, respectively. 
The solid lines show our errors. KIG 637 is
not plotted because the magnitudes are strongly influenced by the nearby
bright star  HD238370 (see \S~\ref{Individual_notes}).}
\label{figure-6}
 \end{figure}
%--------------------------end figure  6 -----------------------

\section{Results}
\label{Results}

Table~\ref{tab-4} reports the seeing-corrected parameters derived from the
S\'ersic and/or the B+D light profile best fit.  The table columns are
described in the table caption. Columns 12 and 13 report the morphological
class we assigned and the notes about the morphology of the residuals after model subtraction.
We adopted the following criteria to assign the morphological class:
We classified as S0 (-2$<$Type$\leq$0) galaxies whose profiles were best fitted by a B+D 
model and have 0.5$<$B/T$\leq$0.7, and we classified as E/S0 (Type=-3) galaxies with 0.7$<$B/T$<$0.8 
\citep[see e.g.][their Figure 23]{Meert2015}. Galaxies whose
luminosity profile was best fitted by a single S\'ersic law  (i.e. by definition, B/T=1)
were classified as E (-5$<$Type$<$-4). We considered that a nearly pure disc galaxy
has a S\'ersic index $n\approx1$ (i.e. by definition, B/T$\approx$0). We added
the notation ``pec'' to the classification when galaxies have structured residuals.

Uncertainties for the bulge and disc effective radii, ellipticity, and position angles for the
adopted models were obtained using the Monte Carlo simulation method. Noise and the uncertainty in the sky level determination were added. 
The errors on ellipticity, position angle, bulge effective radius, etc. quoted in the 
table are based on variables that do not depend on the model, in this case, Poisson noise. 
More realistic  errors are likely larger. This can be verified by comparing the 
values among the bands: the differences may reach 10\%, which is common 
for this type of model fitting.

In Figure~\ref{figure-6} we compare the difference between our total
integrated magnitudes, reported in Table~\ref{tab-4},  and the $g$ and $r$
bands from SDSS DR16  (Table~\ref{tab-2}). Our magnitudes
are brighter on average by 0.10 and 0.09 mag in $g$ and $r$ bands,
respectively. This value is higher than the measurement errors. KIG 637, which is not
included in Figure~\ref{figure-6}, presents integrated magnitude values
that differ largely ($\approx$0.46 mag) from the SDSS DR16
values.  The excess in magnitude in our measures in this case is
partly associated with the diffuse light of HD 238370, a very bright
nearby star that cannot be properly modelled and subtracted.
The extension of the stellar corona is visible in the top panels of 
Figure~\ref{figure-A6}, showing $g$ and $r$ residuals.

Comparisons with other photometric works and/or recent photometries
are difficult for the following reasons: 1) our values refer to either a simple S\'ersic
law  or to a B+D decomposition, that is, they are tied to the adopted model, and 2)
our photometry is deeper, for example, than data sets extracted from
SDSS images, for instance.  The ellipticity, $\epsilon$, and the position angle, PA, of our models
are not directly comparable with the values provided in Table~\ref{tab-2}.
We averaged the $g-$ and $r$ -band values of $\epsilon$ and of PA
for each galaxy. These results are in general comparable with the $\epsilon$ and PA
data  in Table~\ref{tab-2}. Our PA of the bulge, 102.7$\pm$1.1 
and 103.1$\pm$1.1 in $g$ and $r$ band,  and of the disc, 99.8$\pm$1.1 and 100.4$\pm$1.1,
of KIG 637 (NGC 5687) compares well with 103.6$\pm$0.01 and 101.2$\pm$0.1 
provided by the two S\'ersic-law fits by \citet{Costantin2018}. The same holds for
the ellipticity of the bulge, 0.40$\pm$0.04 in both bands, and of the disc
0.36$\pm$0.03 and 0.37$\pm$0.03 in $g$ and $r$ bands compared with
%\LEt{this sentence strikes me as strange because you provide the measures for the ellipticity in parentheses and then compare them to measures that you give as part of the main text. Please consider this and change this either so that all measures are part of the main text or all measures are given in parentheses} 
0.314$\pm$0.001 and 0.338$\pm$0.001 by \citet{Costantin2018}.

The parameters of the 2D galaxy model, provided in Table~\ref{tab-4},
are complemented by Figures~\ref{figure-A1}-\ref{figure-A9} in
Appendix~\ref{AppendixA}. The caption of Figure~\ref{figure-5}
%, shown in the main text,
%\LEt{you repeat the cption in the main text? or do you mean that Fig. 5 is in the main text? If you mean the first, please remove this because the content of the figure caption is not to be repeated in the main text to avoid redundancy. If you mean the second, please remove this because the numbering makes clear whether a figure is part of the main text or is provided in the appendix (the figures have an "A" in front of the number)}, 
details the information  provided by the galaxy
modelling. Our interest in particular is the detection of the
galaxy fine structure, such as stellar tails, streams, fans, and shells. 
The light-profile fit and the residuals after the 2D model subtraction 
are shown for each band in the bottom panels of these figures.

\begin{figure*}
\center
\includegraphics[width=17.8cm]{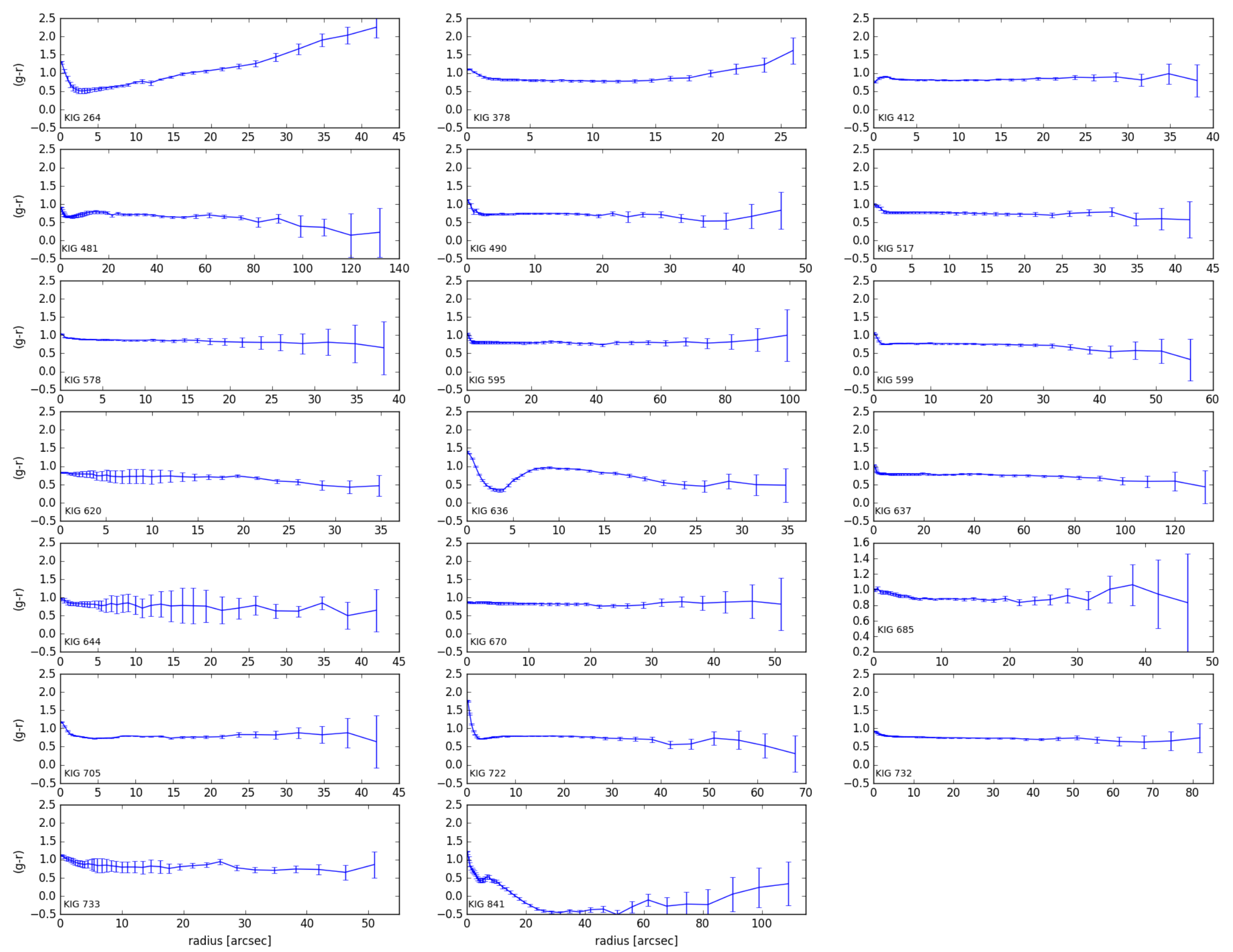}
\caption{Galaxy ($g$-$r$) colour profiles. Single profiles are presented and discussed
 in Section~\ref{Individual_notes}. }
\label{figure-7}
 \end{figure*}

\begin{figure}
\center
\includegraphics[width=8.7cm]{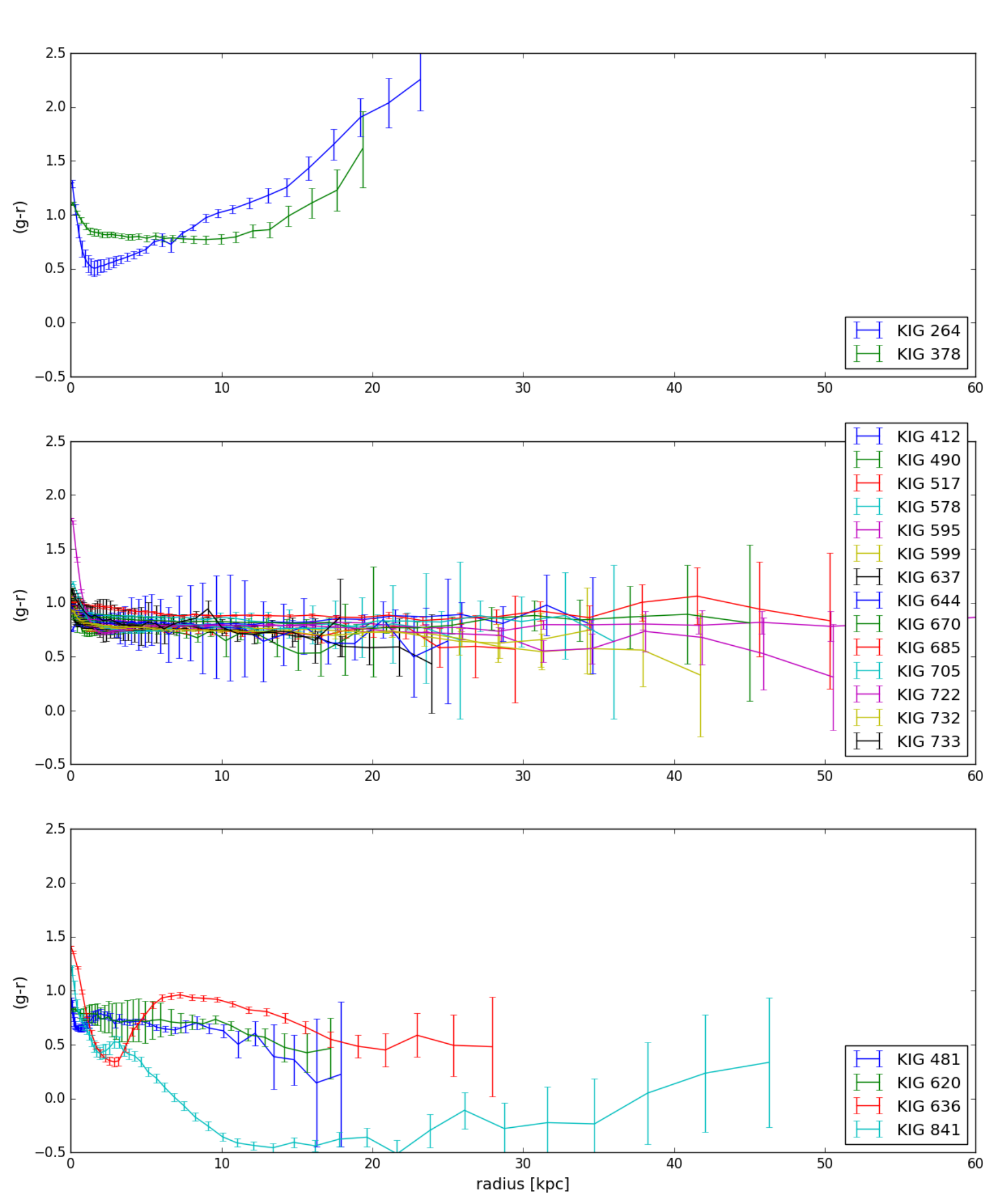}
\caption{($g$-$r$) colour profiles in kpc (using the galaxy distance provided in
Table~\ref{tab-1}). Colour profiles have been subdivided into three classes
according to their behaviour outside the seeing-dominated area: galaxies
with colour profiles become redder with radius ({\it top panel}),  flat
colour profiles ({\it middle panel}), and colour profiles with a peculiar
behaviour or that become bluer in the outskirts ({\it bottom panel}). }
\label{figure-8}
 \end{figure}

The \gmr\ colour profiles are collected in Figure~\ref{figure-7}. In
Figure~\ref{figure-8} the colour profiles are divided into three
categories and plotted using a kiloparsec scale in the abscissa. 
In discussing \gmr\ colour profile, we recall that because the flat field is imperfect, 
we had to fit the background with
 {\tt SWarp} \citep{Bertin2002} (see \S~\ref{Observations}). 
 This might influence the colour profile in the low surface 
 brightness regimes, where the errors are indeed quite large.
Outside the  seeing region, the trend of the vast majority (see the middle panel)  
of the colour profiles is flat, with values typical of ETGs, that is, around 0.7-0.8 mag. 
The colour profiles of KIG 264 and KIG 378 \gmr\ are shown in the top panel. They tend
to become redder with radius. The bottom panel  includes colour profiles either
with peculiar trends, such as those of KIG 636 and KIG 841, or profiles that become bluer with radius.
Single colour profiles are discussed in \S~\ref{Individual_notes}.

%-------------------------------- Figure 7 ---------------------
%--------------------------end figure 7 -----------------------

%-------------------------------- Figure  8---------------------
%--------------------------end figure  8-----------------------

\subsection{Individual notes}
\label{Individual_notes}

In this subsection we discuss the results summarised by the figures in
Appendix A. Detailed surface photometric studies have been dedicated to
some of the iETGs we study here by H-T, \citet{Rampazzo2019} in the K band and by
\citet{Costantin2018} in the SDSS $i$ band. Automatic surface photometry 
and 2D luminosity profile decomposition in these SLOAN bands 
have also been performed for some objects by
%\LEt{again, please check your LaTeX commands here} 
\citet[][]{Simard2011} and by \citet[][]{Meert2015}.
In our discussion of the \gmr\ profile, we exclude values
with a radius below $\text{three times the }$FWHM, which are perturbed by the seeing, and
measurement  in the outskirts with errors larger than 0.3 mag.

\noindent \underbar{KIG 264} ~~~~~ The galaxy is considered a barred
lenticular by \citet{Buta2019} and {\tt HyperLeda}  classifications, as a
mixed E/S0 by \citet{Fernandez2012}, and as an E by H-T. The fit with a
single S\'ersic law provides a poor fit of  the light profile;  a B+D
model (see Figure~\ref{figure-5}) gives a statistically much better
representation. Our surface photometry suggests the presence of a disc,
but we did not find evidence of a bar.  The cross-like shape in the 
$g$ and $r$ residuals in the inner regions (see Figure~\ref{figure-5})
is an artefact due to the boxy shape of real isophotes with
respect to our model. The isophotes boxiness has previously been
noted by H-T. In the NW, the SE and NE residuals (Figure~\ref{figure-5})
compose a shell structure. A fan, likely a wide tail, starting from the galaxy
body, is attached to the NW part of the shell structure. These features have not been detected 
in the study by H-T. The B+D model fits the luminosity profile well up to the fan
that appears as an increase of the disc surface brightness starting at
$\approx$30\arcsec\  in the $g$ and $r$ bands. The B/T=0.68 in both
bands indicates that the bulge component is dominant. We suggest 
that the galaxy is a peculiar unbarred S0.
The \gmr\ colour profile grows monotonically from $\approx$0.5 mag to 
unusually high red values in the galaxy outskirts, which may suggest the
presence of dust.  

\medskip \noindent \underbar{KIG 378} ~~~~~ The galaxy is classified as
an elliptical by \citet{Buta2019} and {\tt HyperLeda,} but it is
considered as an  E/S0 by \citet{Fernandez2012} and as a S0 by H-T. The
luminosity profile is statistically best fitted by a B+D model
(Figure~\ref{figure-A1}). The exponent of  the single S\'ersic-law 
fit is $n\approx3$, although the model tends to overestimate 
the galaxy luminosity in the outskirts. 
The bulge dominates in the adopted B+D best fit. 
It remains unclear why the B/T light ratio, which is 0.63 in $g$ and 0.78 
in $r$,  differ so much although the fit
is good in both cases. A residual structure (a tail?) extending from the nucleus
towards the NW, which is particularly evident in the $r$ band, is revealed after the
model subtraction. We classify this galaxy as S0 pec, taking into 
account the asymmetric residual structure.
The \gmr\ profile is nearly flat at about 0.8 mag up to 17\arcsec\ , and then
it becomes redder and reaches $\approx$ 1 mag at about 20\arcsec\ , 
where there are significant measurement errors, however.

\medskip \noindent \underbar{KIG 412}~~~~~ The galaxy is one of the less
isolated galaxies in the sample (Figure~\ref{figure-2}) for \citet{Verley2007b}, 
but the spectroscopic criterion of \citet{Argudo2013} is fulfilled. 
All classifications in Table~\ref{tab-1} converge in indicating this
galaxy is an E, but it is an S0 galaxy for H-T. The single S\'ersic fit
provides $n=4.38\pm0.11$ and $n=5.80\pm$0.02 in $g$ and $r$ bands,
respectively. However, this model not only overestimates the luminosity
in the galaxy light profile in the outskirts, but the trend of the
residuals also suggests the presence of a second component. The B+D
provides  a statistically better fit (Figure~\ref{figure-A1} bottom
panel), as in the case of KIG 378. The residuals after the B+D model
subtraction show a faint inner ring and a ring or shell-like
structure in the outskirts (similar to KIG 685). 
Although this latter structure is revealed below 2$\sigma$ of the
sky level, it is evident in both original images, especially in the $r$ band,
starting from $\approx$20\arcsec. Our surface photometry supports an E/S0 pec
classification because the bulge is still dominant; B/T is 0.69 and
0.71 in the $g$ and $r$ bands, respectively, and residual structures
are present.
The \gmr\ colour profile is nearly flat at about 0.8 mag.

\medskip \noindent \underbar{KIG 481}~~~~~ The galaxy has a very
uncertain classification. It is an SA (T=1) for \citet{Buta2019}, an Sab
for H-T, a late S0 (T=-0.1) for {\tt HyperLeda,} and a classic S0 (T=-2)
for \citet{Fernandez2012}. \citet{Morales2018} found two classical
shells on either side of the galaxy.
Our best fit (Figure~\ref{figure-A2}) is obtained using a B+D model with
a B/T=0.66 and B/T=0.69 in $g$ and $r$ bands, respectively. Residuals
show a strong irregular dust-lane  and a wide system of three
nearly concentric shells.  We suggest that it is a S0 pec 
\citep[see also][]{Buta2019}.  The galaxy \gmr\ colour profile is quite blue
in the galaxy centre, it is 0.64-0.7 mag, and it tends to become bluer with radius. 

\medskip \noindent \underbar{KIG 490}~~~~~ The galaxy is considered a
late-S0 (Table~\ref{tab-1}) and an Sa for H-T. In Figure~\ref{figure-A2}
we show our B+D best fit  of the galaxy light profile. Residuals show a
wide plume, likely a tail, in the NE part of the galaxy starting at
about 30\arcsec from the nucleus. The feature is reminiscent of the
fan or tail found on the NW side of KIG 264 and connected with the shell. 
The B/T ratio (0.50 in $g$ and 0.53 in
$r$) together with the ripple and tail we detected fully justify the
\citet{Buta2019} classification of a late, un-barred, and peculiar S0. 
The \gmr$\approx$0.74 mag colour profile 
is typical of ETGs (Figure~\ref{figure-8} bottom panel).

\medskip \noindent \underbar{KIG 517} ~~~~~ This galaxy is considered a
classical E by \citet{Buta2019} and H-T and an S0 by {\tt
HyperLeda} and \citet{Fernandez2012}. H-T found that it has a boxy
structure that we do not see in either the original isophotes or as
an artefact in the residuals. The single S\'ersic best fit provides $n$
values near to a de Vaucouleurs law. Our B+D model (Table~\ref{tab-4})
subtraction (Figure~\ref{figure-A3}) shows a faint residual ring in the
range 5-10\arcsec. The B/T values are 0.71 and 0.74 in $g$ and $r,$
respectively, indicating a large bulge contribution. We suggest that the
galaxy is an E/S0. The \gmr\ colour profile is flat around 0.75 mag.

\medskip \noindent \underbar{KIG 578} ~~~~~ The galaxy is classified
E by all authors in Table~\ref{tab-1} and by H-T, who detect
a discy structure in isophotal shape profiles. The best fit is obtained
with a single S\'ersic law, with n$\approx$4; the B+D model best
fit is statistically poorer. The residuals after model subtraction shown in
Figure~\ref{figure-A3} reveal a ring between
$\approx$10\arcsec-20\arcsec and a faint outer ring or shell-like 
structure, detected at 2$\sigma$ of
the sky level, in the galaxy outskirts. KIG 578 is in the bona fide Es sample
of H-T, and they did not detect shells in their SDSS image. We suggest 
that KIG 578 is a peculiar elliptical. The \gmr\ colour profile is flat around 0.86 mag.

\medskip \noindent \underbar{KIG 595} ~~~~~The galaxy is classified an
E by \citet{Buta2019} and {\tt HyperLeda} and as an S0 by
\citet{Fernandez2012} and as an SAB0 by H-T. The light profile is best
fitted by a B+D model with a B/T=0.78 in both bands. The residuals in
Figure~\ref{figure-A4} show a faint inner ring and a series of shells that
connects in the north to a tail (below 2$\sigma$ level) that extends to the NW. No
bar is revealed. We suggest that the galaxy is an E/S0 peculiar.
The \gmr\ colour profile is flat around 0.80 mag.

\medskip \noindent \underbar{KIG 599}~~~~~ Classifications in
Table~\ref{tab-2} see the galaxy as a S0, but  it is an E with discy
isophotes and a system of shells and ring for H-T. A B+D model fits the
galaxy best. Residuals after model subtraction show an arm-like structure
in the inner part, and shells and a faint ripple on the SE side of the
galaxy, emerging at about 40\arcsec\ in both bands
(Figure~\ref{figure-A4}). We suggest that the galaxy is a peculiar S0,
likely late, because B/T=0.39 as in \citet{Fernandez2012}.
The \gmr\ colour profile is flat around 0.77 mag.

\medskip \noindent \underbar{KIG 620}~~~~~The morphological type
of this galaxy ranges from -2 to 1$\pm$1.6, suggesting that  its is a late-S0 or an
early spiral. According to {\tt HyperLeda}, the galaxy hosts a bar. H-T
classifies the galaxy as an early spiral (T=0.69). Our image shows that KIG
620 is composed of a inner disc, seen nearly edge-on, with
$\epsilon$=0.72, which is embedded in a  diffuse nearly round  structure 
(top panels in Figure~\ref{figure-A5}). The S\'ersic and B+D models
both provide a poor fit: the bulge is small, and the parameters are largely
contaminated by the inner ring. We
support the \citet{Buta2019} classification because we do not see a bar.
No fine structures are detected at a significant level. KIG 620 is
reminiscent of 3D early-type galaxies reported in \citet{Buta2015},
whose prototypes are shown in their Figure~23.
The \gmr\ colour profile is flat around 0.72 mag and tends to become bluer 
aound 0.5 in the outskirts, although the errors are large.

\medskip \noindent \underbar{KIG 636}~~~~~ The galaxy is considered an S0
by the tree classifications in Table~\ref{tab-2}. It has a bar both for
\citet{Buta2019} and for the {\tt HyperLeda} classification. H-T
classified the galaxy as E. Residuals from a S\'ersic or a B+D
model show the effect of a small bar or lens (clearly visible in the
original central  isophotes in both bands), and an outer ring 
from which extended arm-like structures (tails?) depart, with arms on the E and W
sides of the galaxy. The bulge parameters are
contaminated by the bar in this case as well. We modelled the galaxy to enhance
the outer galaxy structures. The classification of \citet{Buta2019} 
describes the inner morphology of the galaxy (T=-2.5) well, B/T=0.74 would suggest
T=-3 of \citet{Fernandez2012}. The suffix ``pec'' is necessary because of the arm-like 
residual structures.  The increasing reddening,   from $\approx$0.4 to 1 mag, of the \gmr\ 
colour  profile is connected to structures detected in the residuals map up to 
about 10\arcsec. After this, the \gmr\ trend inverts and becomes increasingly blue. 
It reaches  0.6-0.5 mag, although with large errors, in correspondence 
to the outer ring and arm-like structures. 

\medskip \noindent \underbar{KIG 637}~~~~~ The galaxy is classified as
an elliptical in Table~\ref{tab-1}. Recently, \citet{Costantin2018}
reported the $i$- SDSS band surface photometry of this galaxy (NGC
5687 in the paper) and found that it is an S0 whose bulge can be
unambiguously classified as classical ($n>2$), with an $n=2.58\pm0.0$.
This galaxy was chosen by \citet{Costantin2018} because it is
unbarred.

Several bright stars in the field, in particular, HD238370, which generates
a very extended corona, hampered the surface photometry and the estimate
of the value of the total integrated magnitudes provided in
Table~\ref{tab-4}. The galaxy extends  $\approx$2.5\arcmin\ in radius. We
fit the entire light profile with
a S\'ersic law ($n=3.92$ and $n=4.09$ in the $g$ and $r$ bands,
respectively). The index would suggest a classical E galaxy. However,
the B+D model provides a better fit of the light profile. 
Residuals reveal a ring structure in the galaxy centre. This does not
seem an artefact because this feature is also visible  in $i$ bands by
\citet{Costantin2018} after the subtraction of their
S\'ersic+exponential disc model. Our measured PA=102$^\circ$ and
ellipticity $\epsilon$=0.39 (see Table~\ref{tab-4})
%\LEt{is "reftab" a remainder of some computer command? Please check and remove}
 agrees with those found by the above authors. We suggest that the galaxy is an S0 (B/T=0.57) 
whose inner and outer rings are revealed by the residuals.
The \gmr\ colour profile is nearly flat at $\approx$0.78-0.74 mag, without
a signature of the structure detected in the residuals, and it tends to become
bluer in the outskirts and reaches 0.6 mag with large errors.

\medskip \noindent \underbar{KIG 644}~~~~~The galaxy morphological type
has a wide range of variation $-1 \leq T\leq2.2$. The galaxy is
considered borderline with a spiral: H-T classified it as Sab(s). As in
the case of KIG 620, we failed to fit the light profile: both S\'ersic and
B+D model offer quite a poor fit. In the inner regions lies a ring-like 
structure with a diameter of about 10\arcsec\  that hampered the
fit of a very small bulge. Our images reveal neither spiral
arms, as assumed by {\tt HyperLeda} and H-T, nor fine structures such as
shells and tails in the outskirts. We consider the
classification provided by \citet{Buta2019} adequeate. The authors 
suggested a ring-like feature  that outlines a bar  that might be related to the
$x_1$ family of bar orbits as in NGC 6012 \citep[see e.g.][for
an explanation]{Buta2015}. We assigned S0 as the morphological class.
The \gmr\ colour profile is nearly flat around 0.77 mag with large errors,
without a signature of a ring.

\medskip \noindent \underbar{KIG 670}~~~~~ The galaxy classification
ranges from pure E to classic S0. A B+D model fits the light
profile best. After model subtraction, the residuals show  central asymmetries
that are due to isophote twisting that is not accounted for in the model. In the
outskirts, however, asymmetries (shells or ripples?) are visible in both
bands (Figure~\ref{figure-A6}). We suggest that the galaxy is an E/S0 pec
as a consequence of a B/T=0.73 and of the peculiar outskirts.
The \gmr\ colour profile is nearly flat around 0.82 mag.

\medskip \noindent \underbar{KIG 685}~~~~~ 
The galaxy is considered a bona fide elliptical.  H-T
considered the galaxy an S0, however. \citet{Rampazzo2019} fit the high-resolution
(PSF=0\farcs25) K-band 2D light distribution best with {\tt GALFIT,} adopting
a model composed of two S\'ersic laws representing a pseudo-bulge 
($n=2.95\pm0.10$) plus a disc ($n=0.78\pm0.10$). 
The ellipticity profile, showing two regimes, suggested 
the presence of the disc. KIG 685 K-band residuals
show ring and shell-like structures. 
Our surface photometry is heavily hampered by
bright stars that are located very near the galaxy centre. The PSF (see Table~\ref{tab-3})
is between 3.5 ($g$ band) and 4.8 ($r$ band) times that of 
{\tt ARGOS+LUCI} adaptive optics instrument at the LBT. Our $g$ and $r$ study
confirms the disc, the shell and ring-like structures
that were detected in the previous study,  as shown in
Figure~\ref{figure-A6}. We suggest that the galaxy is a peculiar E/S0.
The \gmr\ colour profiles is nearly flat at $\approx$0.88 mag.

\medskip \noindent \underbar{KIG 705}~~~~~ This galaxy is also
considered an E (see Table~\ref{tab-2}), including H-T, who in
addition reported a discy structure of the isophotes, an inner disc, and shells and ripples. A S\'ersic
model fits the galaxy best, with $n=3.61\pm0.03$ in $g$ and 
$n=4.07\pm0.08$ in $r$, supporting the idea that this  is a classic
E  (top panels of Figures~\ref{figure-A8}). The B+D model, which is statistically
indistinguishable from the previous model, offers a B/T=0.89 in both bands,
suggesting in any case that the bulge component is dominant. The
residuals after model subtraction enhance shells and ripples in the
galaxy outskirts. The galaxy shows a disc-like structure  in the inner 
region. We assigned E pec as the morphological class. 
The \gmr\ colour profiles are nearly flat around 0.77 mag.

\medskip \noindent \underbar{KIG 722}~~~~~ The galaxy is considered a
bona fide E with boxy isophotes according to H-T. Our
best-fit model is a S\'ersic law in both bands with exponents
near the classic de Vaucouleurs law (see Table~\ref{tab-4}).
The residuals (bottom panels of Figures~\ref{figure-A8}) in the 
central part of the galaxy therefore are an artefact created by the combination of the isophote
twisting and boxiness with respect to the model. The excesses of light
between 20\arcsec-40\arcsec\, at the NE and SW sides of
the galaxy appear as real fans and shells  that are also visible in
the original image.
H-T did not detect shells in their SDSS image of KIG 722.
We assigned E pec as the morphological class.
The \gmr\ colour profiles are nearly flat around 0.76 mag without
any obvious connection with the structure that is enhanced by the residuals.

\medskip \noindent \underbar{KIG 732}~~~~~ This galaxy is another bona fide E in
Table~\ref{tab-2}. H-T reported that this E has a boxy structure, 
%\LEt{please check the next part of the sentence. Is this a list: "... boxy structure, dust, and an inner disc"?}
dust, and possibly an inner disc. Our best-fit model is the B+D shown in
Figure~\ref{figure-A9}, although the bulge predominates (B/T=0.88 and
0.78 in $g$ and $r$ bands). Isophotes appear strongly boxy: this is the
reason for the cross-like residuals in the central regions of the galaxy. The residuals
also enhance shells and a spiral-like structure that is likely due to isophote twist and
extends out to the galaxy outskirts. We assigned the E/S0 pec 
morphological class.
There are no traces in the nearly flat (around 0.77 mag) 
\gmr\ colour profile of the structures that are traced by the residuals.
H-T did not detect shells in their SDSS image of KIG 732.

\medskip \noindent \underbar{KIG 733}~~~~~The galaxy is borderline, with
an early spiral (0$\leq T \leq$1.5) in the classifications provided in
Table~\ref{tab-1}. Our B+D model fails to provide an adequate fit to the
light profiles (bottom right panel in Figure~\ref{figure-A9}). The
galaxy has a complex morphology that is well described by the classification of
\citet{Buta2019}, with an outer ring starting from wide open arms. There
is dust along the arms that appears knotty (see the bottom left panel).
Considered as a whole, the galaxy is reminiscent of the 3D early-type
objects in \citet{Buta2015}, similarly to KIG 620. We suggest that
the correct classification is provided \citet{Buta2019}, so we assigned
S0 pec as the morphological class.
The \gmr\ is red, nearly flat around 0.8 mag, with a mini peak of 0.94 mag
at about 26\arcsec \ in correspondence of the outer ring.

\medskip \noindent \underbar{KIG 841}~~~~~The galaxy is considered
late-S0 (T=0) in \citet{Fernandez2012}. For \citet{Buta2019}, is also an
S0, but with several peculiar features, such as an inner ring and plume, while
it is an E/S0 for {\tt HyperLeda}. Our surface photometry confirms the
peculiar features in the \citet{Buta2019} classification. Furthermore, we
find evidence of a wide systems of shells that is particularly evident  on the NE
side. Our best fit is obtained with a B+D model. The outer
shells appear as an increase in luminosity in the model at
$\approx$70\arcsec\ at about 27 mag arcsec$^{-2}$ in $g$ band and  26
mag arcsec$^{-2}$ in $r$ band. We classify this galaxy as a late-S0 pec
because B/T$\approx$0.6. 
The \gmr\ colour profile (Figure~\ref{figure-8}) is entirely unexpected
for ETGs. Outside the seeing, it has values of about 0.6 mag, then it becomes
increasingly bluer up to 25\arcsec\ , with a flat part at about -0.4 mag up to 50\arcsec.
In the galaxy outskirts, the \gmr\ trend turns to red, but the values are still
below 0 mag. 

\bigskip 
\bigskip
To summarise, our light profiles reached on
average $\mu_g$=28.11$\pm$0.7 mag~arcsec$^{-2}$ 
and $\mu_r$=27.36$\pm$0.68
mag~arcsec$^{-2}$. The results of the light-profile analysis
decomposition are listed in Table~\ref{tab-4}. A minority, 15\%
(3 out of 20), of our iETGs can be considered bona fide Es that are fitted best by a S\'ersic law
with the exponent $n\approx$4, that is, by a de Vaucouleurs law. For the vast
majority, the B+D model fits the galaxy light profile best. The bulge 
flux is dominant, with an average B/T$\approx$0.66 in both bands. 
Three out of 20 iETGs, that is, NGC 620, NGC 644, and NGC 733, 
are late S0s at the border with spirals, some showing arm-like
structures in their central region, as in the \citet{Buta2019} classification. 
On the whole, their morphology is reminiscent of the 3D ETG class 
described  in \citet{Buta2015}. We conclude that our sample
is composed of bona fide ETGs.

\medskip 
\noindent After the  2D model was subtracted,
most of the galaxies showed fine structures that emerged in their
inner regions and outskirts. In particular, 12 out of 20 iETGs (60\%) showed
clear evidence of shell and ripple signatures (Table~\ref{tab-4},
Figure~\ref{figure-5}, and Figures~\ref{figure-A1}--\ref{figure-A9}). None
of the late S0s mentioned above show shells, ripples, or tails. 
In \S~\ref{isolation} we discussed the revised isolation criteria for the
galaxy in the sample performed by \citet{Verley2007b} and \citet{Argudo2013}. 
Figure~\ref{figure-2} shows that all iETGs with shells except for KIG 412 and KIG 595
 lie within the fiducial range of isolation in the $\eta_k$-Q plane. 
 This means that the fraction of shell galaxies within the isolation fiducial range reaches 
62\% (10 out of 16). 
%Assuming the ``strict'' spectroscopic isolation
%criteria adopted by \citet{Argudo2013}, that leave as isolated 
%10 iETGs in the present sample (remind that for 4 iETGs no spectroscopic
%information about neighbours are available), 6/10 galaxies have shells (60\%).

Most of the colour profiles, 13 out of 20 (65\%), are flat with values in
\gmr\ of about 0.7-0.8 mag, which is typical of ETGs. Ten of the 12 shell 
galaxies have red \gmr\ colour profiles; 2 galaxies, KIG 481 and KIG 841,
have blue (and peculiar, in particular, KIG 841) colour profiles. 

\section{Discussion}
\label{Discussion}

By definition,  iETGs are early-type galaxies without  obvious companions.
Considering the above results, the question is how long  isolated early-type 
galaxies have been isolated. The answer, which involves both the timing 
and the physical mechanisms driving the iETG evolution, can be approached 
using the various fine structures that are detected as markers, together with other indicators
that we discuss in the following sections.

\subsection{Shapes of the residuals}
\label{Residuals}

Simulations suggest that the fine structures revealed by our surface photometry are
connected to past interacting and merging history.  We review the residual shapes 
and discuss them in the light of the literature.

\subsubsection{Shells} \label{shells}

In the literature, shells have been widely associated with both minor and major and 
both wet and dry merging episodes 
\citep[see e.g.][and references therein]{Dupraz1987,Weil1993,Mancillas2019,Pop2018}. 
Several studies have suggested that the fraction of shells, and in general, the fraction
of tidal features, indicate a strong environmental effect.
\citet{Reduzzi1996} found that about 16.5\% of  ETGs in the 
field showed shells, which is at odds with the hypothesis that 4\% of ETGs 
are members of physical pairs.
Only 4\% of the shell galaxies in the original \citet{Malin1983} list are
located in clusters or rich groups. An incidence of  50\% of shells has been found 
by  H-T in 18 isolated galaxies that they considered bona fide Es, using the SDSS DR6 data set. 

Our set of galaxies, however, needs to be compared with results from deep surface
photometry because the fraction of detected features certainly depends  on this 
factor. \citet{Tal2009} investigated a complete  volume
-limited  (15-50 Mpc) sample of bright ($M_B$<-20) ETGs and detected
12 out of 55 (22\%) objects with shells. Their surface brightness reached
27.7 mag arcsec$^{-2}$ in the V band. The authors noted that the fraction of
galaxies with tidal features increased and did not include cluster members. 
We found that about 60\% of our iETGs shows shell structures. 
With respect to H-T, we revealed shells also in KIG 264, 
KIG 578, KIG 722, and KIG 732, which are all included in their sample of isolated Es. 
 Recently, \citet{Pop2018} investigated shells in 220 
 of the most massive galaxies, regardless of their environment,
in the {\tt Illustris} simulation \citep{Genel2014,Vogelsberger2014a,Vogelsberger2014b}.
 They reported shells in 39 galaxies, which is 18$\pm$3\%. This fraction is consistent 
 with the fraction found by \citet{Tal2009}.
The fraction of our iETGs that shows shells is three 
times greater than that found by \citet{Tal2009}.

Most (9 out of 13) iETGs with shells have flat and red \gmr\ colour profiles.
This is consistent with the general finding that shells are more common in red 
 than blue (spiral) galaxies \citep[see e.g.][]{Atkinson2013}. KIG 481 and KIG 841, 
 which have blue colour profiles, are still in the green  valley (see \S~\ref{NUVr}).

The \citet{Mancillas2019}  simulations suggested that shells are a long-lasting 
feature \citep[see also][]{Longhetti1999,Rampazzo2007}.  The estimated
survival time of shells is $\approx$3Gyr.
Lacking possible perturbers, shells in our iETGs were generated 
by the last merging event and set  the isolation time, which is of that order. 
Shells basically disappear in cluster ETGs, which likely is a consequence of the galaxy
transformation due to continuous harassment processes
\citep{Moore1998}. In less isolated ETGs 
(see \S~\ref{isolation}), the fraction of galaxies with shells
is still higher than the fraction quoted by \citet{Tal2009}.  
Further indications about the age of the interaction
are developed in \S~\ref{NUVr}.

 Two of the iETGs with shells deserve further attention.
KIG 264 shows a system of shells and a ripple that is reminiscent of a tail
(Figure~\ref{figure-5}). The galaxy has been studied at radio wavelength
by \citet{Wong2015}. They described KIG 264 (identified as J0836+30 in
their paper) as the more passive object in their sample. They did not
detect nuclear radio emission in the galaxy at 1.4 Ghz, but
observed two radio lobes at 88.4 kpc north-west and 102.5 kpc south-east
of  the galaxy centre. In the direction connecting the two radio
lobes, they detected an extragalactic \Hi\ cloud midway between the galaxy
centre and the lobe on the north-west side (see their Figures 2 and 5).
\citet{Wong2015} suggested that an active central engine has the
required energy to expel gas from the galaxy because two radio lobes are located
in the same direction as the \Hi\ cloud.  KIG 264 likely hosted
a radio AGN that may have blown out the observed  gas cloud. 

Our surface photometry shows that  KIG 264 has  no bar that would drive gas
to the centre. The wide system of shells extends to almost include the
\Hi\ cloud. Because shells and ripples are the scars of a recent merging 
\citep[see e.g.][]{Mancillas2019}, we suggest that this
event has likely involved gas-rich galaxies (wet merging) 
and might have activated the past AGN activity in KIG 264.

KIG 264 shows the remains of the AGN activity,
%\LEt{should this be "the remains of the AGN burning"?}, 
and the processes
may occur during a merging episode. \citet{Bennert2008} found several 
cases of (active) AGN host galaxies with an underlying shell system. 
On the other hand, \citet{Ibarra2013} found low AGN-level
activity in our iETGs sample: only three galaxies, KIG 378, KIG
595, and KIG 705, have a low ionization nuclear emission line region
%\LEt{you use this abbreviation only here, please spell it out} 
(LINER) in the nucleus.

The other interesting case is KIG 841 (NGC 6524). The type Ia pec 
supernova SN2010hh exploded in its shell \citep{Marion2010,Silverman2010} 
(SN1991bg-like).  The progenitors of this type of supernova\ are old stars,
that is, stars that were likely re-distributed from the parent galaxies during the 
merging episode.

\subsubsection{Tails, plumes, and fans}

Tails of stellar matter are generated by interaction and merging phenomena
\citep[see e.g.][and references therein]{Mancillas2019}. Broad stellar fans 
are also generated by interaction \citep[see e.g.][]{Tal2009}.
The tail survival time is $\approx$ 2 Gyr, while streams remain visible in
all phases of galaxy evolution. Streams, however, are detected in simulations 
at very low levels of surface brightness. The \citet{Mancillas2019} simulations revealed 
between two and three times more streams with a surface brightness cut of 
33 mag arcsec$^{-2}$ than with 29 mag arcsec$^{-2}$.

 \citet{Tal2009} found that tails are less frequent than  shells in ETGs: 
 they found tails in  13\% (7 out of 55) of the ETGs. Tails exist for shorter 
 times than shells \citep{Mancillas2019}.  Tails tend to indicate a recent 
 encounter and/or merging. Furthermore, tails are evidence for 
 a dynamically cold component (e.g. a disc).
 % also in dry mergers/encounters  \citep[see e.g.][]{Combes1995}. 
 
We consider as tails, fans and plumes the features found in KIG 264, KIG 378, KIG 490, 
KIG 595, and KIG 636, that is, 25\% (5 out of 20) of our sample. KIG 378, KIG 490, and KIG 636
were considered isolated by both \citet{Verley2007b} and \citet{Argudo2013}. 
 All these galaxies are best fitted by a B+D model, 
which supports the idea described above that a cold disc structure is present.

\subsubsection{Spiral arm-like residuals}

KIG 599 and KIG 732  show shells and very wide spiral 
arm-like residuals on a very large scale. Arm-like structures are generated
 by an encounter that ends in a merger. 
The arm-like structure during the first phase, when the two nuclei are still visible,
of both an encounter and a merging episode was shown in \citet{Combes1995}, for instance. 
Clearly, KIG 732  is in an advanced merging phase. Residuals like this have been 
shown as residuals in other deep photometry 
\citep[see e.g.][in the case of NGC 1533]{Cattapan2019}.
 They were discussed as a signature of  the disc instability during a merging  episode 
 \citep[see e.g.][]{Rampazzo2018,Mazzei2019}.

%-------------------------------- begin Table 4 --------------------------------

\begin{sidewaystable*}[h]
\scriptsize
%\begin{table*}
\caption{KIG photometric data from the adopted S\'ersic and B+D best-fit models and morphological notes}
\centering
%% \tablesize{} %% You can specify the fontsize here, e.g.  \tablesize{\footnotesize}. If commented out \small will be used.
\begin{tabular}{llccccccccccl}
%\rowcolor{gray!50}
\hline
\textbf{KIG} & \textbf{Filter}  &$\epsilon_{bulge}$ &$\epsilon_{disc}$ & PA$_{bulge}$ & PA$_{disc}$ & $r_{e,bulge}$ & $r_{scale,disc}$ &   \textbf{$n $} &   \textbf{$m_T$}  & \textbf{B/T}  &  \textbf{Morphological} & \textbf{Notes on residual} \\
\textbf{}          &                         &                       &                      & [deg]       &  [deg]     & [\arcsec]    & [\arcsec]    &                        & [ABmag]               &   \textbf{}              & \textbf{class} & \textbf{morphology}     \\
\textbf{} (1  )&    (2)                  &   (3)                &    (4)              & (5)           &  (6)         & (7)             & (8)             &   (9)                 & (10)                        & (11)             & (12)  &        (13)            \\
\hline
 \rowcolor[gray]{0.8} 264   &  $g$ &0.65$\pm$0.06 & 0.21$\pm$0.03  & 91.4$\pm$1.4 &  107.8$\pm$1.7 &  1.00$\pm$0.10  &   6.01$\pm$0.30  & \dots &  14.66$\pm$0.03  &  0.68  & S0 pec & NW fan, tail and shells\\
 \rowcolor[gray]{0.8}          &  $r$ &0.48$\pm$0.05 & 0.13$\pm$0.02 & 92.8$\pm$1.4 &  105.4$\pm$1.8 &  1.25$\pm$0.11  &   7.46$\pm$0.30 & \dots  &  13.83$\pm$0.05  &  0.62    &  &\\
378                                    &  $g$ &0.28$\pm$0.03 & 0.27$\pm$0.02 & 170.3$\pm$2.7& 166.8$\pm$2.6 &  4.14$\pm$0.51  &  5.36$\pm$0.31  &\dots &  15.06$\pm$0.02  &  0.63    & S0 pec & NW asymmetric residuals (tail?) \\
                                         &  $r$ &0.25$\pm$0.02 & 0.31$\pm$0.03 & 168.8$\pm$2.6& 165.7$\pm$3.9 &  5.05$\pm$0.57  &  6.17$\pm$0.11  & \dots &  14.17$\pm$0.02  &  0.78  &  \\
 \rowcolor[gray]{0.8} 412  & $g$ & 0.14$\pm$0.03 & 0.12$\pm$0.02 &  98.8$\pm$1.3 & 67.9$\pm$1.1    & 3.55$\pm$0.45  &  6.84$\pm$2.91 & \dots&   15.40$\pm$0.03 &   0.69  & E/S0 pec & inner and outer rings, shell-like structures \\
  \rowcolor[gray]{0.8}       & $r$ &0.15$\pm$0.02  & 0.11$\pm$0.02 &  95.7$\pm$1.5 &   51.0$\pm$1.1 &   2.86$\pm$0.14 &  8.25$\pm$0.63  & \dots &    14.53$\pm$0.04 &  0.71  &      & \\
                                 481  & $g$ & 0.48$\pm$0.05  &0.30$\pm$0.03 & 97.0$\pm$1.0  & 81.1$\pm$1.0  &   10.18$\pm$0.21  & 27.12$\pm$0.67 & \dots  &   12.56$\pm$0.04*&  0.66 & S0 pec & dust, concentric shells\\
                                        & $r$ &  0.47$\pm$0.05 & 0.31$\pm$0.02 & 97.1$\pm$1.0 & 81.3$\pm$1.0 &   10.76$\pm$0.10   &  24.81$\pm$0.21 & \dots &   11.88$\pm$0.02* &  0.69 &  & \\
\rowcolor[gray]{0.8} 490  & $g$ &  0.36$\pm$0.04 & 0.17$\pm$0.02 & 33.9$\pm$1.4  & 30.7$\pm$1.4&   4.21$\pm$0.03    &    8.92$\pm$0.42  & \dots &  14.39$\pm$0.05&  0.53 & S0 pec & NE plume-tail\\
\rowcolor[gray]{0.8}       &  $r$ &  0.36$\pm$0.02 & 0.17$\pm$0.02 & 31.5$\pm$1.4  & 34.3$\pm$1.5 & 3.60$\pm$0.40   &    8.55$\pm$0.11  & \dots &   13.66$\pm$0.02&   0.50 &   & \\
                               517  &  $g$ & 0.29$\pm$0.03 &0.47$\pm$0.03 &  63.7$\pm$1.1 &  65.5$\pm$1.3 & 4.65$\pm$0.20  &    9.35$\pm$0.22 &  \dots &   14.82$\pm$0.12&  0.71 & E/S0   &  inner ring, outer irregular residuals\\
                                      &    $r$ & 0.21$\pm$0.03 &0.50$\pm$0.04 &  64.5$\pm$1.1 &  65.1$\pm$1.1 & 3.86$\pm$0.10 &     9.07$\pm$0.38 &  \dots &  14.05$\pm$0.09&  0.74  &    & \\
  \rowcolor[gray]{0.8} 578 & $g$   &  0.09$\pm$0.01 & \dots& 78.1$\pm$1.5 & \dots  &   6.86$\pm$1.50  &  \dots                                                &   4.19$\pm$0.58 & 14.96$\pm$0.11& \dots  & E pec  & inner ring, outer ring or shell\\
  \rowcolor[gray]{0.8}       & $r$   & 0.10$\pm$0.02   & \dots & 79.3$\pm$1.2& \dots  &   6.40$\pm$1.60 & \dots  &    4.18$\pm$0.45  & 14.27$\pm$0.02&  & &    \\
                                  595 &$g$ & 0.41$\pm$0.05&   0.38$\pm$0.03&    51.3$\pm$1.1  & 40.1$\pm$1.1  &    20.50$\pm$0.21 & 19.39$\pm$2.21  & \dots&  14.01$\pm$0.05 &  0.78   & E/S0 pec & shells, NW tail at 2$\sigma$\\
                                       &$r$&  0.42$\pm$0.04&   0.38$\pm$0.04& 51.5$\pm$1.1    &  40.8$\pm$1.1 &    19.82$\pm$0.30 &  19.82$\pm$3.24 &  \dots &  13.21$\pm$0.06 & 0.78      &  &  \\
\rowcolor[gray]{0.8} 599  &  $g$   & 0.24$\pm$0.03 &  0.15$\pm$0.02 & 86.2$\pm$1.3 & 91.3$\pm$2.3  & 2.58$\pm$1.19  &    8.60$\pm$1.01  & \dots   &    14.06$\pm$0.06 & 0.39   & S0 pec & shells, spiral arm-like residuals\\
\rowcolor[gray]{0.8}        &  $r$    & 0.25$\pm$0.03 &  0.15$\pm$0.02 & 89.1$\pm$1.3 & 90.8$\pm$2.2  & 2.48$\pm$0.29  &    8.27$\pm$0.29   & \dots   &    13.29$\pm$0.06 &0.39    &  & \\
                               620 & $g$  &  \dots & \dots  & \dots   & \dots& \dots  & \dots    &\dots       & 14.70$\pm$0.02 &  & S0 & inner ring, disc embedded into halo\\
                                     &  $r$ &  \dots &  \dots &   \dots & \dots & \dots  & \dots     & \dots   & 13.99$\pm$0.01   &  &  & \\ 
\rowcolor[gray]{0.8} 636 &  $g$&  0.40$\pm$0.04 &  0.11$\pm$0.05 & 115.8$\pm$1.1 & 171.2$\pm$1.7 &   4.27$\pm$0.20 &  3.91$\pm$0.50 &\dots &  14.92$\pm$0.02 & 0.74  &  SB0 pec & inner bar(?), ring, tails or arm-like structures \\
 \rowcolor[gray]{0.8}       &   $r$ & 0.42$\pm$0.04 & 0.10$\pm$0.17  & 116.2$\pm$1.1  & 170.8$\pm$1.7  &  4.79$\pm$0.17 &   3.86$\pm$0.25  & \dots & 14.21$\pm$0.01  &  0.73  &  & \\
                                637  & $g$ &0.40$\pm$0.04  & 0.36$\pm$0.03  & 102.7$\pm$1.1    & 99.8$\pm$1.1&   11.53$\pm$0.13   &   29.64$\pm$1.15   & \dots &  12.52$\pm$0.11  &  0.56  & S0  & inner and outer ring\\
                                        & $r$ &0.40$\pm$0.04  &  0.37$\pm$0.05 &  103.1$\pm$1.1  & 100.4$\pm$1.1&   10.88$\pm$0.15   &   28.93$\pm$1.32    & \dots &  11.71$\pm$0.11 &  0.57  &      &\\
 \rowcolor[gray]{0.8}644  &  $g$ & \dots  & \dots  &  \dots  & \dots &  \dots       &   \dots     &  \dots     &   14.74$\pm$0.03 &  \dots  & S0 & inner ring and lens\\
 \rowcolor[gray]{0.8}   & $r$   & \dots  &  \dots &   \dots  & \dots &  \dots       &    \dots   &   \dots       & 13.91$\pm$0.02& \dots &   & \\ 
                                 670  & $g$  & 0.36$\pm$0.03  & 0.38$\pm$0.04  &  148.9$\pm$1.5 &148.4$\pm$1.3&  4.78$\pm$0.12   &   10.91$\pm$0.16  &  \dots & 14.52$\pm$0.04 & 0.72 & E/S0 pec & shells, asymmetries \\
                                      &  $r$  & 0.36$\pm$0.04  & 0.38$\pm$0.03    &   148.7$\pm$1.4 &148.8$\pm$1.4 &  4.80$\pm$0.11  &   10.85$\pm$0.48  &  \dots& 13.69$\pm$0.01 & 0.73 &  &  \\
 \rowcolor[gray]{0.8}  685  &   $g$  &0.18$\pm$0.05   &0.16$\pm$0.02 &   99.1$\pm$2.5 & 148.4$\pm$1.5  &  3.55$\pm$0.12   &   7.53$\pm$0.60   &\dots&  15.31$\pm$0.03 & 0.74 & E/S0 pec  & shells, rings \\
\rowcolor[gray]{0.8}     &   $r$  & 0.17$\pm$0.05      & 0.10$\pm$0.02 &  100.2$\pm$1.3   &  150.2$\pm$1.3&   3.26$\pm$0.91   &    8.33$\pm$0.50  & \dots&  14.34$\pm$0.04  & 0.65 &             &  \\ 
                                  705  & $g$ &  0.01$\pm$0.01  & \dots                 & 16.7$\pm$2.5  & \dots  & 3.67$\pm$0.12    &  \dots   &   3.61$\pm$0.30 &  14.62$\pm$0.05  & \dots & E pec&  inner ring, dust ?, shells\\
                                         & $r$  &  0.01$\pm$0.01  & \dots                 &  14.7$\pm$3.2  & \dots   & 3.48$\pm$0.13   & \dots     &   4.08$\pm$0.48 &  13.79$\pm$0.02 &  \dots&  &  \\
\rowcolor[gray]{0.8} 722 &  $g$ &   0.08$\pm$0.03 & \dots  &   96.6$\pm$2.0&  \dots&   9.33$\pm$0.17  &  \dots   &   4.23$\pm$0.37 &14.13$\pm$0.09  & \dots &  E pec & shell and fans \\
 \rowcolor[gray]{0.8}       &  $r$  &  0.09$\pm$0.03 & \dots  &   86.2$\pm$2.0 &  \dots & 8.90$\pm$0.15  &  \dots   &  3.58$\pm$0.39 & 13.28$\pm$0.04 & \dots   &  & \\ 
                                732  & $g$ &0.15$\pm$0.03 & 0.04$\pm$0.01  &  61.8$\pm$2.0 &  91.8$\pm$2.2 &  8.18$\pm$1.3  & 13.91$\pm$1.78  & \dots &  13.11$\pm$0.02 & 0.77  & E/S0 pec & shells, spiral arm-like residuals \\
                                        & $r$  &0.15$\pm$0.03 & 0.01$\pm$0.01  &  64.3$\pm$2.2  &  90.1$\pm$1.0  &  7.95$\pm$1.4  & 13.50$\pm$1.45  & \dots &  12.43$\pm$0.01 &0.79 &  &\\
\rowcolor[gray]{0.8}  733  &  $g$   & \dots   & \dots   & \dots     & \dots  &  \dots     &     \dots            &   \dots            &       14.44$\pm$0.16            &           \dots                          &  S0 pec & open arms, ring  \\
\rowcolor[gray]{0.8}      &      $r$   & \dots    &  \dots  &  \dots  &  \dots  &    \dots       &      \dots            &    \dots            &    13.60$\pm$0.14             &        \dots                          &        &  \\
                                 841 &$g$  &0.49$\pm$0.04 &0.21$\pm$0.02 & 157.4$\pm$1.1  & 148.4$\pm$1.1 &  10.57$\pm$1.14   &    17.45$\pm$1.25  & $\dots$ & 13.22$\pm$0.01*  &  0.59 & S0 pec & dust-lane, shells\\
                                          &$r$  & 0.46$\pm$0.05& 0.24$\pm$0.02 & 158.5$\pm$1.1 & 148.9$\pm$1.1  &  10.07$\pm$1.14  &   17.47$\pm$1.32   & $\dots$  & 12.43$\pm$0.03*  &  0.62 &  &  \\
\hline
\label{tab-4}
\end{tabular}
\tablefoot{ The elllipticity and position angle for the bulge, ($\epsilon_{bulge}$, PA$_{bulge}$), and disc 
components, ($\epsilon_{disc}$, PA$_{disc}$), are given in col.s 3--6. 
The bulge, $r_{e,bulge}$, and disc, $r_{scale,disc}$ effective radii and disc scale length, in col.s 7 and 8, 
refers to B+D decomposition. When the best fits is obtained with a single S\'ersic law fit, 
whose index $n$ is given in col. 9, $r_{e,bulge}$, $\epsilon_{bulge}$ and PA$_{bulge}$ refers to the whole galaxy.  
The value of $m_T$ in column 10 refers to  the galaxy magnitude from the aperture photometry integration. 
When the 4K-CCD gap is present, we report the total integrated magnitude of the model (indicated with the asterisk). 
Col. 11, reports the ratio between the bulge and the total flux of the galaxy obtained 
from the B+D model of the light profile. Col. 12  provides  the morphological class according to the criteria 
explained in the first paragraph of \S~\ref{Results}. Col. 13 summarises the morphology of the residuals after the model
subtraction. Detailed comments about the galaxy morphology are given in Section~\ref{Individual_notes}.}
%\end{table*}
\end{sidewaystable*}

%------------------------------- end Table 4 -------------------------------------------

% ------------------------------------------begin Table 5-------------------------------------
\begin{table}
\center
\tiny
\caption{iETG \Hi\ content}
\begin{tabular}{lccccc}
\hline\\
KIG &  H{\sc{i}} flux &  S/N &  $W_{50}$ &  $\log M_\mathrm{\Hi}$ & $\log M_\mathrm{\Hi\ - pred}$\\
& [Jy \kms] & & [\kms] &  [$\mathrm{M_\odot}$] & [$\mathrm{M_\odot}$]\\
\hline\\
 \rowcolor[gray]{0.8}264 &              $1.29 \pm 0.10$ &    7.6 &         $440 \pm 10$ &  9.50 & 9.54\\
%                                378 &             \dots         & \dots  &  \dots       &        \dots                 &  $<$10.18  & 9.74\\
% \rowcolor[gray]{0.8}412 &         \dots  &    \dots    &      \dots        &   \dots            &  $<$10.01  & 9.66\\
                                481 &              $2.29 \pm 0.25$ &    8.8 &         $278 \pm 39$ &       8.42 & 8.98\\
\hline
\end{tabular}
\label{tab-5}
\tablefoot{The galaxy ID is listed in col. 1. The measured \Hi\ flux and the signal-to-noise ratio (S/N) are provided
in col.s 2 and 3. Col. 4 provides W50, the velocity width at the 50\% level. Columns 5 and 6 list
the measured \Hi\ mass and the mass predicted by relation  17 in
\citet{Jones2018} using the B-band luminosity, respectively.} 
\end{table}
%-----------------------------------------end Table 5--------------------------------------------

\subsubsection{Rings}
\label{rings}

When possible artefacts are excluded (see \S~\ref{Individual_notes}), several types 
of inner features become visible that are revealed in the residuals after
model subtraction. Rings are one of the most frequent of these features,
while bars are barely detected in the present sample. 

Rings (inner and outer) are considered in the \citet{Buta2019}
classification of isolated galaxies.  Table~\ref{tab-1} reports inner rings for four galaxies,
KIG 481, KIG 490, KIG 620, and KIG 841, and a outer ring for KIG 733. The ring in
KIG 481 (NGC 3682) is included in the \citet{Comeron2014} catalogue among
resonance rings in 
%\LEt{again, please switch this so that the spelled-out version becomes part of the main text}
the  {\tt Spitzer} Survey of the Stellar Structure
in Galaxies (S$^4$G), where the galaxy is classified as SA(r)0$^+$). After our model
subtraction, the galaxy shows a strong central dust lane that perturbs
the fit. It also shows concentric shell structures.

We detect inner ring-like structure in the KIG 412, KIG 517, 
KIG 578, KIG 620, KIG 636, KIG 637, KIG 644, KIG 685, and KIG 733 galaxies,
of which seven fulfill the strict isolation spectroscopic criterion by \citet{Argudo2013}. 
In the cases of KIG 620, KIG 733, and KIG 733, which do not show obvious signatures
of either interaction or merging, the rings may have been caused by resonances \citep{Buta2010,Laurikainen2011,
Laurikainen2010}. However, most of the detected rings are associated with other residual structures, 
such as shells or fan,  that the literature assumes are merging signatures.
%\LEt{again, please check your LaTeX commands for the next references} 
\citet[][]{Eliche2018} and \citet[][and references therein]{Mazzei2019} showed that the existence 
 of embedded inner components,  such as inner discs, rings, pseudo-rings, inner spirals, 
 and nuclear bars at the centres of many S0s, which commonly are attributed to internal secular evolution, 
 disc fading, or environmental processes, are compatible with a major-merger origin, 
 including the relaxed and ordered discs of  present-day S0s. Most of our iETGs show all
 these features, in combination with specific signatures of accretions/merging, such as
 shells. This corroborates this view.
  
%-------------------------------- Figure 9 ---------------------
\begin{figure}
\center
\includegraphics[width=8.7cm]{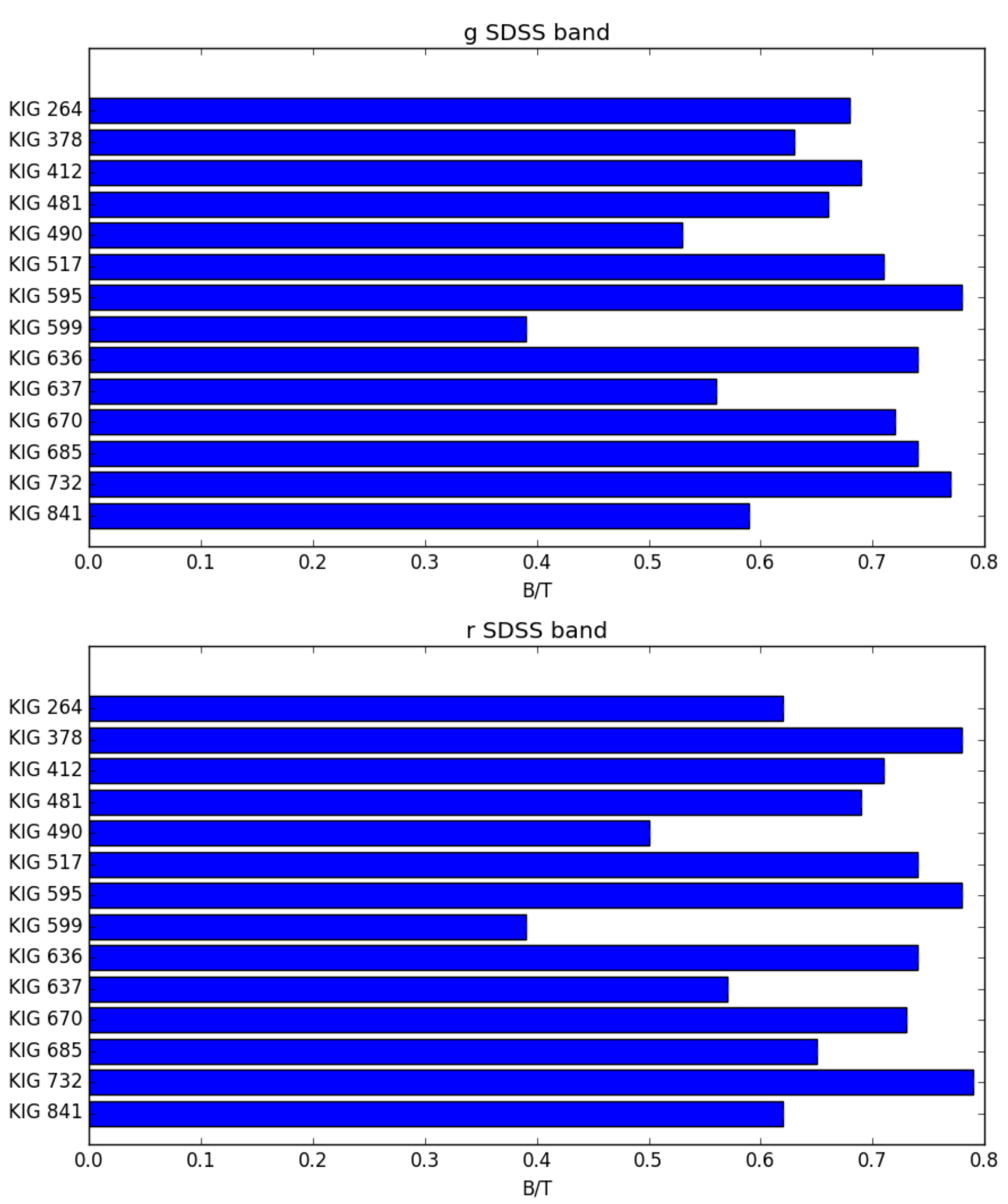}
\caption{Distribution of B/T  in the $g$ (top panel) and 
$r$ bands. The average  B/T values is 0.66 in both bands.}
\label{figure-9}
 \end{figure}
%--------------------------end figure 9 -----------------------

\subsection{\Hi\ content}
\label{HI}

The \Hi\ content of iETGs is important for at least two reasons. 
On the one hand, this content is connected to the
ETGs environment, and on the other, to their evolutionary scenario. More specifically, 
because we found so many fine structures, the \Hi\ content  is essential to distinguish  
between the wet versus dry version of the merging episode that iETGs appear to come from.

It has been known since the early 1980s that cluster spirals are gas an\ae mic with respect to 
their counterparts in a less dense environment \citep[see the pioneering work of][]{Giovanelli1982}. 
 ETGs, which as a family contain less \Hi\  than spirals,
have only recently been found to share this property with spirals. 
This has been possible through high-sensitivity surveys. A significant fraction of ETGs in a low-density
environment are more \Hi\ rich than cluster ETGs (10\% in Virgo vs. 40\% in the field)
\citep[see e.g.][]{Serra2012}. ETGs in a low-density environment may contain as much \Hi\  
as a spiral galaxy.

\citet{Serra2012}
%\LEt{you cannot refer to a reference given in parenthesis with a pronoun in the main text. Please repeat the reference here for correctness} 
found that as far as \Hi\ properties are concerned, the main difference between ETGs 
with large amounts of \Hi\ and spirals is that the former miss the high column density \Hi\
 that is typical of the bright stellar disc of the latter. The star formation rate 
 per unit area is much higher in spirals than in ETGs. Instead,  \citet{Serra2012} 
 concluded that the column density of \Hi\ found in ETGs is similar to the densities observed
 in the outer regions of spirals, which from a  morphological and kinematic point of view 
 show warps and other peculiarities that are connected to gas accretion. \citet{Serra2012}
 asserted that this is what is detected in \Hi-rich ETGs, that is, \Hi-rich ETGs and spirals may 
 look very similar in their outskirts
 %\LEt{direct questions work well in an oral presentation, but not in a formal paper. Please rephrase the two questions below and all others so that they become proper text. Moreover, a single sentence does not constitute a paragraph. Please either merge the short single sentence or add to it}.

We investigated if  those of our iETGs that 
 %\LEt{or did you mean "those of our iETGs that show"?} 
 show many morphological disturbances are gas rich. The  sample of observed iETGs is very small.
\citet{Jones2018} recently presented the largest catalogue of \Hi\ single-dish observations 
of isolated galaxies as part of the multi-wavelength study of the AMIGA sample of 
isolated galaxies. Only two galaxies in our sample were detected in 
\Hi\ (CIG 264 and 481):  KIG 685, and KIG 841. They lack \Hi\ data. The others have 
just 5$\sigma$ upper limits.  Based on the distances given in Table~\ref{tab-1}, 
Table~\ref{tab-5} reports the measured \Hi\ masses (col. 5) and the expected \Hi\ mass (col. 6) 
according to the relations provided in  \citet{Jones2018}.  With the iETG upper limits, 
the $\log M_\mathrm{\Hi\ - pred}$ 
 provides values in the range of KIG 264 and KIG 418, that is, similar to
 median values for spirals (log M$_{\Hi}/M_{\odot}$=9.47, 9.84, and 9.59
for T$\leq$3, 3$<T<$5, and T$\geq5$, respectively). We are aware that the relations 
in  \citet{Jones2018} were derived from a sample that is dominated by spirals because they 
are the vast majority of  the AMIGA galaxies. 

In the HI survey of  ATLAS$^{3D}$ ETGs,  \citet{Serra2012} observed KIG 637 
(NGC 5687). They reported an upper limit that after adjustment to our distance 
(Table~$\ref{tab-1}$) and homogeneisation to \citet{Jones2018} is log M$_\HI<$8.12 M$_\odot$.  
 \citet{Serra2012} only detected about 40\% of field ETGs, therefore this tight limit by itself does 
not suggest that other iETG are necessarily poor in \HI.
 
 We conclude that 1) for most of these iETGs, an \Hi\ 
 component cannot be excluded because the expected HI content is generally 
below the upper limits of the surveys, and 2) the two existing detections in shell galaxies
are in the range of the \Hi-rich ETGs in the \citet{Serra2012} data set.

\subsection{iETGs in the (NUV - r) versus M$_r$ colour magnitude diagram}
\label{NUVr}

Colour magnitude diagrams (CMDs hereafter), in particular those built in
UV versus optical bands, have been shown to be a powerful tool for understanding the
evolutionary phases of galaxies. In the plane M$_r$ versus (NUV - $r$),
galaxies occupy well-defined positions \citep[see e.g.][]{Wyder2007}. 
Evolved galaxies, ETGs, and star-forming late-type  galaxies
are located in two well-populated and separated areas of the CMD called
the red sequence and the blue cloud, respectively.  The
intermediate area, less populated by galaxies, has been labelled
green valley \citep{Kaviraj2007,Shawinski2007}. This
area is believed to be populated by transforming galaxies. 
\citet[][]{Mazzei2019}  showed the
evolutionary path followed by galaxies when either encounter or mergers 
drive their evolution. Starting their evolution  in  the
blue cloud as disc galaxies, they reach a point of maximum brightness before they start to
quench their star formation and cross the green valley. They finally reach the red
sequence and in the meantime modify their morphology and kinematic 
properties to become mature ETGs. The crossing time of the green valley 
depends on their mass (brightness). It lasts  $>4$ Gyr for the faintest ETGs. 

\begin{figure}
\center
\includegraphics[width=8.7cm]{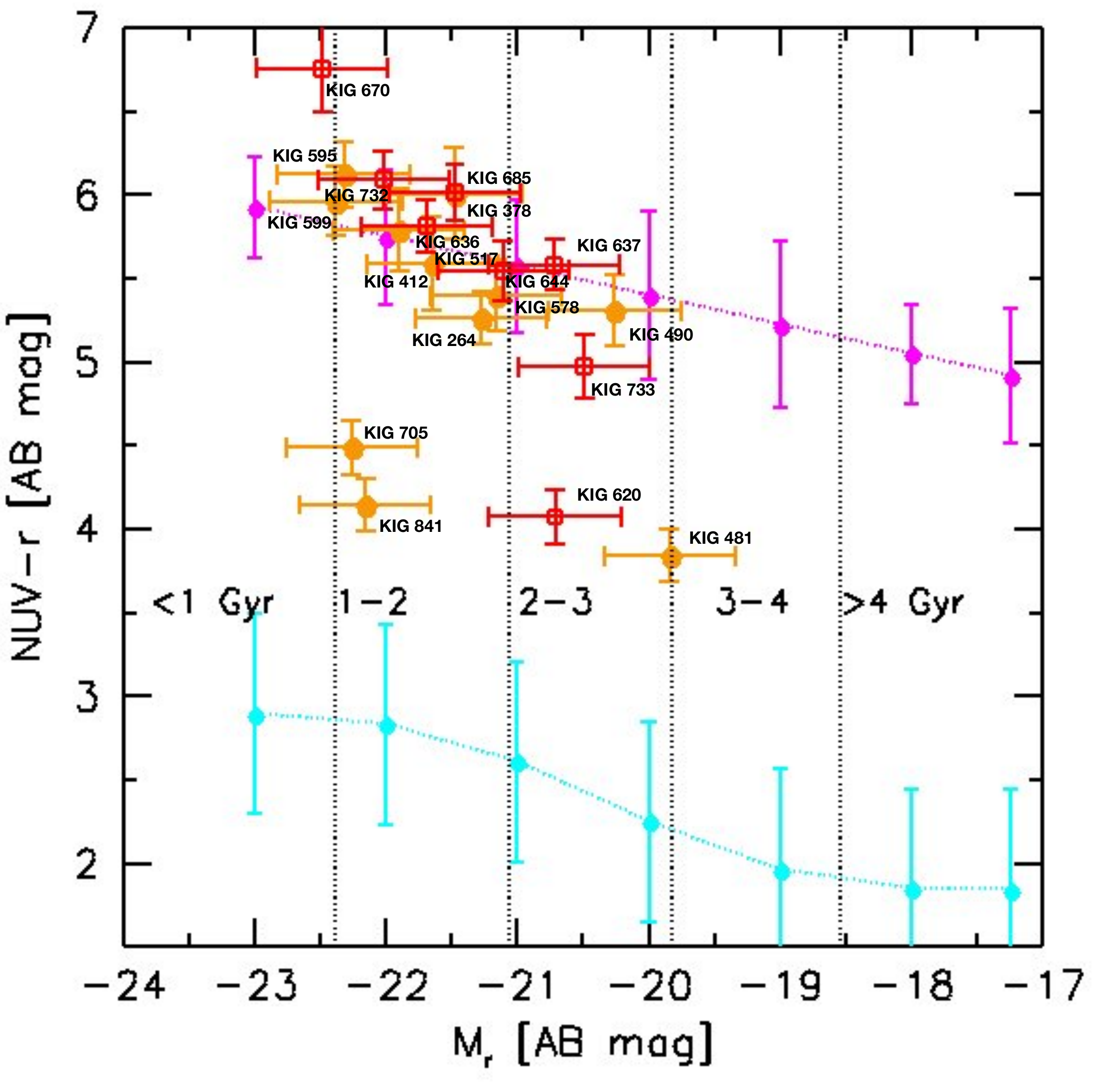}
\caption{UV-optical CMD of iETGs. In the M$_r$ vs. (NUV-$r$) plane we plot the
\citet{Wyder2007} fits to the red sequence (magenta dots and dotted line) and to the 
blue cloud (cyan dots and dotted line). iETGs with a shell system are
indicated with orange dots. KIG 722 is missing from the iETGs sample because no
NUV measures are available. The dotted vertical lines provide an indication of the
green valley crossing time as a function of the galaxy luminosity (see text). The horizontal
error bar accounts for a distance uncertainty of 10\%. The vertical error bar accounts for the 
NUV ZP uncertainty, 0.15 mag \citep{Bai2015}.}
\label{figure-10}
 \end{figure}

Figure~\ref{figure-10} shows the   M$_r$ versus $(NUV - r)$ CMD of our targets.  
The dotted vertical lines, as derived from Fig.8  in \citet[][]{Mazzei2019}, show the  crossing 
time of the green valley as a function  of the intrinsic brightest magnitude 
reached by simulated ETGs  during their evolution in the blue cloud  
(see the quoted paper for a quantitative definition of different regions of the
UV CMD and the evolutionary paths of the galaxies).  
All our iETGs have left the blue cloud. Some of them, both with and without
a shell  (red squares and orange dots, respectively),  are still found in 
the green valley, but they mainly populate the red sequence.

The \citet{Mazzei2019} smoothed particle hydrodynamics
%\LEt{you use SPG twice. Please spell this out both times} 
simulations with chemo-photometric implementation, 
anchored to global properties of  11+8 ETGs, showed that the  luminosities  of these ETGs
fade away by 0.5 mag on average,  after they reached the brightest point 
in the blue cloud, and   the galaxies  enter in the green  valley.
On this basis, the four iETGs in the green valley  will spend between  
1 up to 3 Gyr in this area before they reach the red sequence. 
 iETGs that are located 
in the red sequence on average took less than 3 Gyr to reach it. However,
it is unknown how long they have been and will remain in the red sequence.
 A careful analysis and a match of their global multi-wavelength properties 
are required to constrain our smoothed particle hydrodynamics simulations 
with chemo-photometric implementation, to understand their 
formation mechanisms and their evolutionary path  \citep{Mazzei14b,Mazzei2019}. 
This will be the subject of a forthcoming paper.

We investigated if   iETGs in the red sequence are so-called red-and-dead systems.
It is known that the star formation in ETGs in the red sequence 
cannot be entirely quenched because {\it secondary} episodes of star
formation can still be revealed. Their signatures have been found from 
optical to UV and mid-IR wavelengths \citep[see e.g.][and references
therein]{Annibali2007,Rampazzo2007,Panuzzo2007,Panuzzo2011,
Rampazzo2013,Jeong2009,Salim2010,Thilker2010,Marino2011a,Marino2011b,
Cattapan2019}. The \citet{Mazzei2019} simulations suggest that these phenomena
are activated during the post-merging evolution of iETGs.
In this context,  KIG 264, a former AGN, has already reached the red
sequence, where most of our iETGs are located, and it still shows the remains 
of this activity.

%-------------------------------- Figure 10 ---------------------
%--------------------------end figure 10 -----------------------

\section{Summary and conclusions}

We presented the morphological and photometric study of 20 iETGs from the
AMIGA catalogue \citep{Verdes-Montenegro2005} that are bona fide ETGs according to the 
revised classification of \citet{Fernandez2012}. The revised classification contains 
about 100 ETGs. Although the morphological classification has been revised both
visually \citep{Buta2019} and quantitatively \citep{Hernandez2008} using
SDSS, several discrepancies remain. One of our task has been 
to test the reliability of the above morphological classifications in the 
light of deep imaging in a quantitative way.

The AMIGA isolation criteria have been refined by \citet{Verley2007b} and by 
\citet{Argudo2013}. Four of the iETGs considered in this paper 
fulfil the \citet{Verley2007b}  isolation criteria with a trend to present a 
larger number of minor companions (higher $\eta_k$ value),
while 10 out of 20 are strictly isolated for  \citet{Argudo2013} 
(who lack spectroscopic information for 4 galaxies in the sample, however),
the other is expected to be only minimally affected by any neighbours.
The iETGs in our sample are hence isolated from major companions 
and located in poorly populated environments. 

Galaxies were observed in the $g$ and $r$ SLOAN bands at the 1.8m
VATT telescope with the 4KCCD. The light profiles on the average reach 
$\mu_g$=28.11$\pm$0.70 mag~arcsec$^{-2}$ and 
$\mu_r$=27.36$\pm$0.68 mag~arcsec$^{-2}$,
which makes this sample the deepest observations of iETGs so far. We used the {\tt
AIDA} package for the 2D photometric analysis and accounted for PSF effects
during the decomposition of the light profiles up to the galaxy outskirts. 
We used the current literature to discuss the morphology of the residuals 
obtained after the 2D model subtraction from the original image.
We list our results below. 
\begin{itemize}

\item All the galaxies in the sample are bona fide ETGs, from Es to late S0s. None
is a misclassified spiral.

\item Fourteen out of 20 iETG light profiles are best fit by B+D model.
The average B/T is 0.66 in both bands, indicating that the bulge dominates 
the galaxy light distribution.

H-T found that bona fide Es are rare among iETGs.  KIG 578,
KIG 705, and KIG 722 are best fit by a simple S\'ersic law whose exponent is
near to a classic de Vaucouleurs law (see Table~\ref{tab-4}). These galaxies are
in the Es set of H-T.  However, we fit 
KIG 264, KIG 599, and KIG 732 best with a B+D model. 
These galaxies are all in the H-T Es bona fide sample.

Our B+D models failed to provide an acceptable  description 
of the light distribution of  iETGs with type close to 0, 
%\LEt{it is unclear what you meand by the double-lined arrow, please use words}
KIG 620, KIG 644, and KIG 733 because strong additional features
such as one or more rings and wide arms are present in their central region. 
For the latter, \citet{Buta2019} provided the best classification. 
Their one or more rings may be interpreted in either the framework of resonances 
or other phenomena that have been described as a consequence of secular evolution 
\citep[see e.g.][]{Comeron2014,Buta2010,Laurikainen2010,Laurikainen2011} 
or as a consequence of merging
\citep[see e.g.][and references therein]{Eliche2018,Mazzei2019}.

\item Most of our iETGs show fine structures such as shells, 
tails, residual faint rings, stellar fans, and a residual spiral arm-like structure after model 
subtraction. These features are considered signatures of accretion and merging events 
in the literature.
 
\item In 12 out of 20 (60\%) of the iETGs, we detected shells. They compete with rings (10 out of 20) 
for the most frequently dectected fine structures \citep[see also][]{Tal2009}. This percentage
is confirmed also when we considered the set of iETGs that was smaller because stricter isolation
criteria were used. H-T also found very many shell galaxies in isolated ETGs. 
However, in  Table~6 of H-T, the detection of 4 out of 9 shells and ripples 
was uncertain, while we revealed shells in KIG 264, KIG 578, 
KIG 722, and KIG 732, which are listed in that table.  This is expected because H-T used 
more shallow images \citep[see the considerations in][]{Mancillas2019}.

\item Most ($\approx$ 80\%) of the iETGs  are located in the red sequence in the 
(NUV-$r$) versus M$_r$ CMD diagram.  Seven out of 15 (KIG 722 lacks NUV data) 
show a shell structure. Four galaxies,  that is, KIG481, KIG 620, KIG 705, and
KIG 841, are still located in the green valley. Three of them have shells.  
Several other features connected to accretion and merging events  are found in iETGs 
that are located in the red sequence. 

\item Fourteen iETGs have nearly flat \gmr\ colour profiles at $\approx$0.7-0.8 mag. These are 
normal values for early-type galaxies. We measured  \gmr\ colour profiles
that become increasingly blue with radius for KIG 481, KIG 620, and KIG 841, 
at least in some regions. In contrast, quite red colours are measured
in the outskirts for KIG 264 and KIG 378, suggesting the presence of dust. Dust is 
normally expected in the galaxy centre.

In  a small number of iETGs, the blue \gmr\ colour profile may suggest a recent 
star formation episode  by the high fraction of accretion and merging signatures. 
 However, the time required to cross the (NUV-$r$) versus M$_r$ 
 plane from the green valley to the red sequence is relatively long, as shown in
 Figure~\ref{figure-10}. This suggests that iETGs have had enough time to quench their star formation
and AGN activity (as in KIG 264) in the case of wet mergers as well.
This is in contrast to H-T and \citet{Tal2009}, who  suggested that dry mergers 
might provide a possible explanation of the evolution of these galaxies.

\item The prediction by \citet{Jones2018} about the \Hi\ content of our iETGs needs 
to be confirmed with deep 21 cm observations because they are based on a 
relation fit with a sample dominated by spiral galaxies. 
The wide  morphological disturbance in the galaxy outskirts and the measure of  M$_{\Hi}$ 
for KIG 264 and KIG 481 suggest that \Hi\ might have fuelled galaxy rejuvenation and AGN activity,
 as has been found in low-density environments \citep[see e.g.][]{Annibali2007,Annibali2010}.

\end{itemize}

The subsample of iETGs we studied presents a different behaviour from isolated late-type 
 galaxies from the same AMIGA sample, showing a significantly higher number   
 of perturbation signatures. Consistently, \citet{Fernandez2013} found no difference 
 in the stellar mass size  relation between iETGs and counterparts in more dense 
 environments, unlike late-type galaxies.
We conclude that a significant fraction of our iETGs are a by-product of a merger.
The high fraction of shells, a long-lasting fine structure in isolated and poorly populated environments,
indicates that iETGs have not been perturbed for a long time (up to 3 Gyr). This is consistent 
with their degree of isolation. The crossing time required to move galaxies
from the green valley to the red sequence in the M$_r$ versus (NUV-$r$) CMD
is consistent with this view. Shell galaxies that are found in the green valley will
accumulate in the red sequence.
Because of their isolation, iETGs appear as the cleanest environment for
 investigating a wide phenomenology (structure and sub-structure formation, duration,
evolution, AGN feeding and fading, and star formation ignition and quenching)  
 induced by both accretion and merging episodes.
 %\LEt{again, this is a single-sentence paragraph. Please either add to it or merge}.

%%%%%%%%%%%%%%%%%%%%%%%%%%%%%%%%%%%%%%%%%%
\begin{acknowledgements}
A.O. and R.R. are deeply grateful to Fathers Richard Boyle SJ, Paul
Gabor SJ and Jean-Baptiste Kikwaya SJ for assistance at the telescope.
RR thanks the Specola Vaticana Director Brother Guy  Consolmagno, Father
Christopher Corbally and Father Thomas Williams SJ for the kind
hospitality in Tucson (AZ -USA). R.R. and P.M. acknowledge funding from
the INAF PRIN-SKA 2017 program 1.05.01.88.04. 
%L.V.M. acknowledgessupport from the grant AYA2015-65973-C3-1-R (MINECO/FEDER, UE). 
MGJ is supported by a Juan de la Cierva formaci\'{o}n fellowship (FJCI-2016-29685). 
MGJ and LVM also acknowledge support from the grants AYA2015-65973-C3-1-R 
and RTI2018-096228-B-C31 (MINECO/FEDER, UE). This work has been supported by the 
Spanish Science Ministry ``Centro de Excelencia Severo Ochoa'' program 
under grant SEV-2017-0709.
{\tt IRAF}  is distributed by the National Optical Astronomy Observatories,
which are operated by the Association of Universities for Research in
Astronomy, Inc., under cooperative agreement with the National Science
Foundation. This research has made use of the NASA/IPAC Extragalactic
Database (NED), which is operated by the Jet Propulsion Laboratory,
California Institute of Technology, under contract with the National
Aeronautics and Space Administration. We acknowledge the usage of the
{\tt HyperLeda} database  (http://leda.univ-lyon1.fr).
\end{acknowledgements}

%=====================================
% References, variant A: internal bibliography
%=====================================

\begin{appendix}
%All appendix sections must be cited in the main text. In the appendixes, Figures, Tables, etc. should be labeled starting with `A', e.g., Figure A1, Figure A2, etc. 

\section{iETG surface photometry}
\label{AppendixA}

In this section we show the photometric analysis we performed on each galaxy.  
The figures show the $g$ (left panel) and $r$ (right panel) results described in detail
in Section~\ref{Results} and discussed in Section~\ref{Discussion}.

Each panel shows the model of the galaxy we adopted, the residual after the model
subtraction, the luminosity profile in the filter, and the residual (O-C) from the 
model we adopted.

%-------------------------------- Figure A1 ------------------------
\begin{figure*}
\center
{\includegraphics[width=7.8cm]{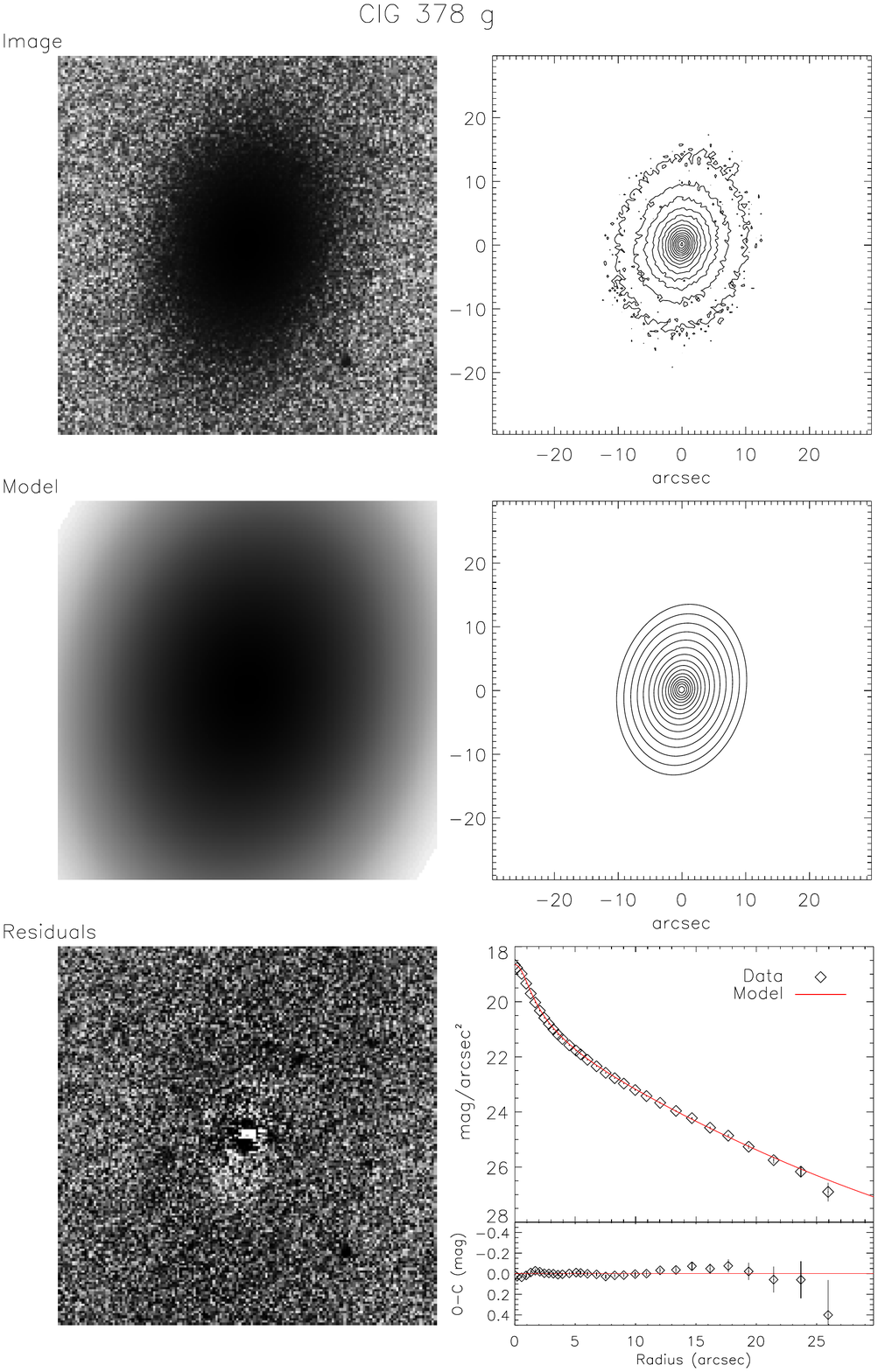}
\includegraphics[width=7.8cm]{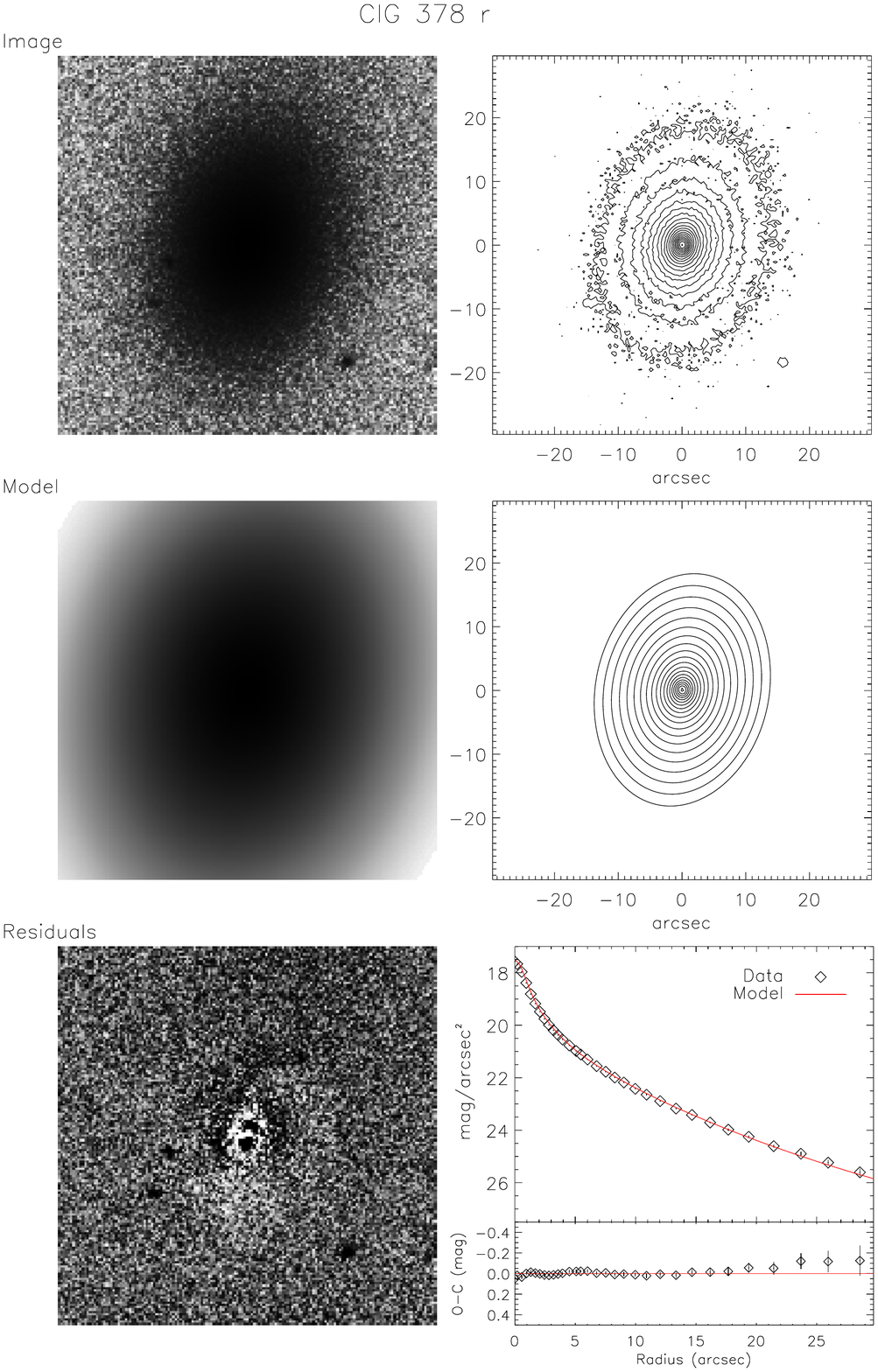}}
{\includegraphics[width=7.8cm]{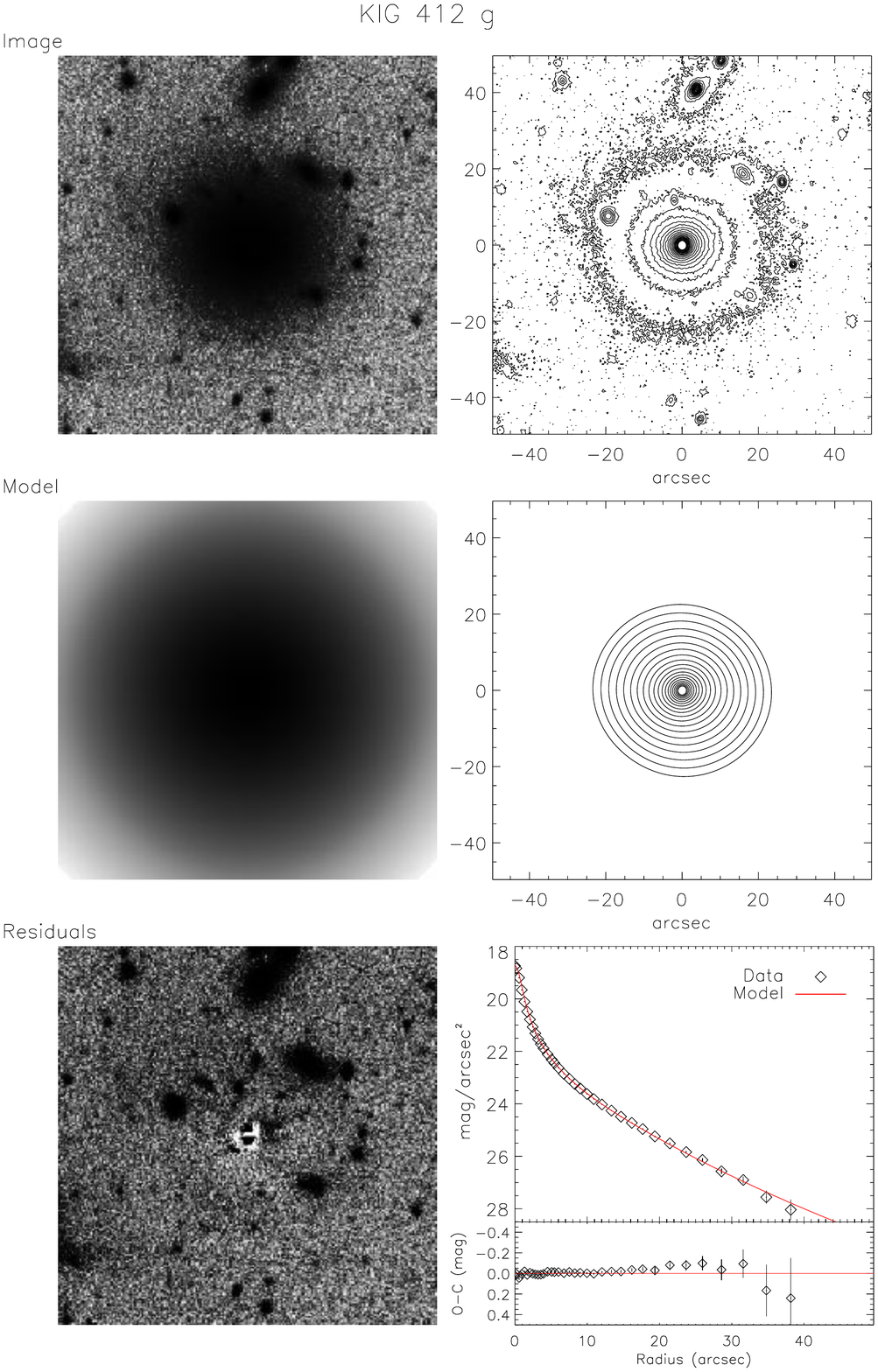}
\includegraphics[width=7.8cm]{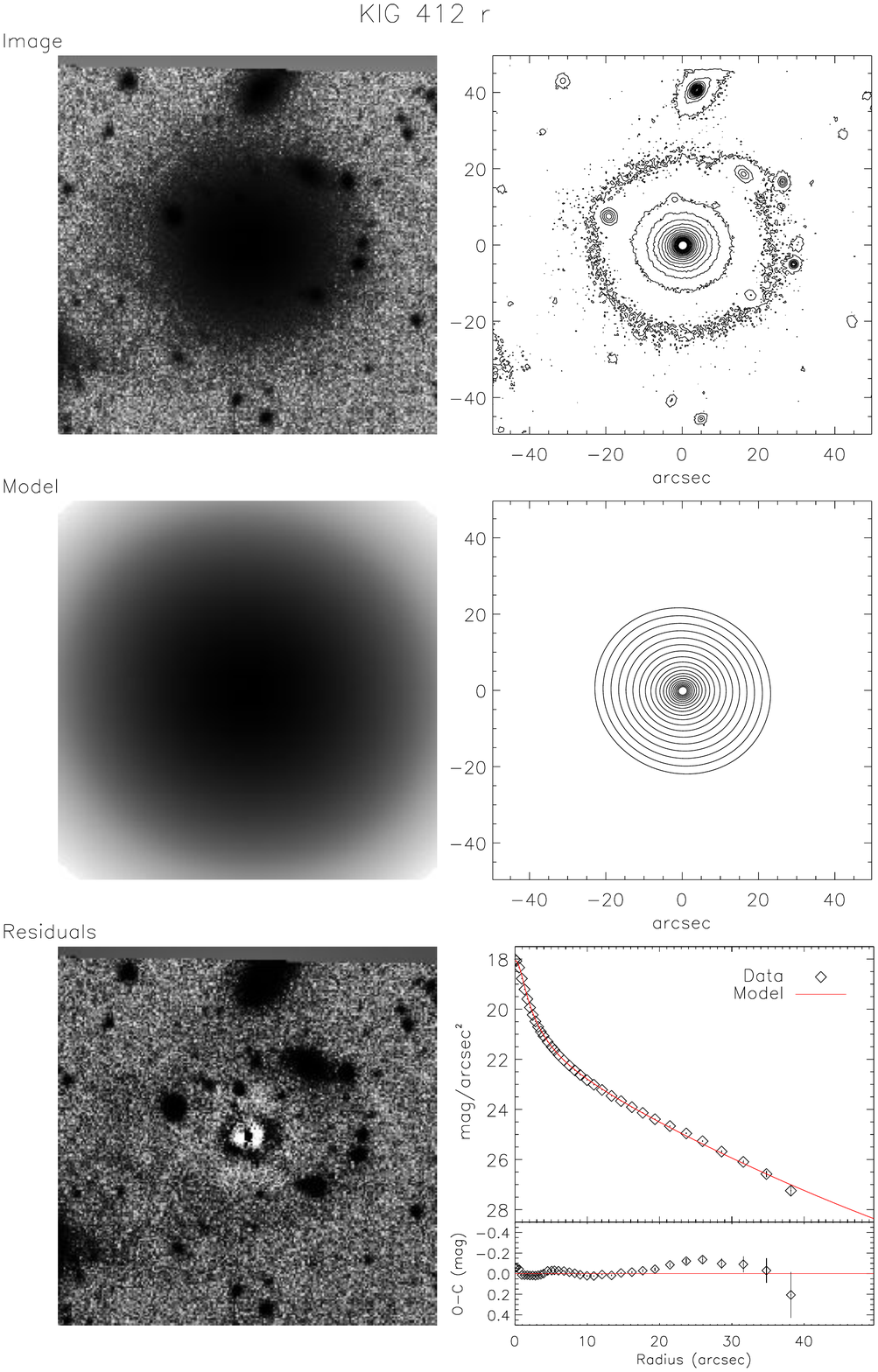}}
\caption{As in Figure~\ref{figure-5}. Summary of the surface photometric analysis of KIG 378 ({\it top panels}) and 
KIG 412  ({\it bottom panels}). In both bands we show the B+D best-fit models. We show 20 isophote levels, between 500 and 
2 $\sigma$ of the sky level,   for the original and model images.}
\label{figure-A1}
 \end{figure*}
%--------------------------end figure A1 -----------------------

%-------------------------------- Figure A2 ------------------------
\begin{figure*}
\center
{\includegraphics[width=7.8cm]{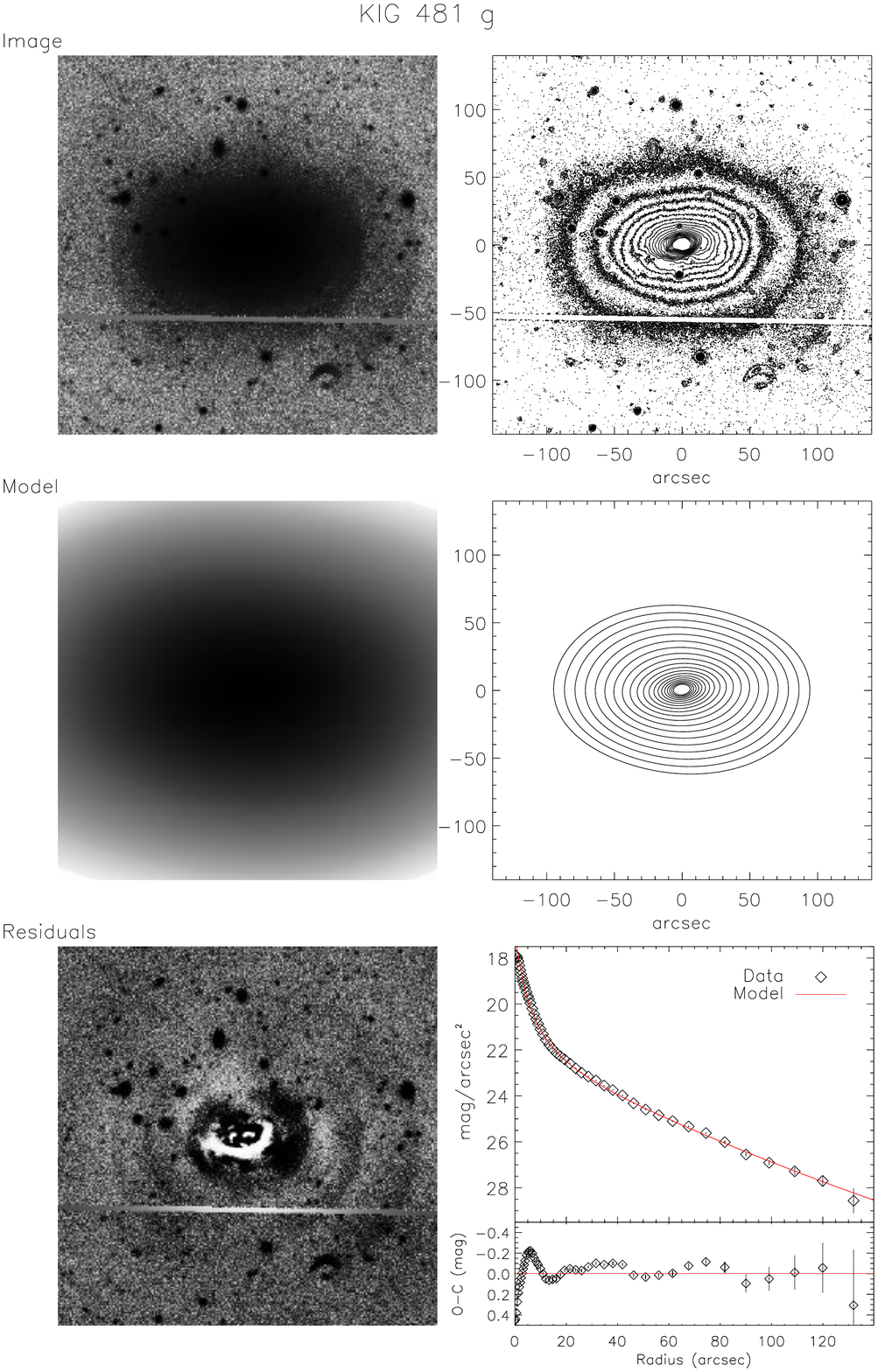}
\includegraphics[width=7.8cm]{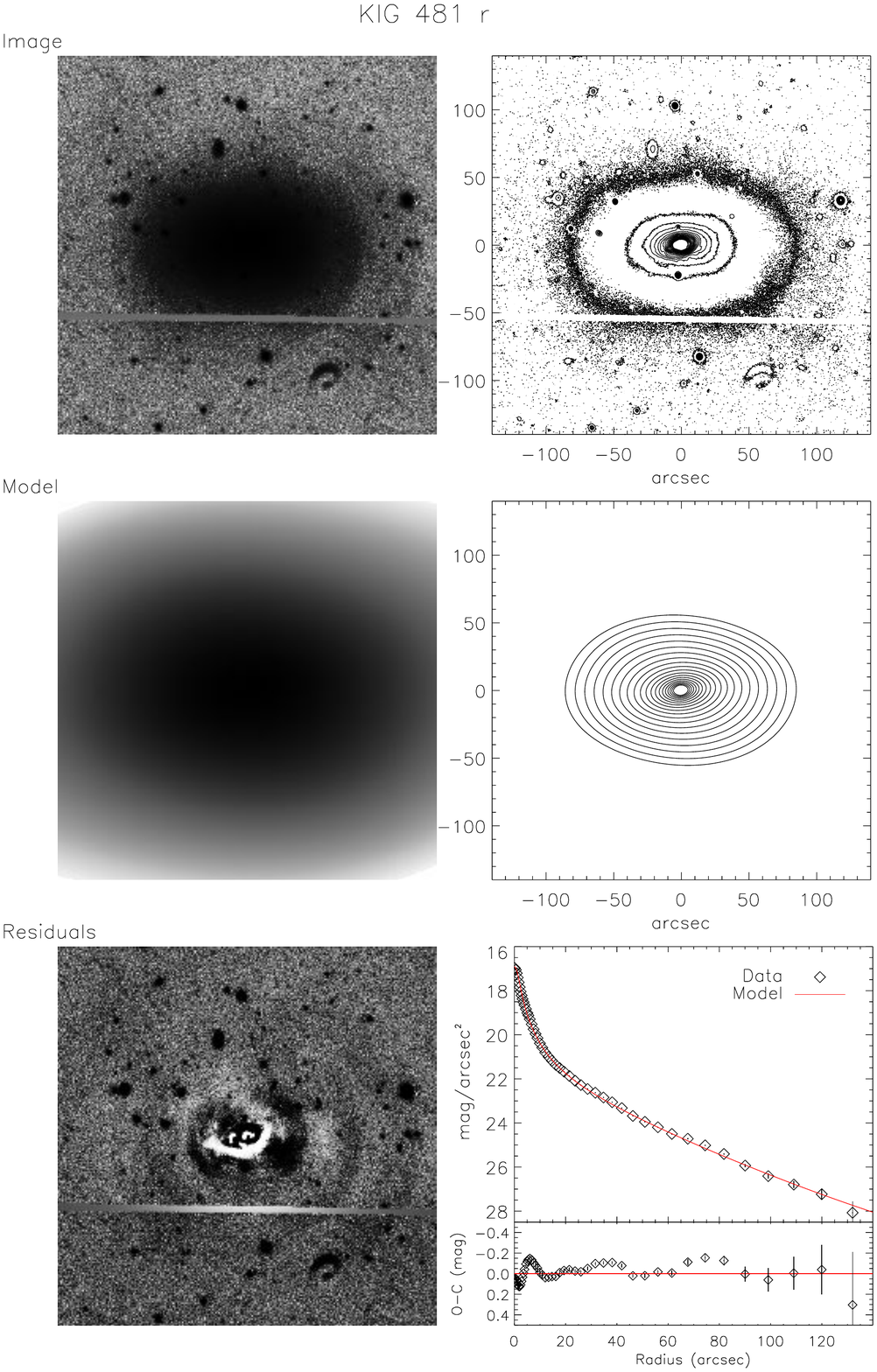}}
{\includegraphics[width=8.2cm]{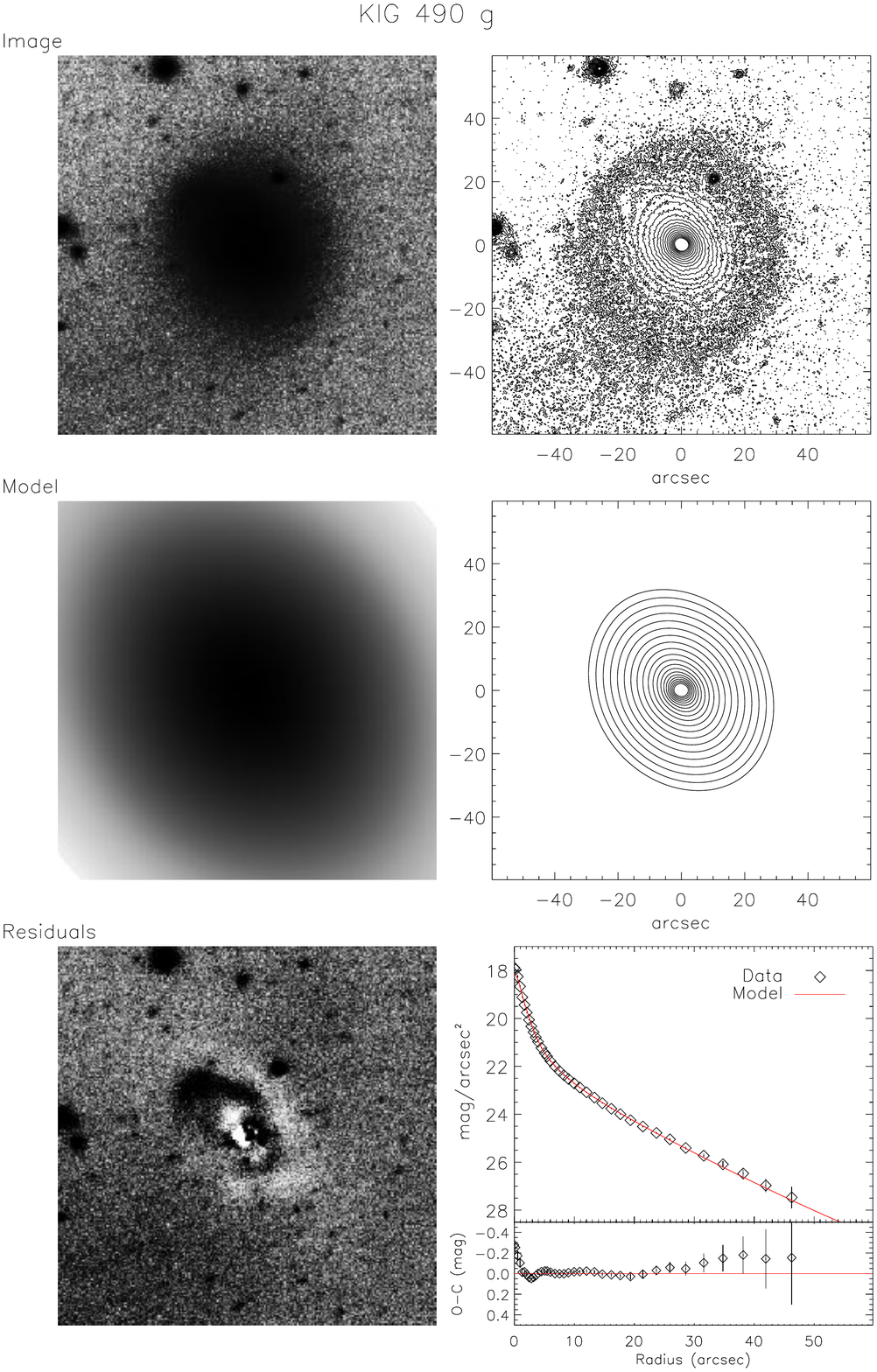}
\includegraphics[width=7.8cm]{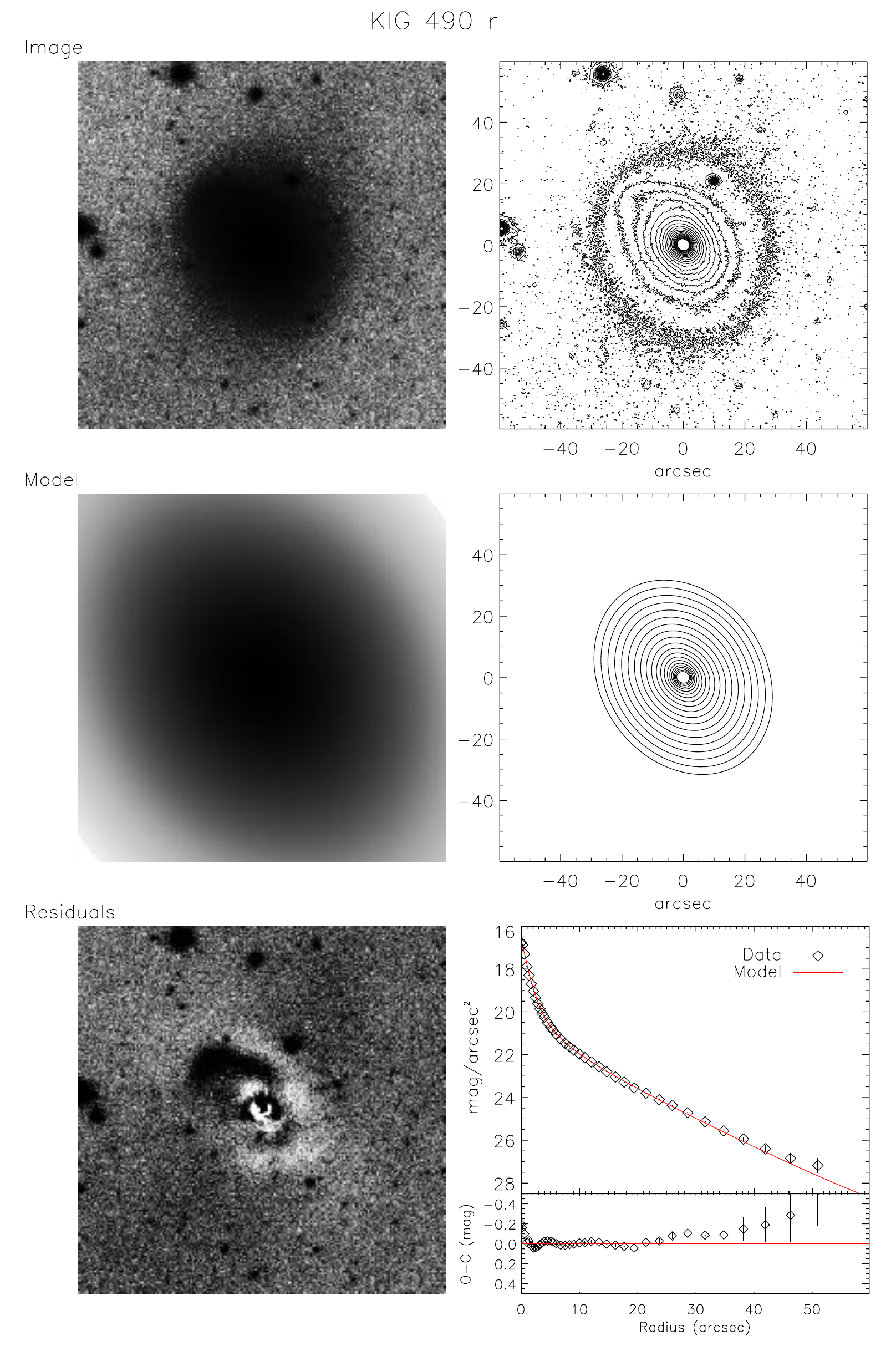}}
\caption{As in Figure~\ref{figure-5}. Summary of the surface photometric analysis of KIG 481 ({\it top panels})
and KIG 490  ({\it bottom panels}). In both bands the B+D best-fit models are shown. 
The KIG 481 image shows the gap in the 4K CCD and a dust feature that was not removed 
by the flat-fielding. We show 20 isophote levels, between 500 and 
2 $\sigma$ of the sky level,   for the original and model images.}
\label{figure-A2}
\end{figure*}
%--------------------------end figure A2 -----------------------

%-------------------------------- Figure A3 ------------------------
\begin{figure*}
\center
{\includegraphics[width=7.8cm]{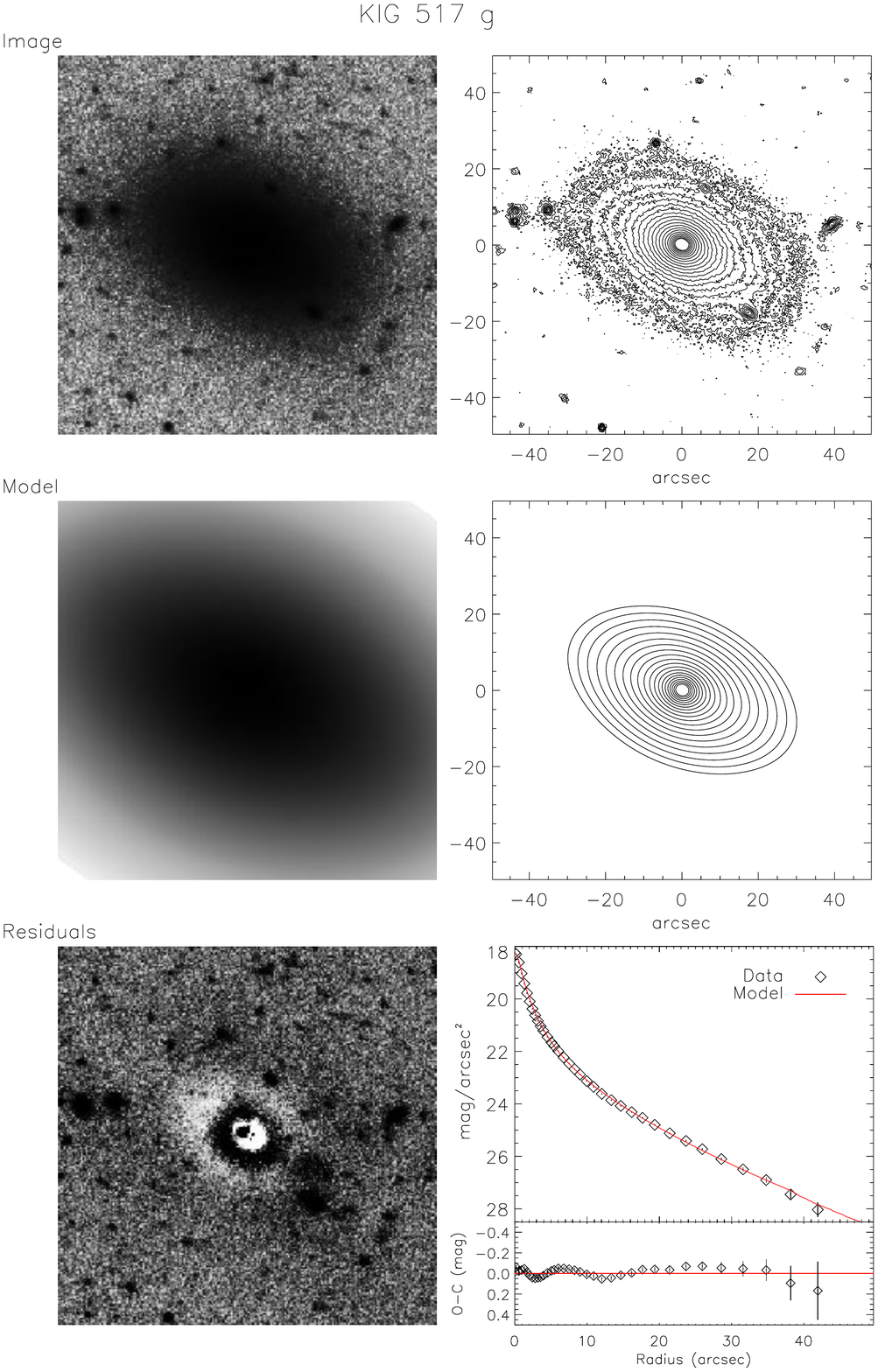}
\includegraphics[width=7.8cm]{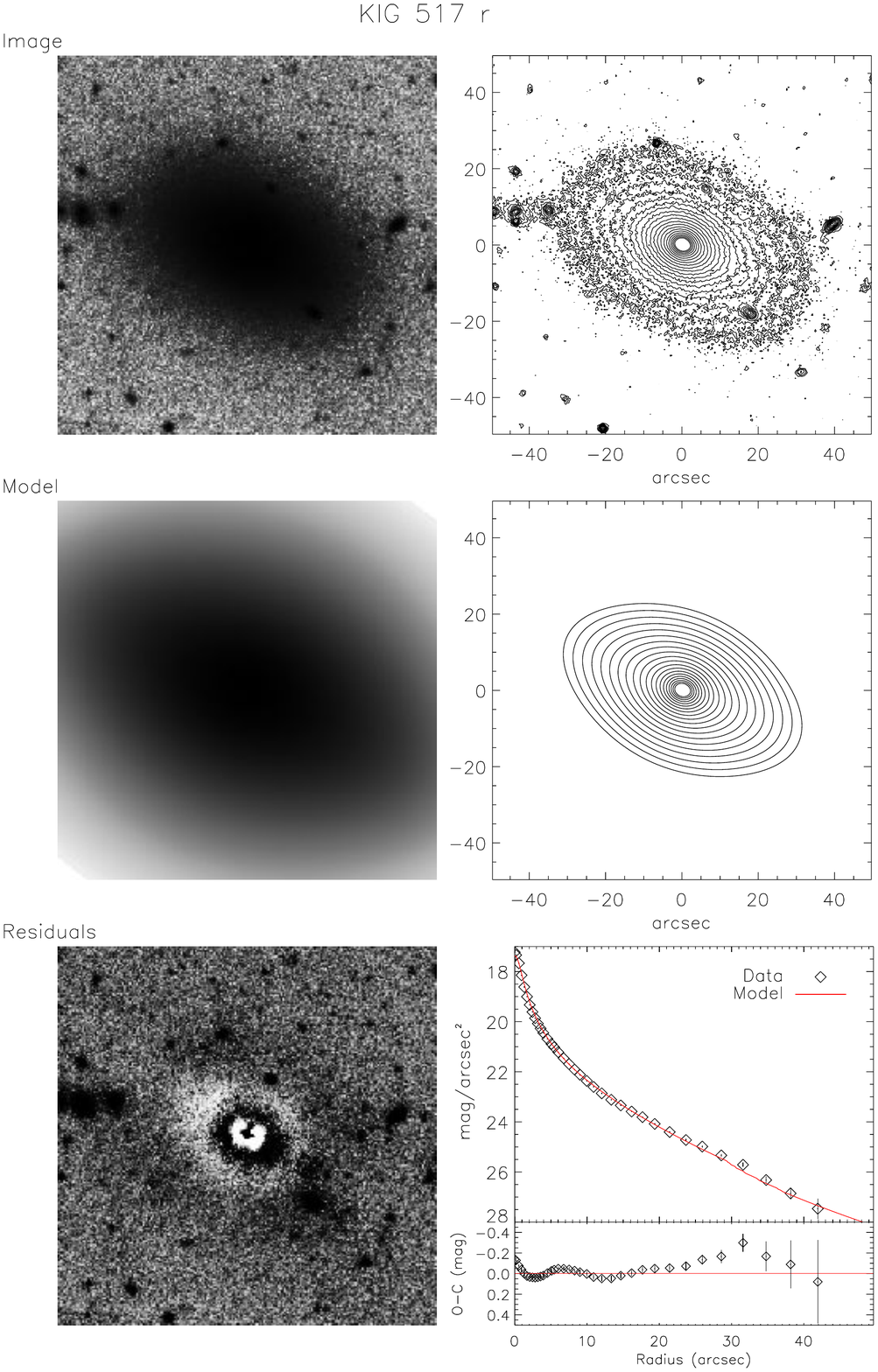}}
{\includegraphics[width=7.8cm]{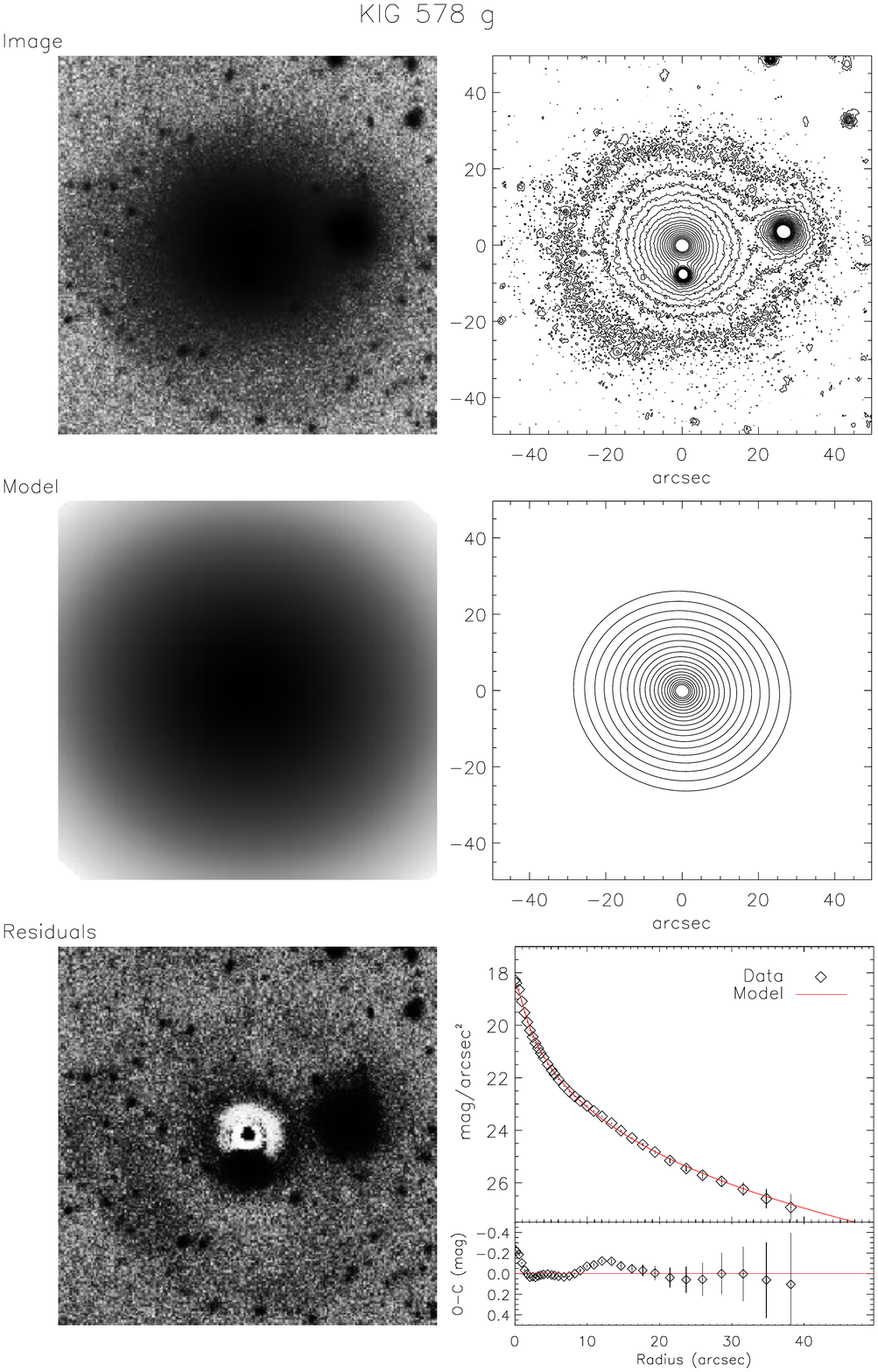}
\includegraphics[width=7.8cm]{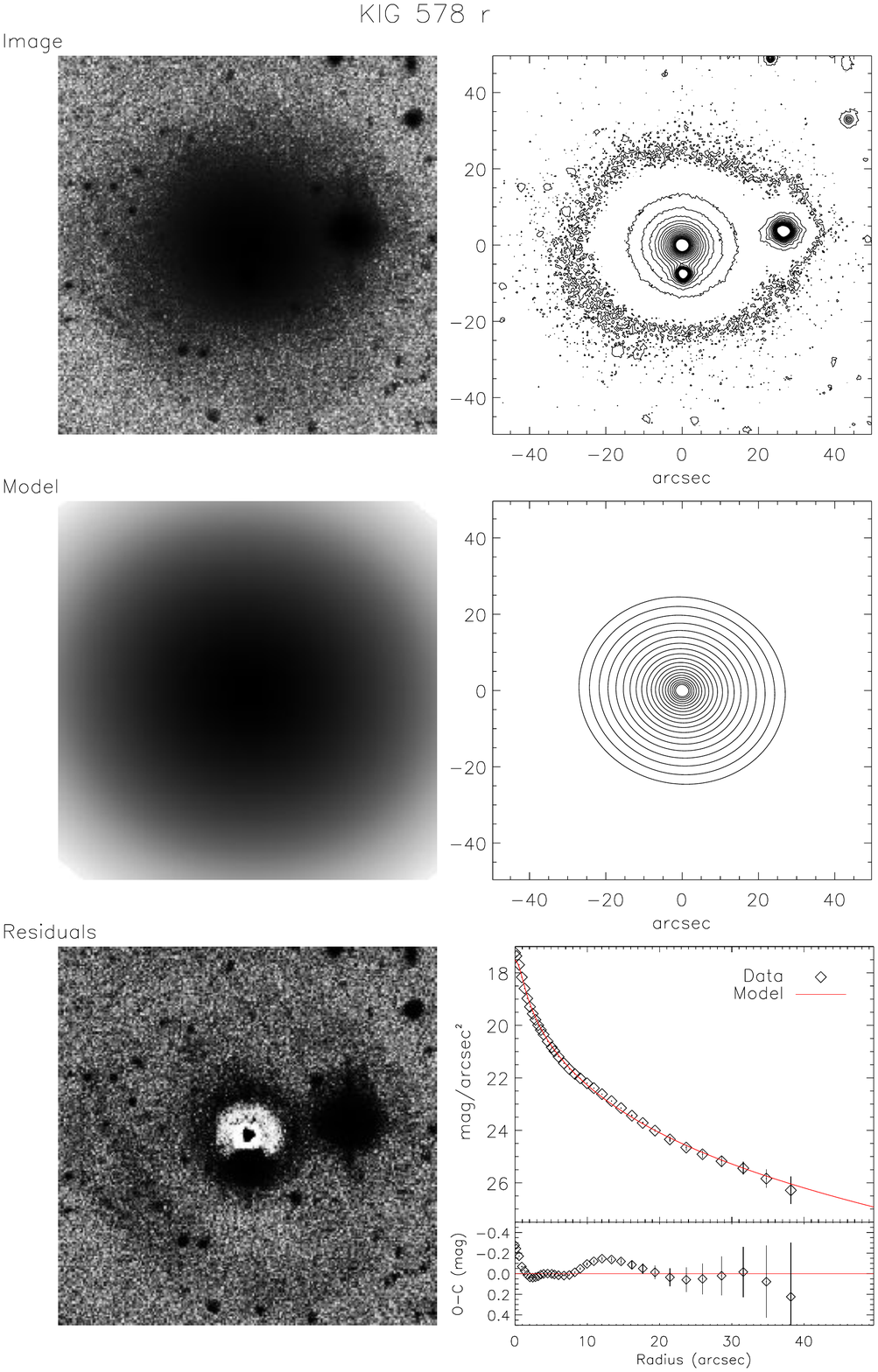}}
\caption{As in Figure~\ref{figure-5}. Summary of the surface photometric analysis of KIG 517 ({\it top panels})
and KIG 578  ({\it bottom panels}). The B+D and S\'ersic best-fit models are 
shown for KIG 517 and KIG 578, respectively.
We show 20 isophote levels, between 500 and 2 $\sigma$ of the sky level,   
for the original and model images.}
\label{figure-A3}
\end{figure*}
%--------------------------end figure A3 -----------------------

%-------------------------------- Figure A4 ------------------------
\begin{figure*}
\center
{\includegraphics[width=7.8cm]{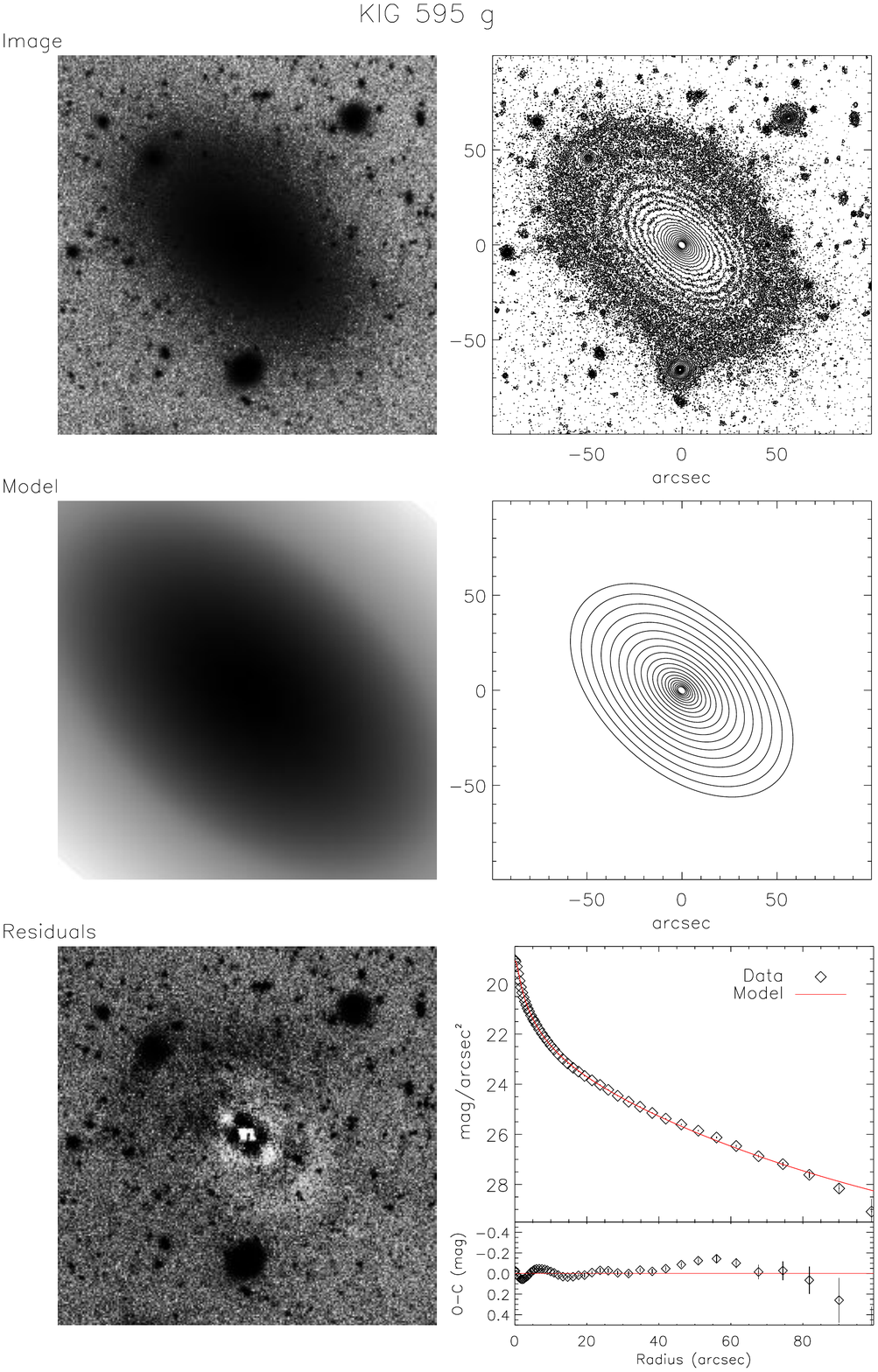}
\includegraphics[width=7.8cm]{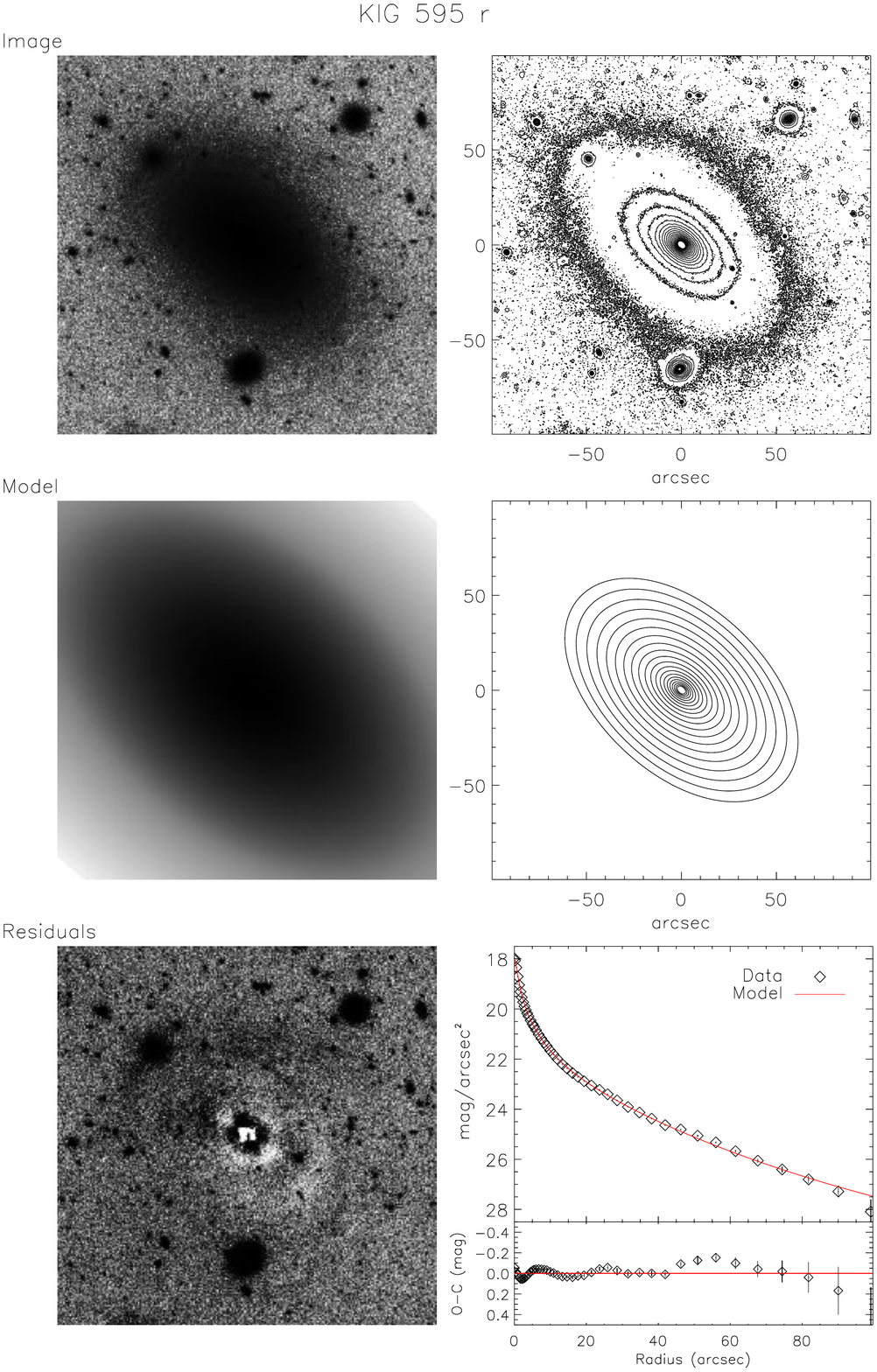}}
{\includegraphics[width=7.8cm]{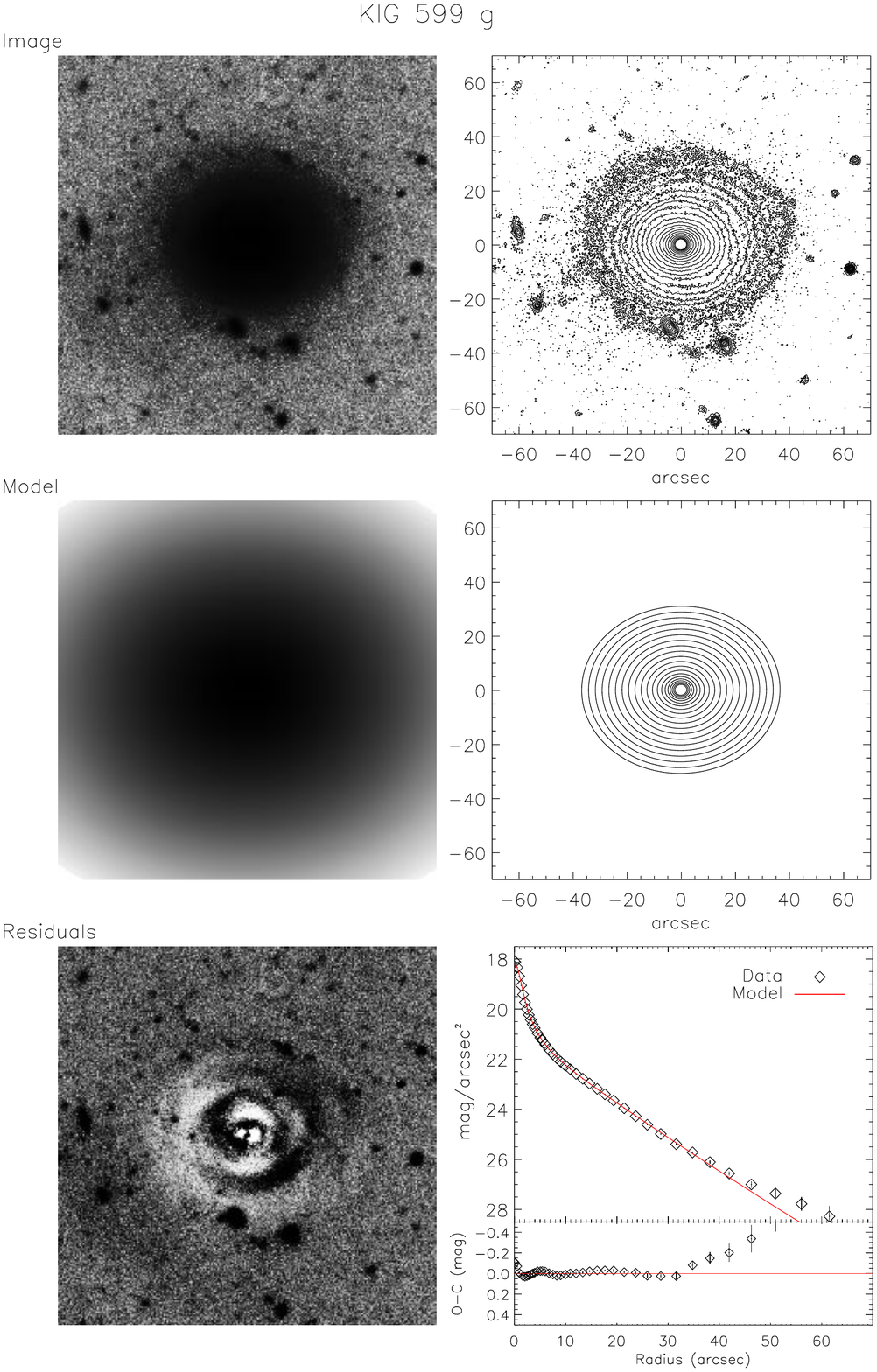}
\includegraphics[width=7.8cm]{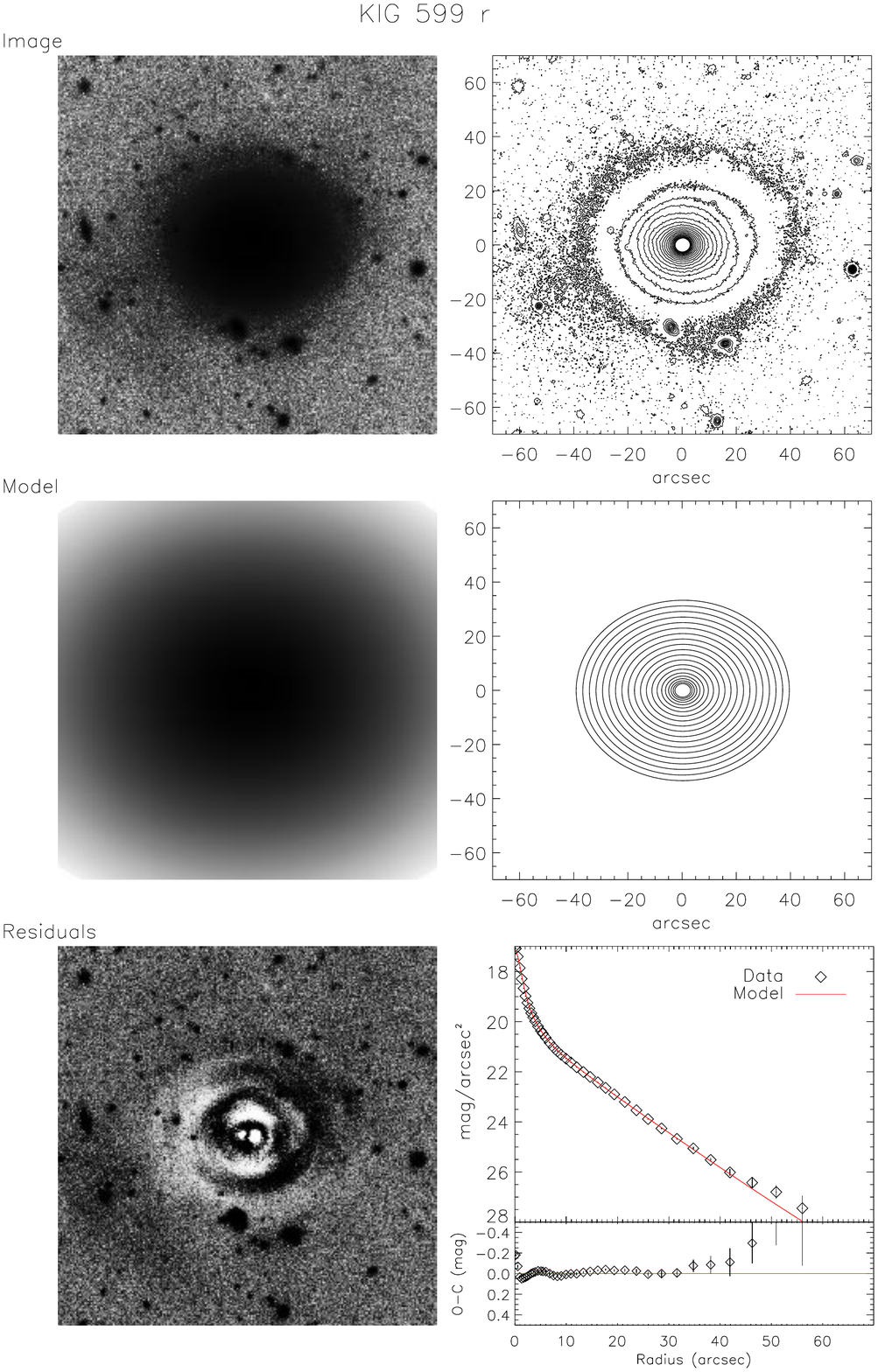}}
\caption{As in Figure~\ref{figure-5}. Summary of the surface photometric analysis of KIG 595 ({\it top panels})
and KIG 599  ({\it bottom panels}). In both bands we show the B+D models. 
We show 20 isophote levels, between 500 and 
2 $\sigma$ of the sky level,   for the original and model images.}
\label{figure-A4}
\end{figure*}
%--------------------------end figure A4 -----------------------

%-------------------------------- Figure A5 ------------------------
\begin{figure*}
\center
%{\includegraphics[width=8.0cm]{KIG620g_B+D.pdf}
%\includegraphics[width=8.0cm]{KIG620r_B+D.pdf}}
\includegraphics[width=15.6cm]{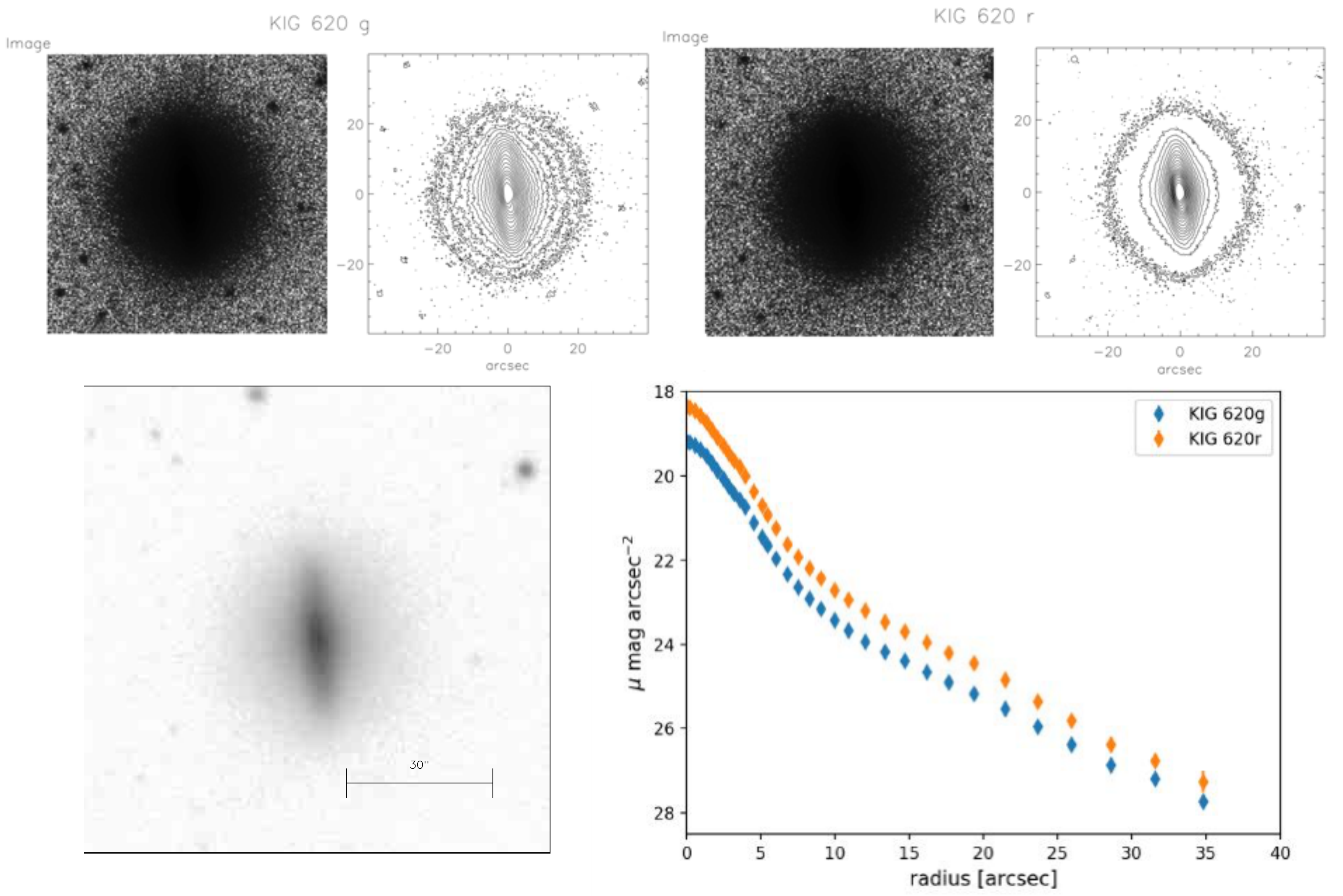}
{\includegraphics[width=8.0cm]{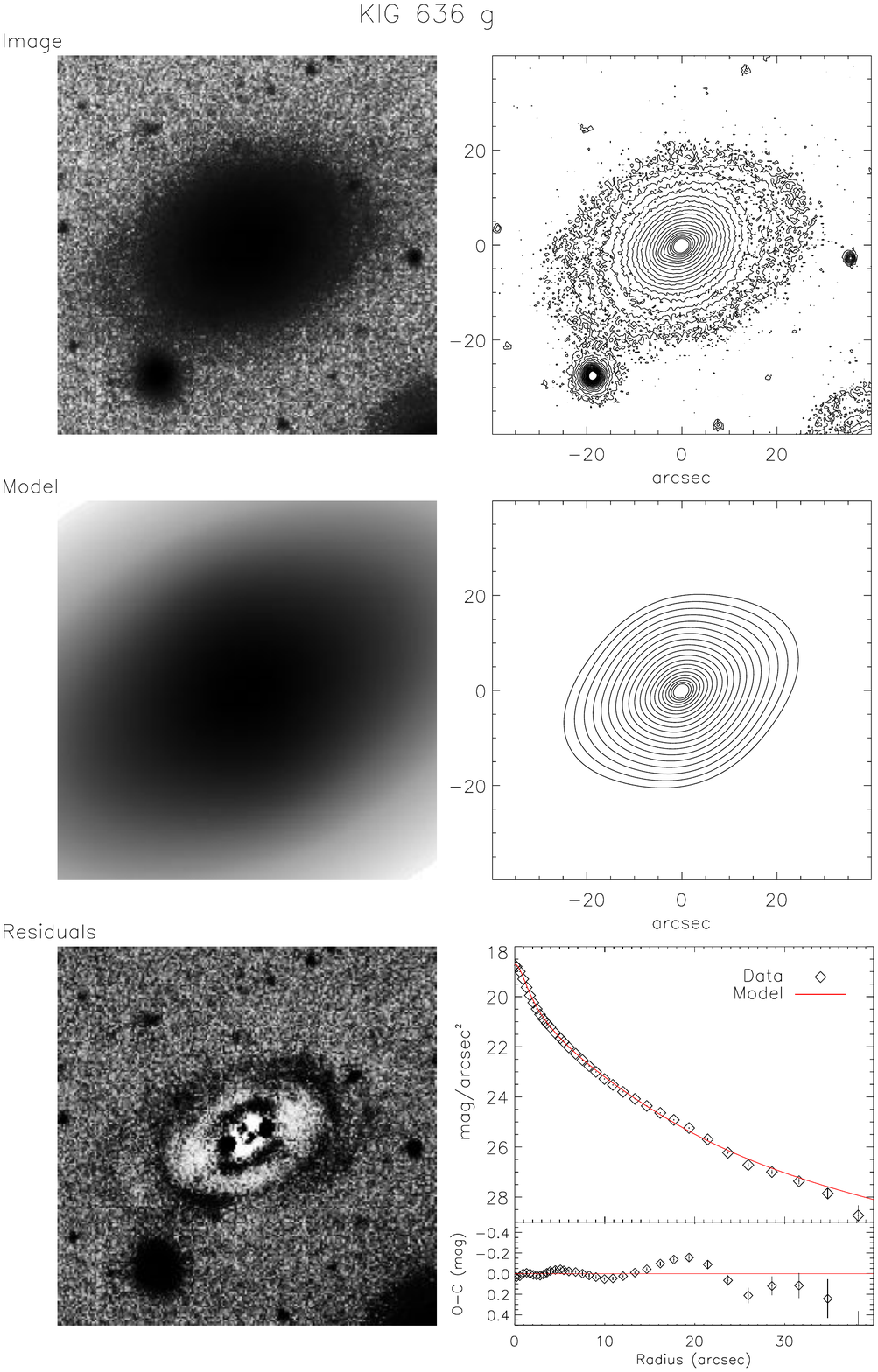}
\includegraphics[width=8.0cm]{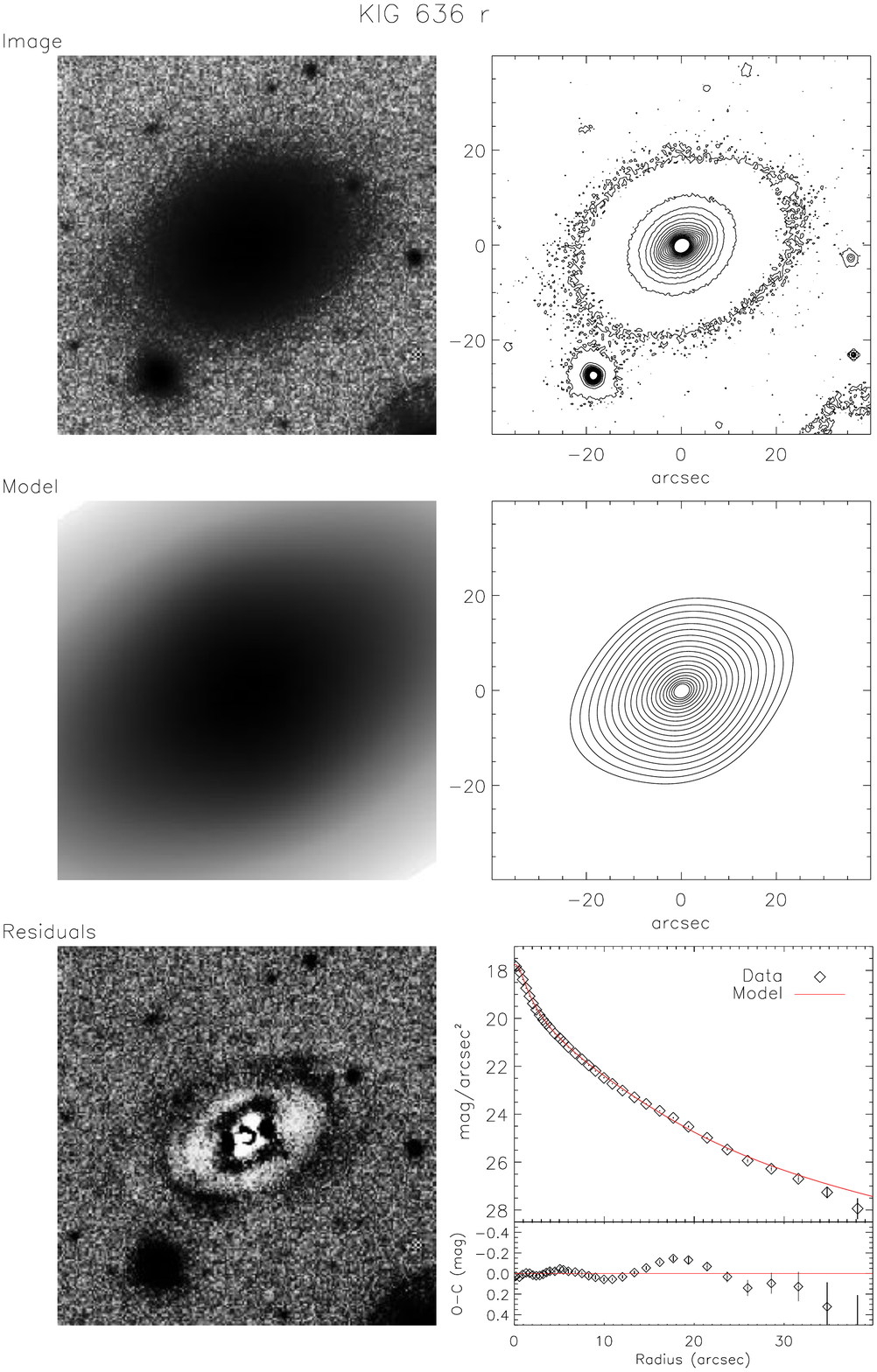}}
\caption{Summary of the surface photometric analysis of KIG 620 ({\it top panels}).  Deep $g$ and $r$
images are shown, with isophote contours down to the 2$\sigma$ level of the sky.
 ({\it Middle panels}) The right $g$ image enhances
the central part of KIG 620, showing the ring structure, and the left panel shows 
the $g$ and $r$ light profiles. ({\it Bottom panels})
As in Figure~\ref{figure-5}, but  for KIG 636. The  B+D model is shown. 
We show 20 isophote levels, between 500 and 
2 $\sigma$ of the sky level,   for the original and model images.}
\label{figure-A5}
\end{figure*}
%--------------------------end figure A5 -----------------------

%-------------------------------- Figure A6 ------------------------
\begin{figure*}
\center
{\includegraphics[width=7.8cm]{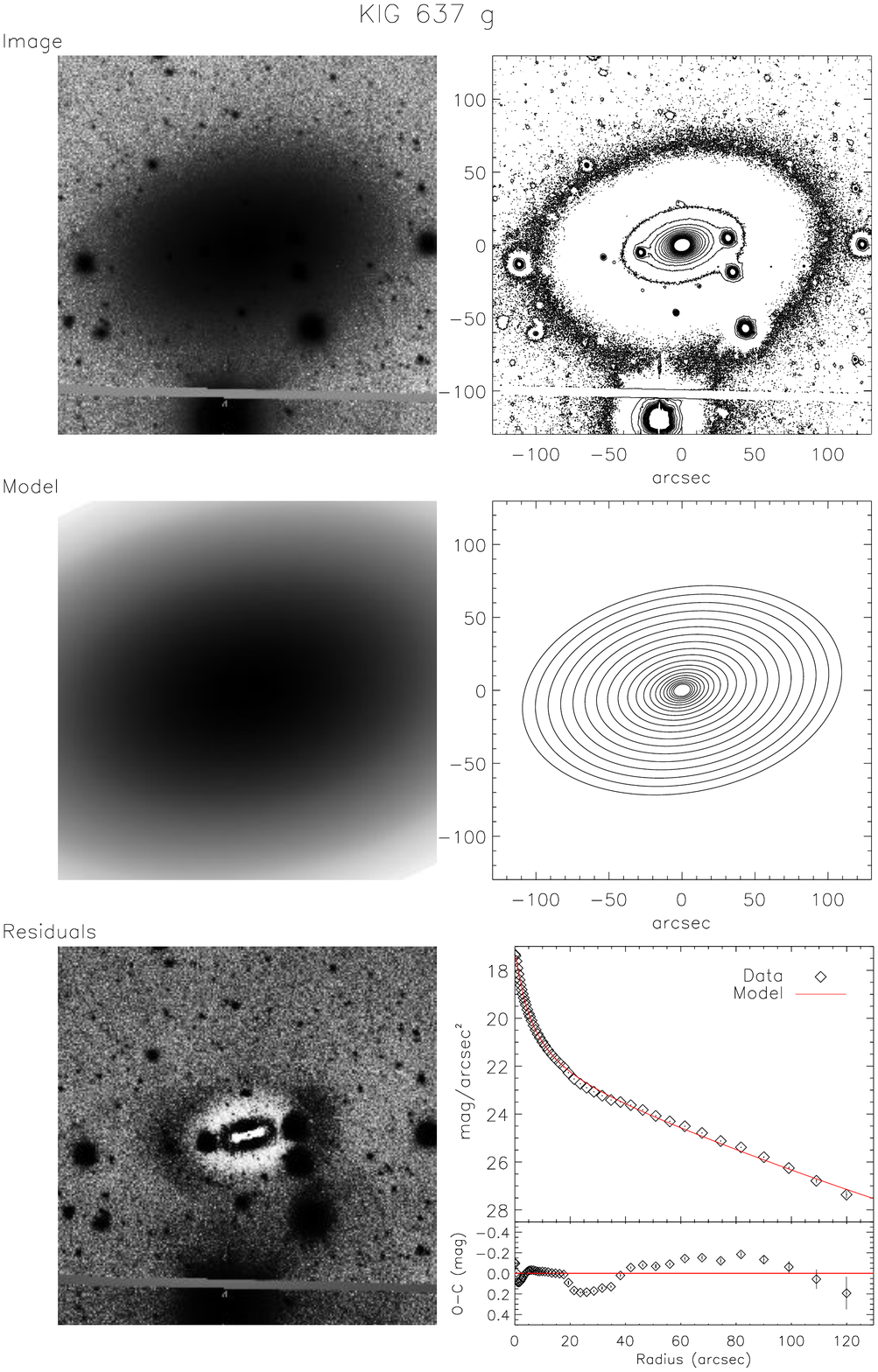}
\includegraphics[width=7.8cm]{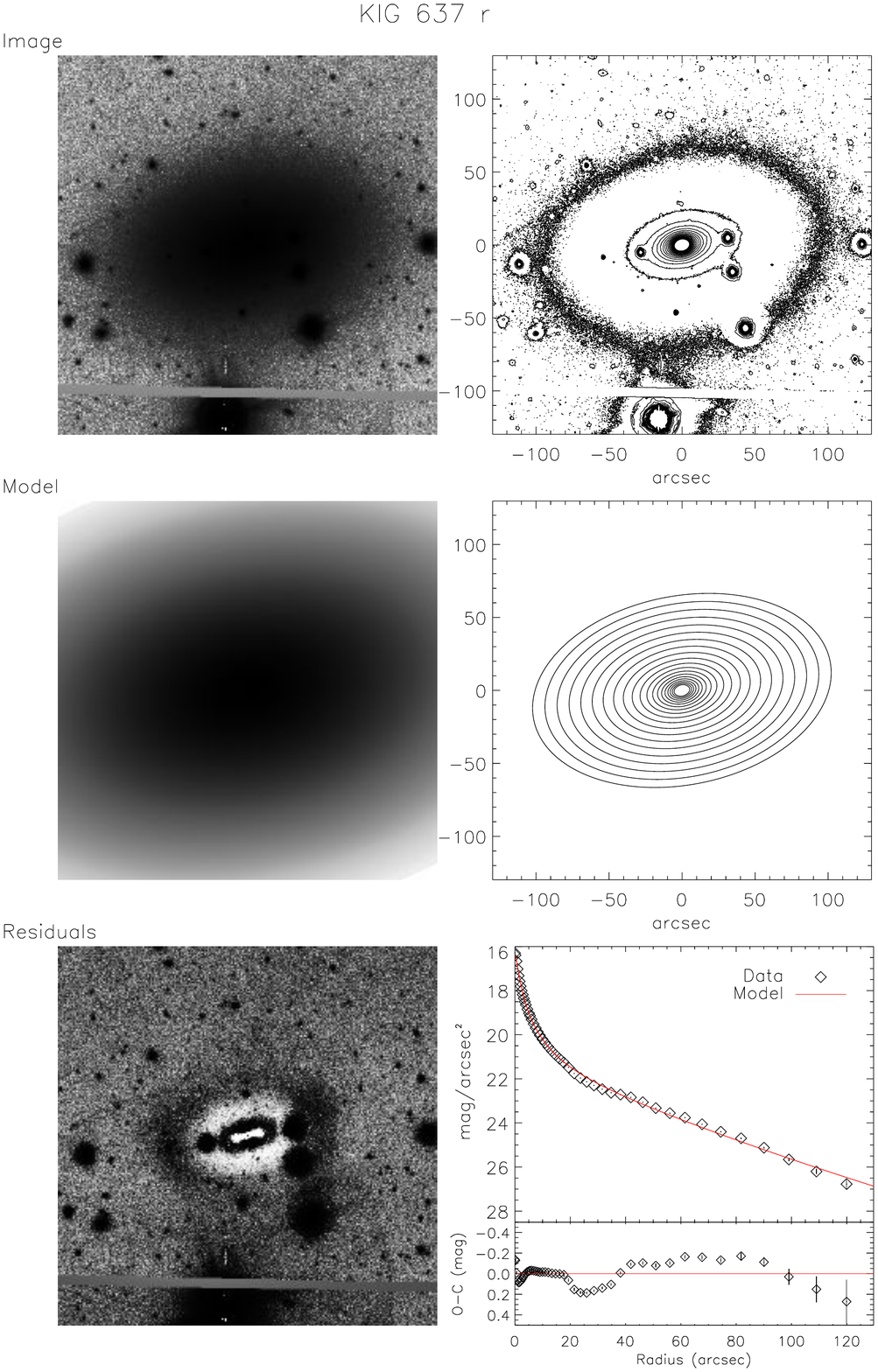}}
%{\includegraphics[width=8.0cm]{KIG644g_B+D.pdf}
%\includegraphics[width=8.0cm]{KIG644r_B+D.pdf}}
\includegraphics[width=15.6cm]{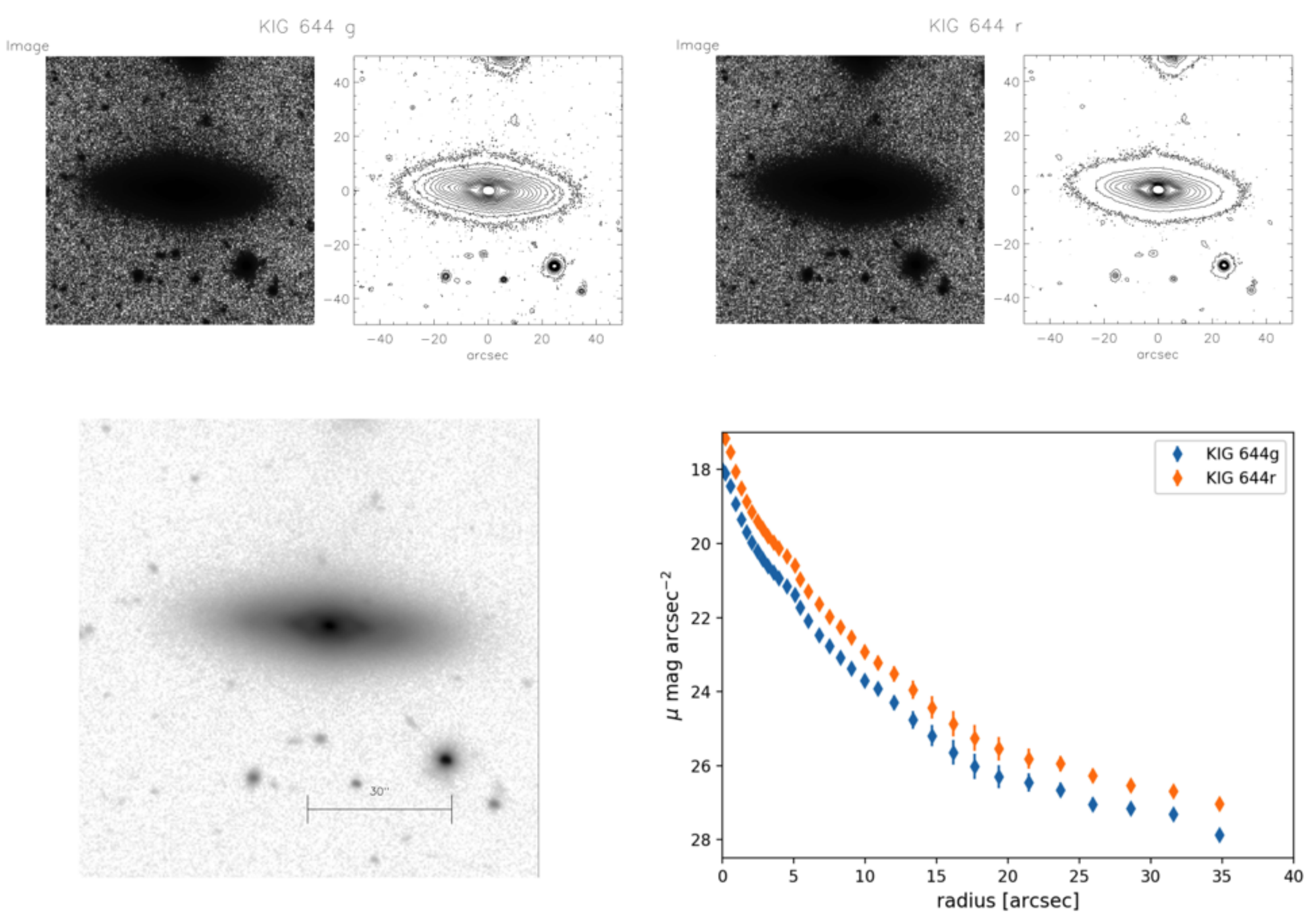}
\caption{As in Figure~\ref{figure-5}. Summary of the surface photometric analysis of KIG 637 ({\it top panels}).
In both bands we show the B+D models. The image of KIG 637 shows the gap 
in the 4K CCD. The corona of the bright star HD 238370, south of the galaxy nucleus,
 extends across the entire galaxy and strongly affects the photometric measures. 
 We show 20 isophote levels, between 500 and 2 $\sigma$ of the sky level,  for the original and model images. 
Summary of the surface photometric analysis of KIG 644 ({\it middle and bottom panels}). Deep $g$ and $r$
images are shown with isophote contours down to the 2$\sigma$ level of the sky. The right $r$ image enhances the central part of KIG 644, showing the ring and the lens structures; the left panel shows the $g$ and $r$ light profiles.}
\label{figure-A6}
\end{figure*}
%--------------------------end figure A6 -----------------------

%-------------------------------- Figure A7 ------------------------
\begin{figure*}
\center
{\includegraphics[width=7.8cm]{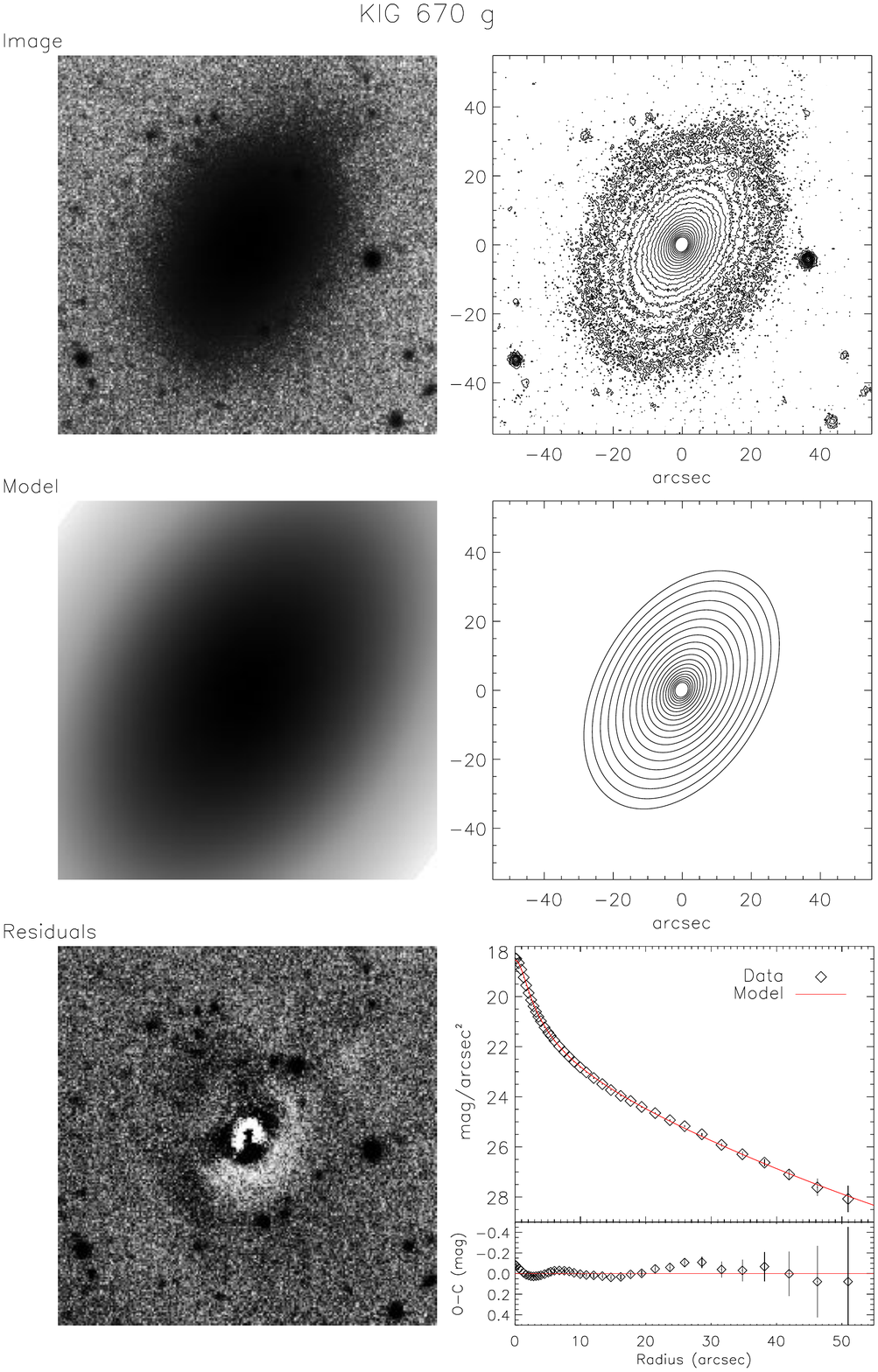}
\includegraphics[width=7.8cm]{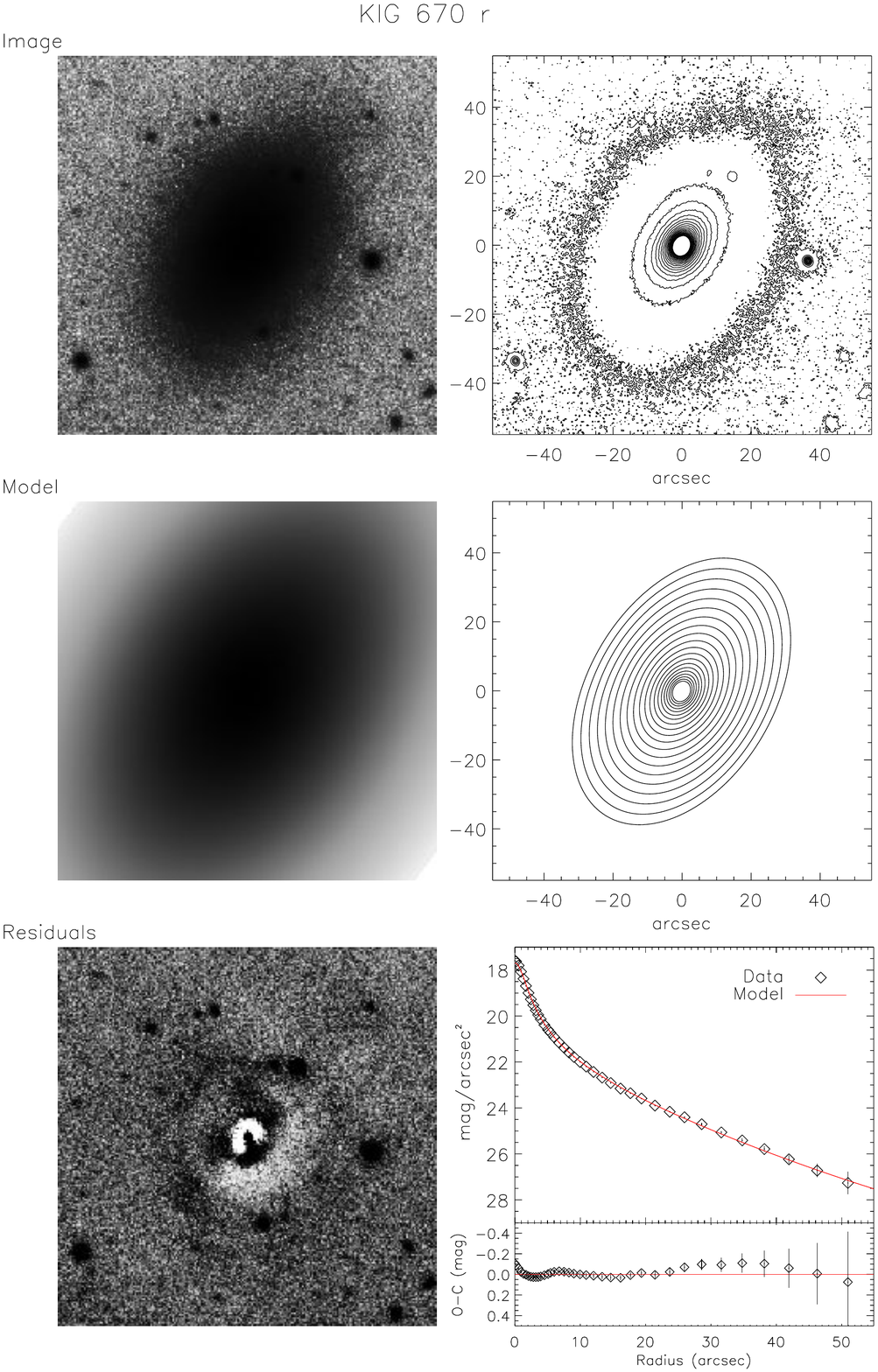}}
{\includegraphics[width=8.2cm]{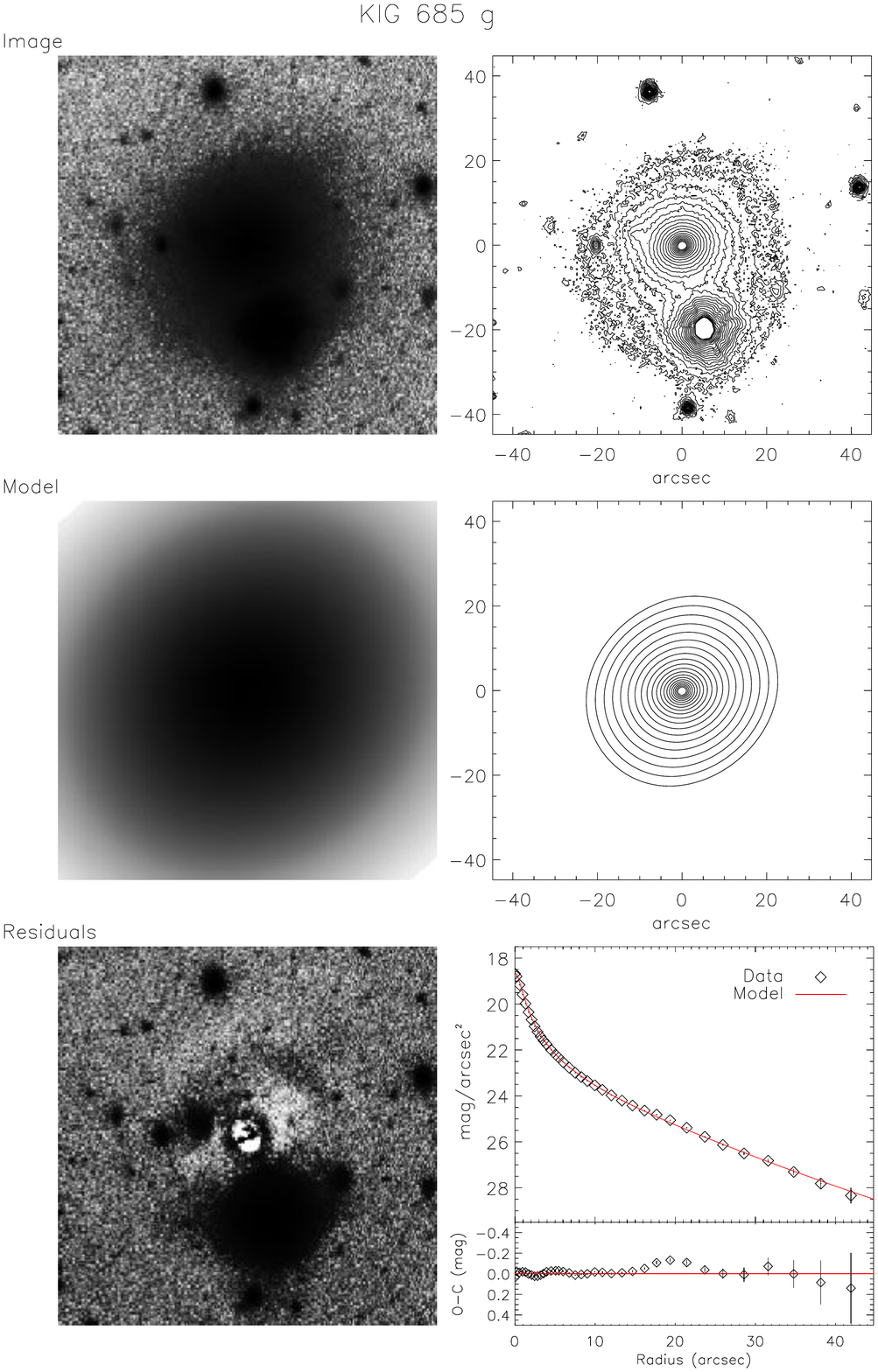}
\includegraphics[width=7.8cm]{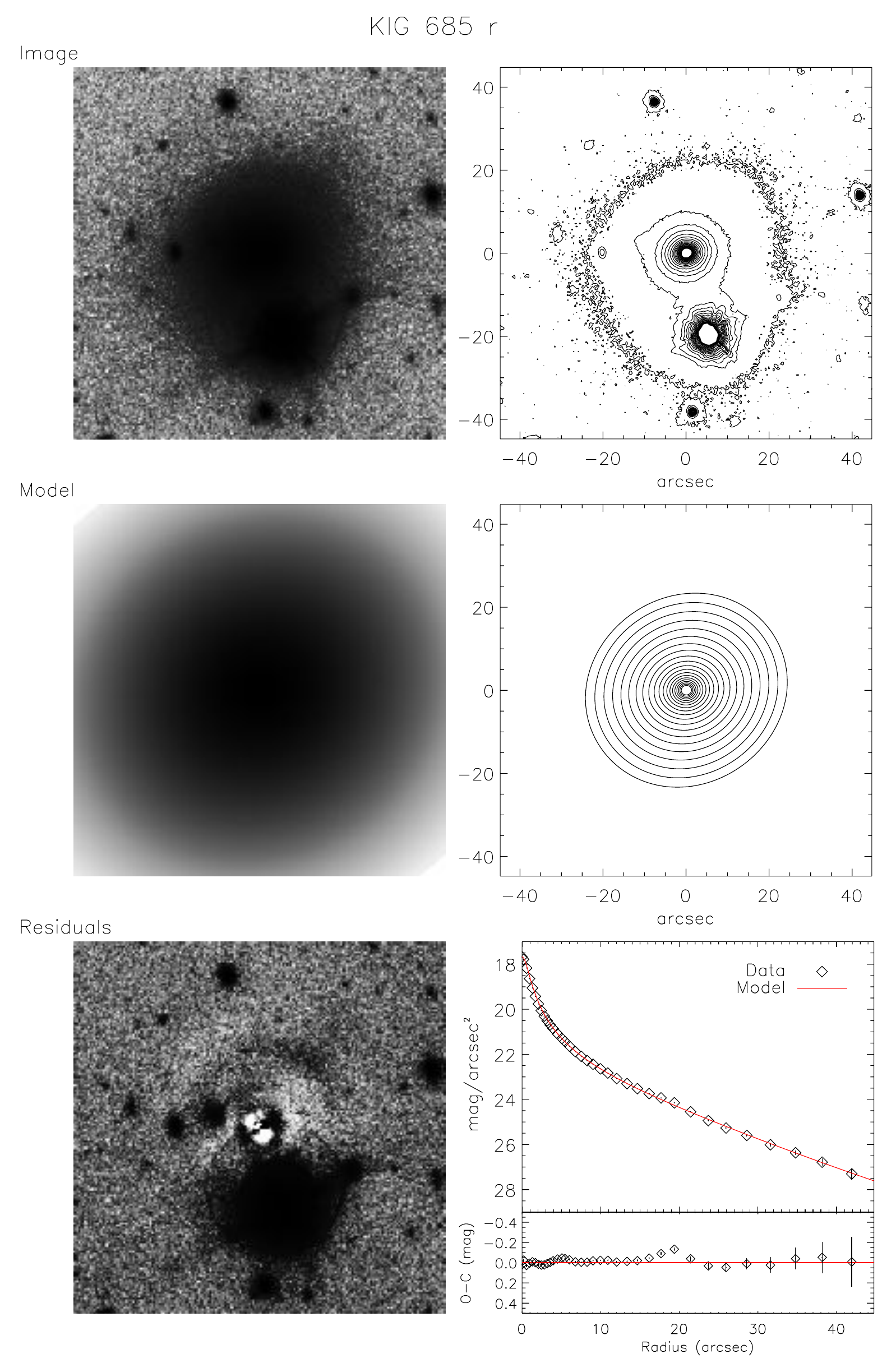}}
\caption{As in Figure~\ref{figure-5}. Summary of the surface photometric analysis of KIG 670 ({\it top panels})
and KIG 685  ({\it bottom panels}). In both bands we show the B+D models. We show 20 isophote levels, between 500 and 
2 $\sigma$ of the sky level,   for the original and model images.}
\label{figure-A7}
\end{figure*}
%--------------------------end figure A7 -----------------------

%-------------------------------- Figure A8 ------------------------
\begin{figure*}
\center
{\includegraphics[width=7.8cm]{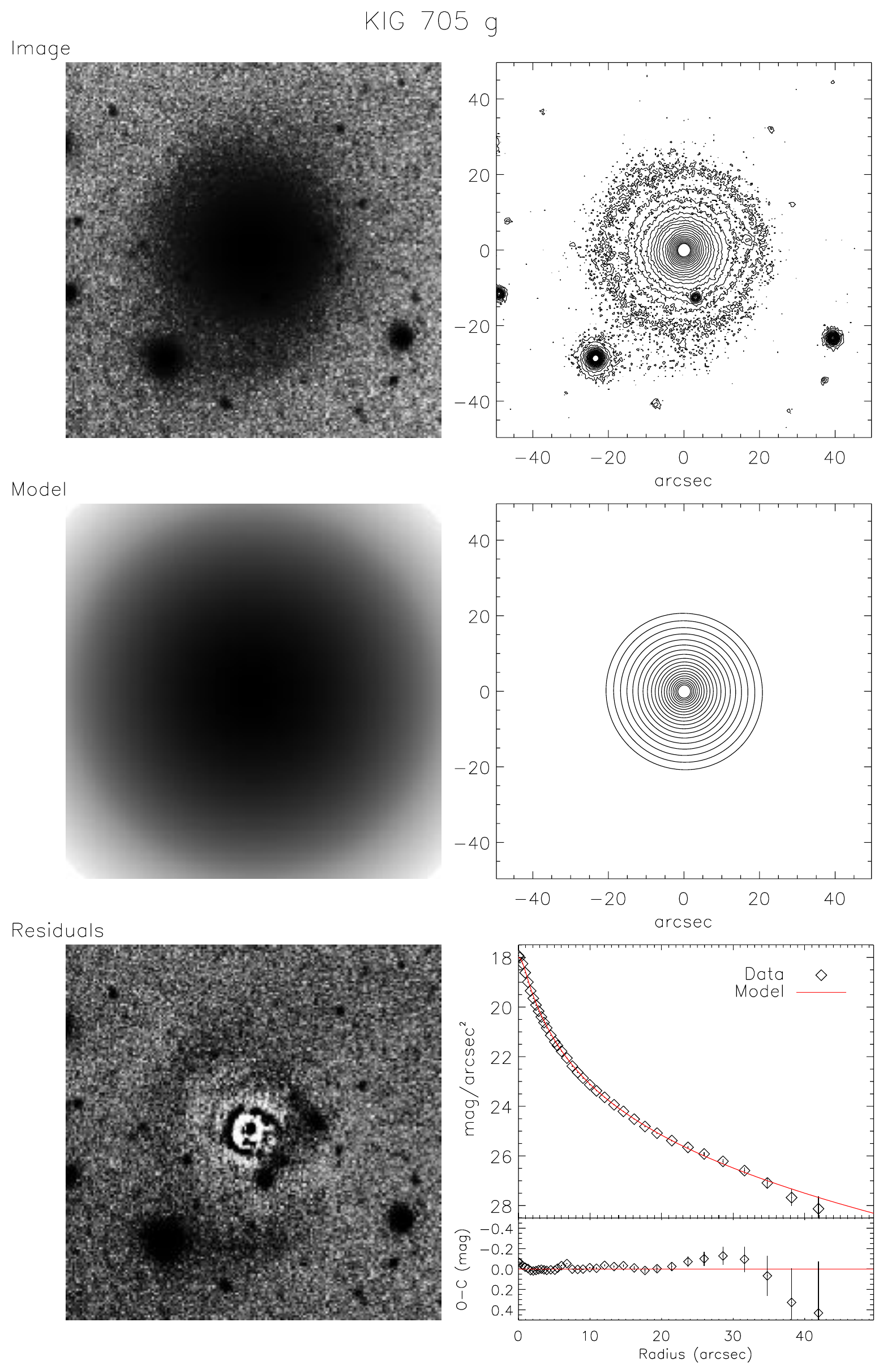}
\includegraphics[width=7.8cm]{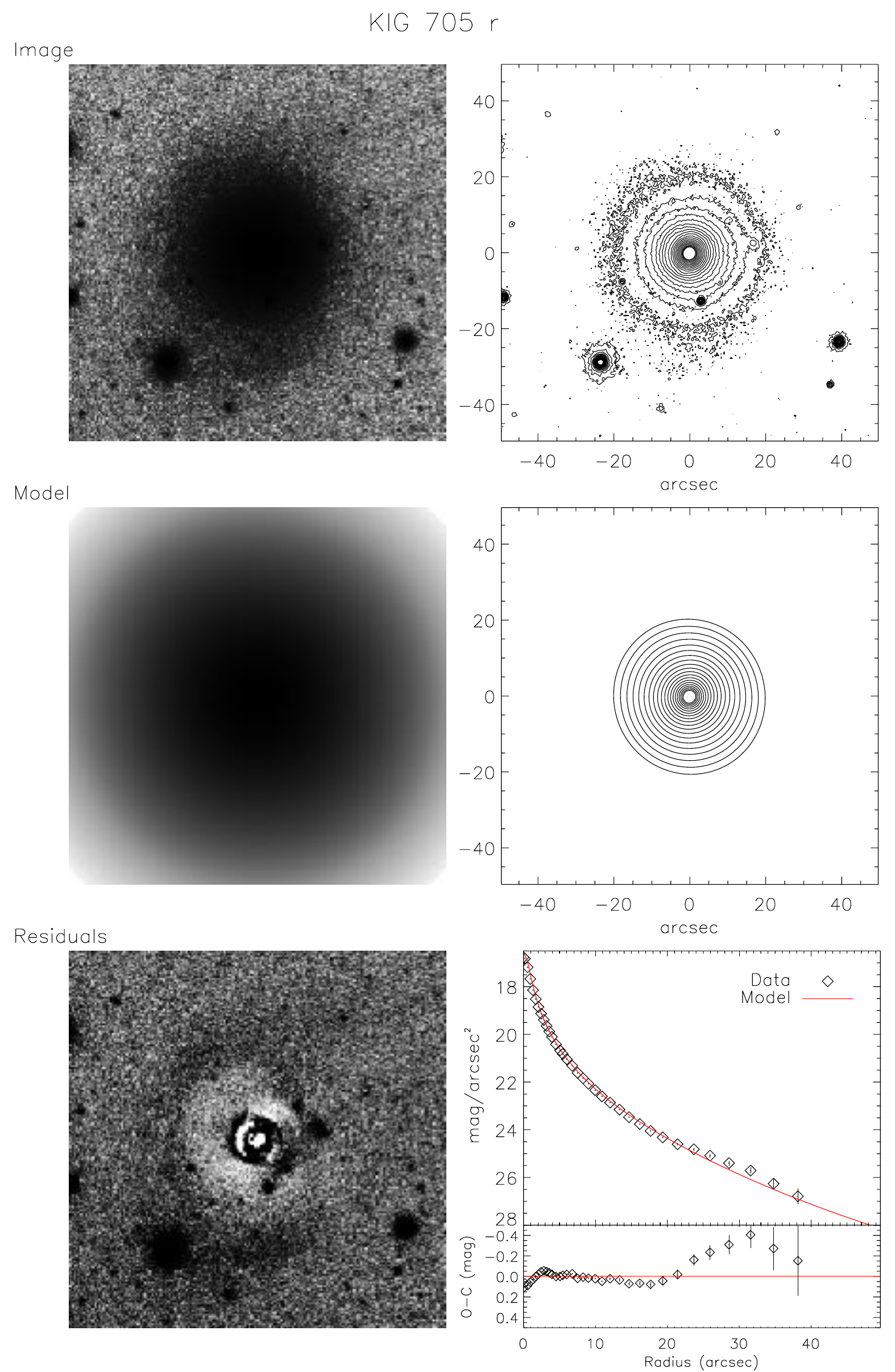}}
{\includegraphics[width=7.8cm]{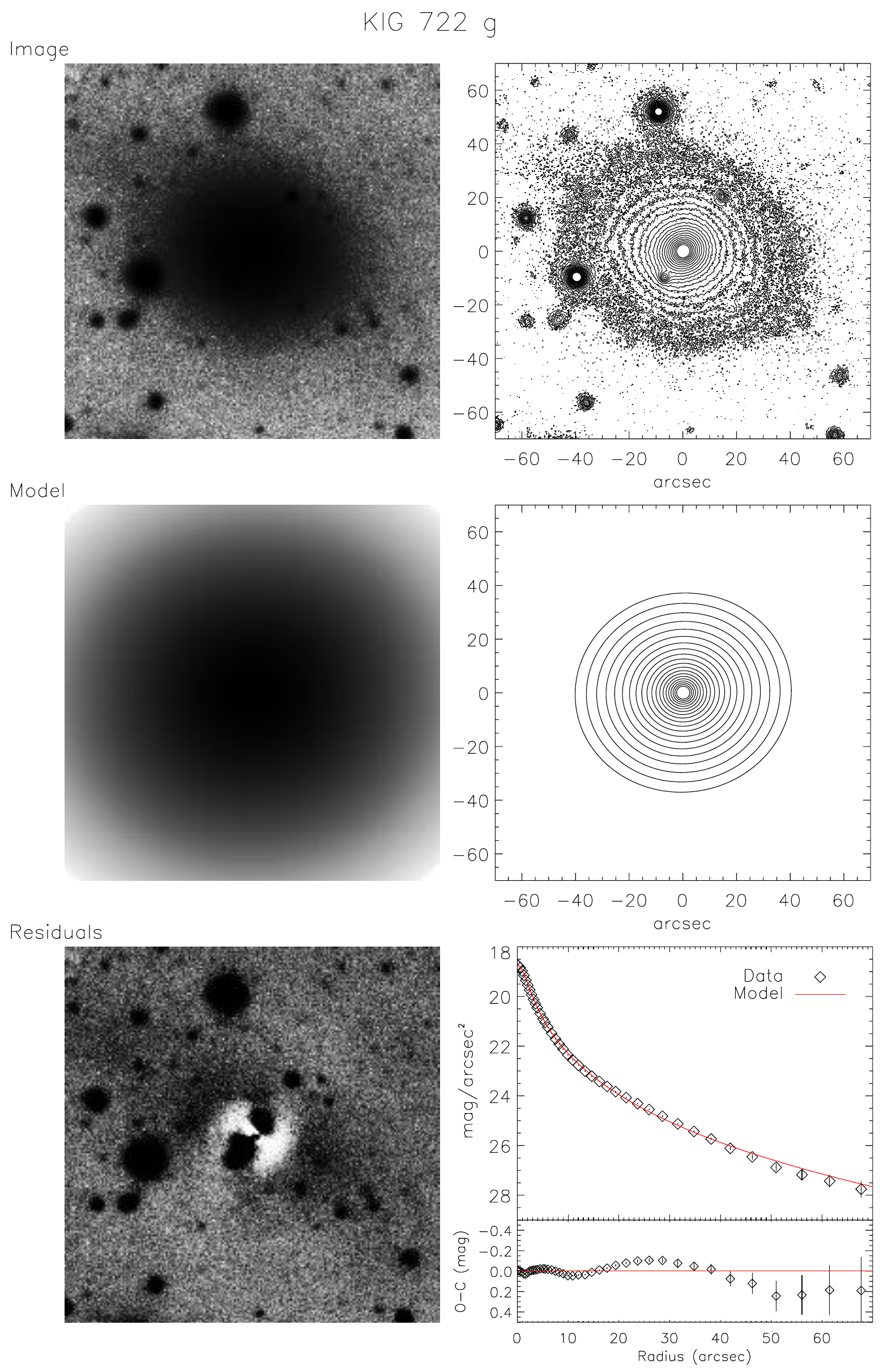}
\includegraphics[width=7.8cm]{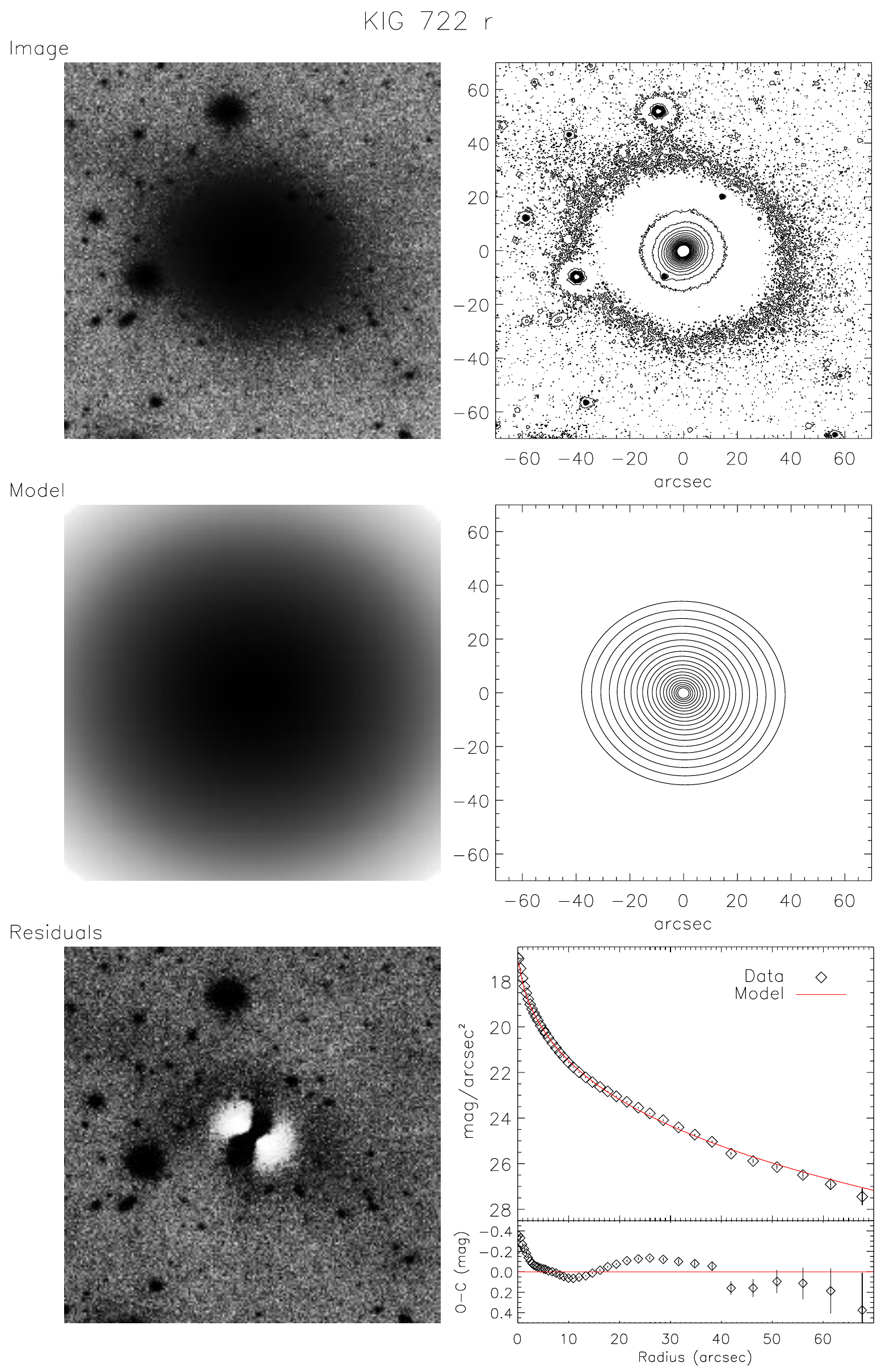}}
\caption{As in Figure~\ref{figure-5}. Summary of the surface photometric analysis of KIG 705 ({\it top panels})
and KIG 722  ({\it bottom panels}). In both bands for KIG 722 we show the S\'ersic models. We show 20 isophote levels, between 500 and 
2 $\sigma$ of the sky level, for the original and model images.}
\label{figure-A8}
\end{figure*}
%--------------------------end figure A8 -----------------------

%-------------------------------- Figure A9 ------------------------
\begin{figure*}
\center
{\includegraphics[width=7.8cm]{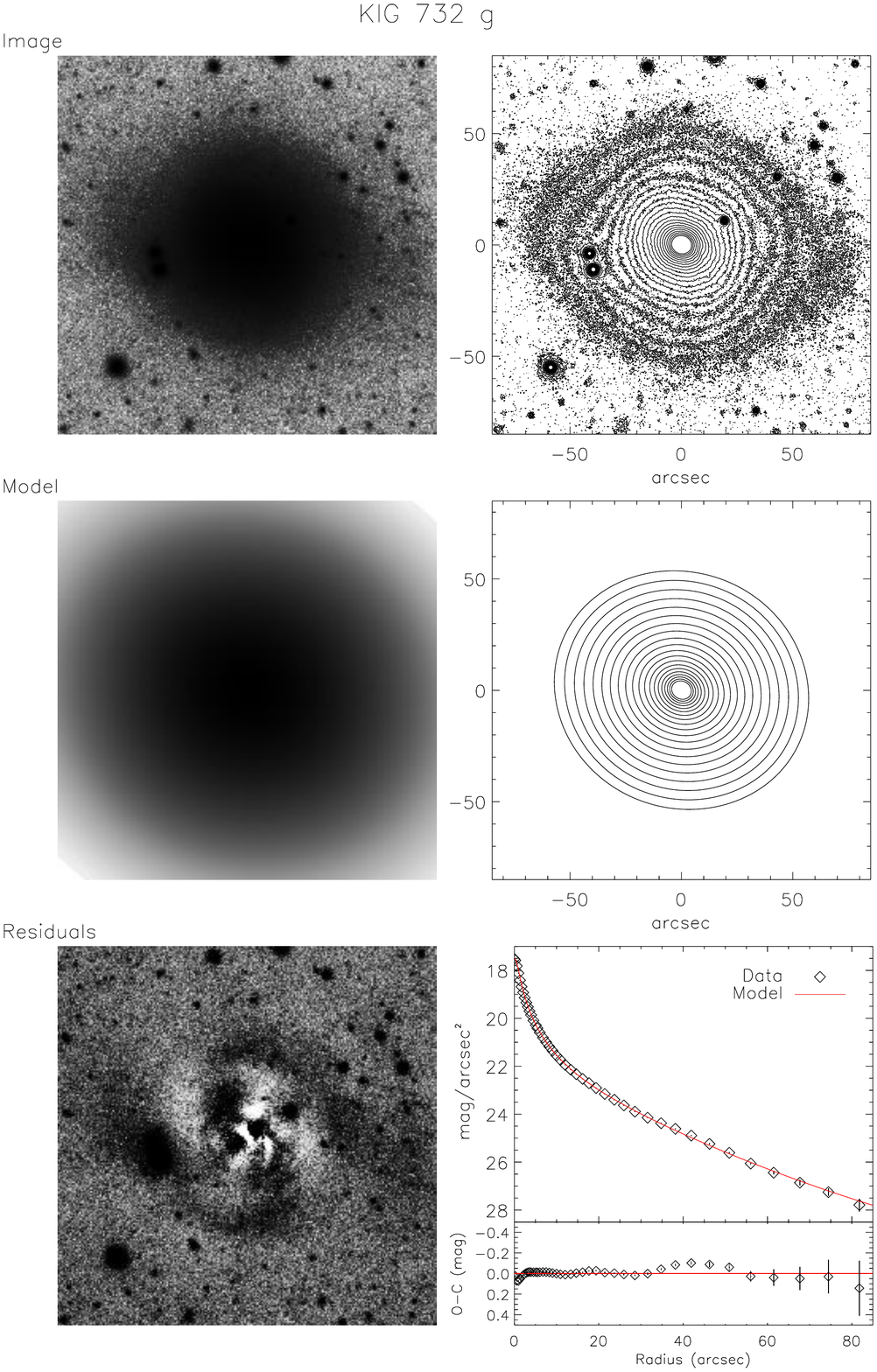}
\includegraphics[width=7.8cm]{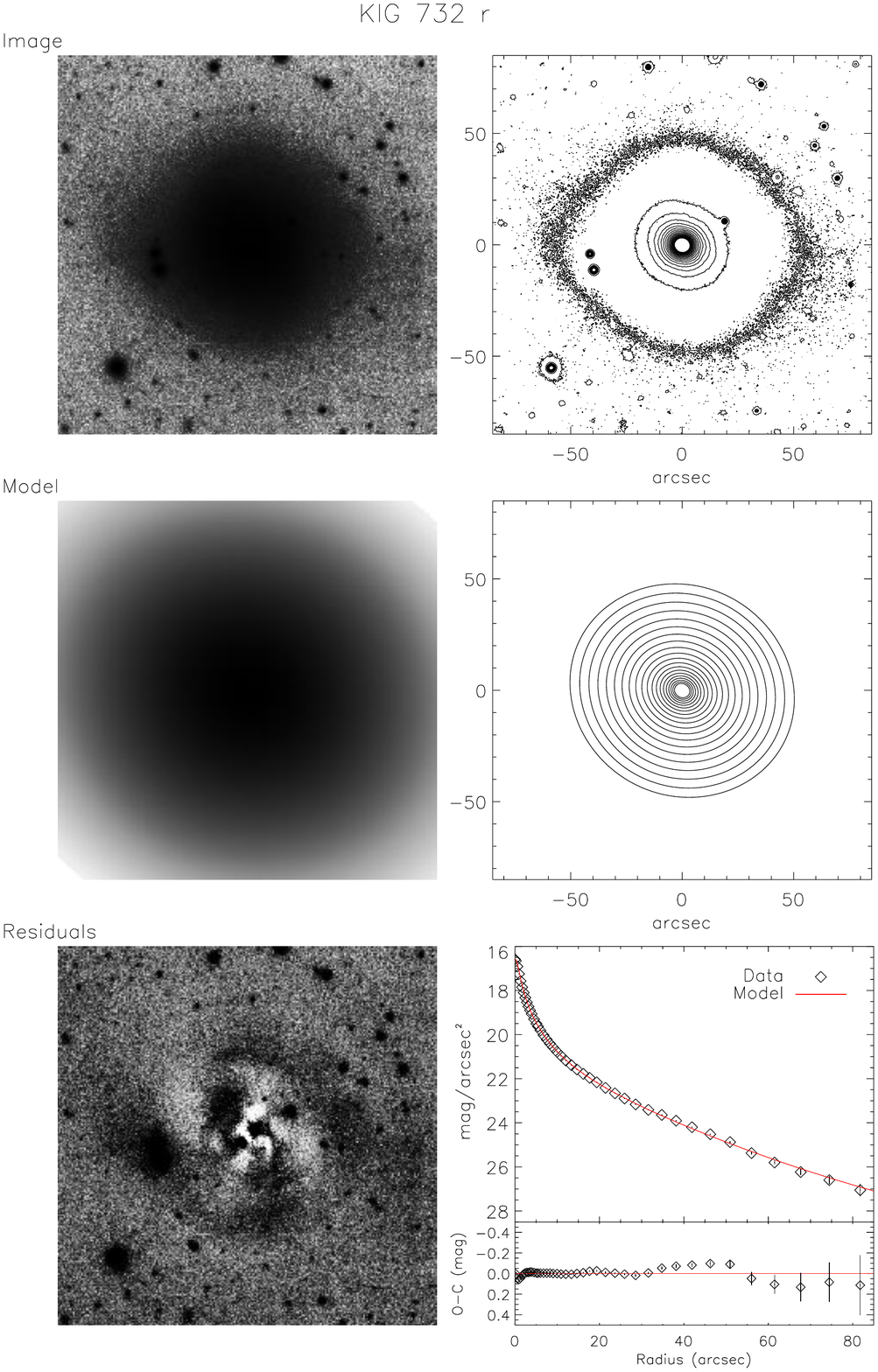}}
%{\includegraphics[width=8.0cm]{KIG733g_B+D.pdf}
%\includegraphics[width=8.0cm]{KIG733r_B+D.pdf}}
\includegraphics[width=15.6cm]{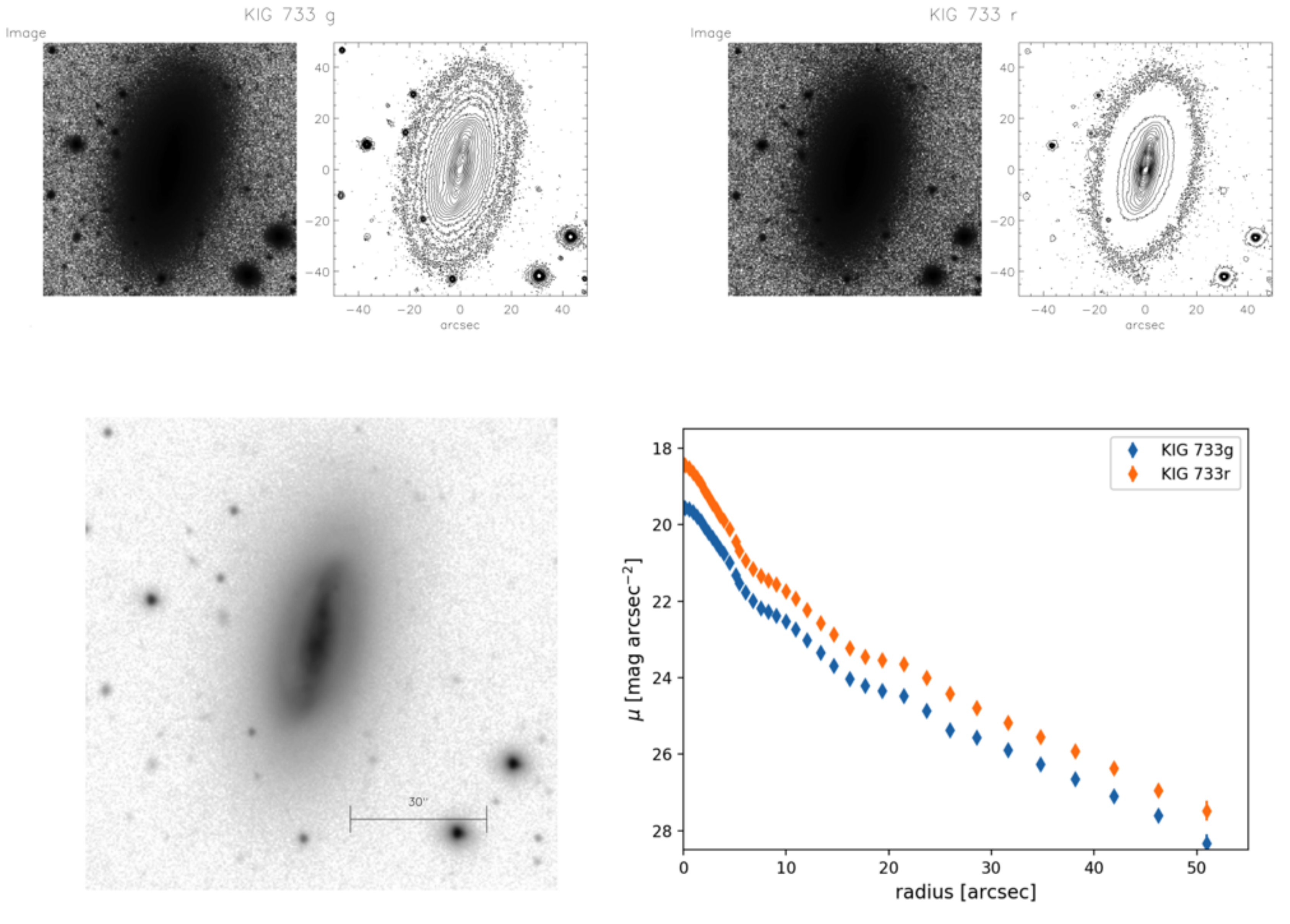}
\caption{As in Figure~\ref{figure-5}. Summary of the surface photometric analysis of 
KIG 732 ({\it top panels}). We show the B+D model. We show 20 isophote levels, between 500 and 
2 $\sigma$ of the sky level, for the original and model images.
Summary of the surface photometric  analysis of KIG 733 ({\it middle panels}). Deep $g$ and $r$
images are shown with isophote contours down to the 2$\sigma$ level of the sky. ({\it Bottom panels})
The right $r$ image enhances the central part of KIG 733, showing the ring and the arm-like structure, and 
the left panel shows the $g$ and $r$ light profiles.}
\label{figure-A9}
\end{figure*}
%--------------------------end figure A9 -----------------------

%-------------------------------- Figure A10 ------------------------
\begin{figure*}
\center
{\includegraphics[width=7.8cm]{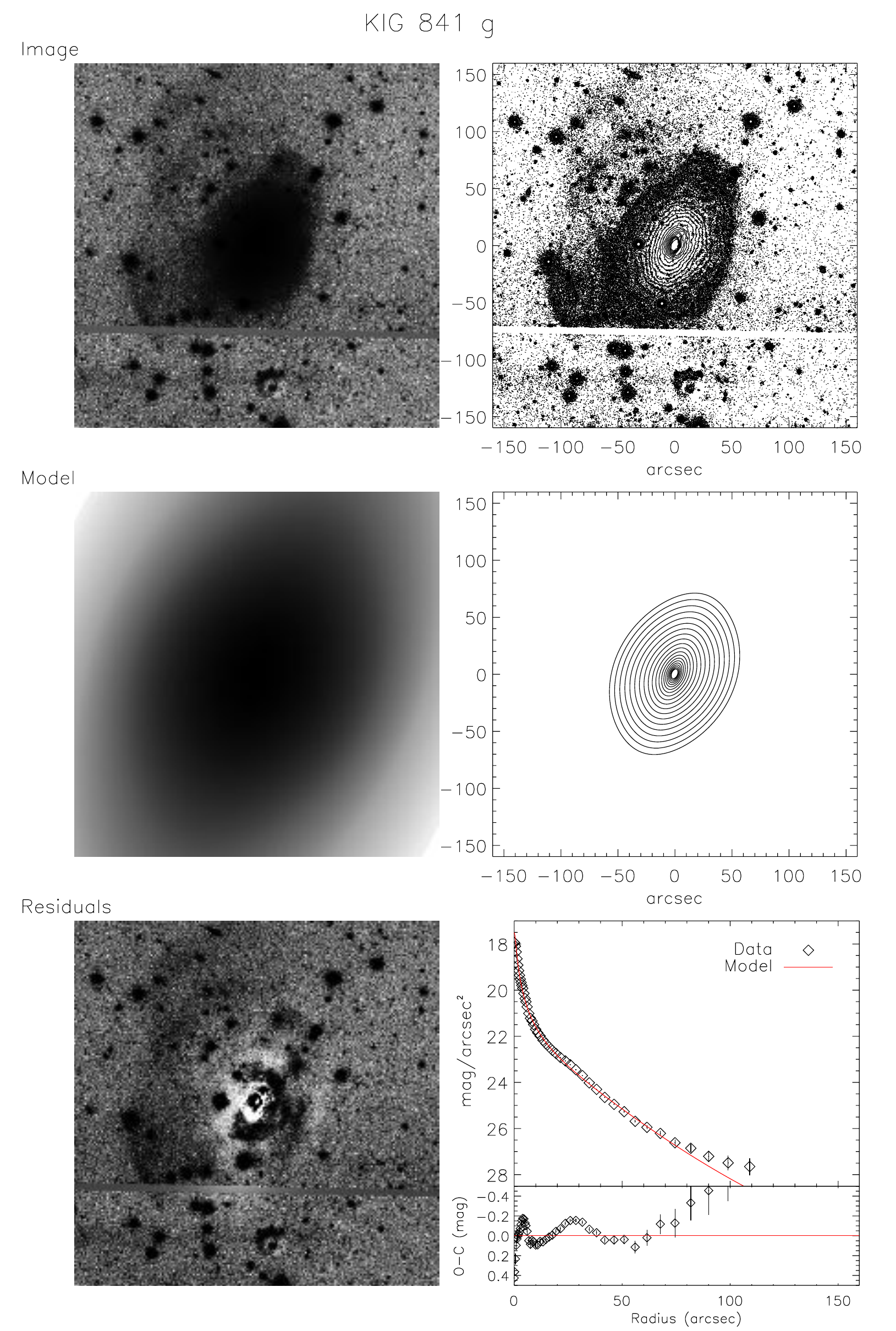}
\includegraphics[width=7.8cm]{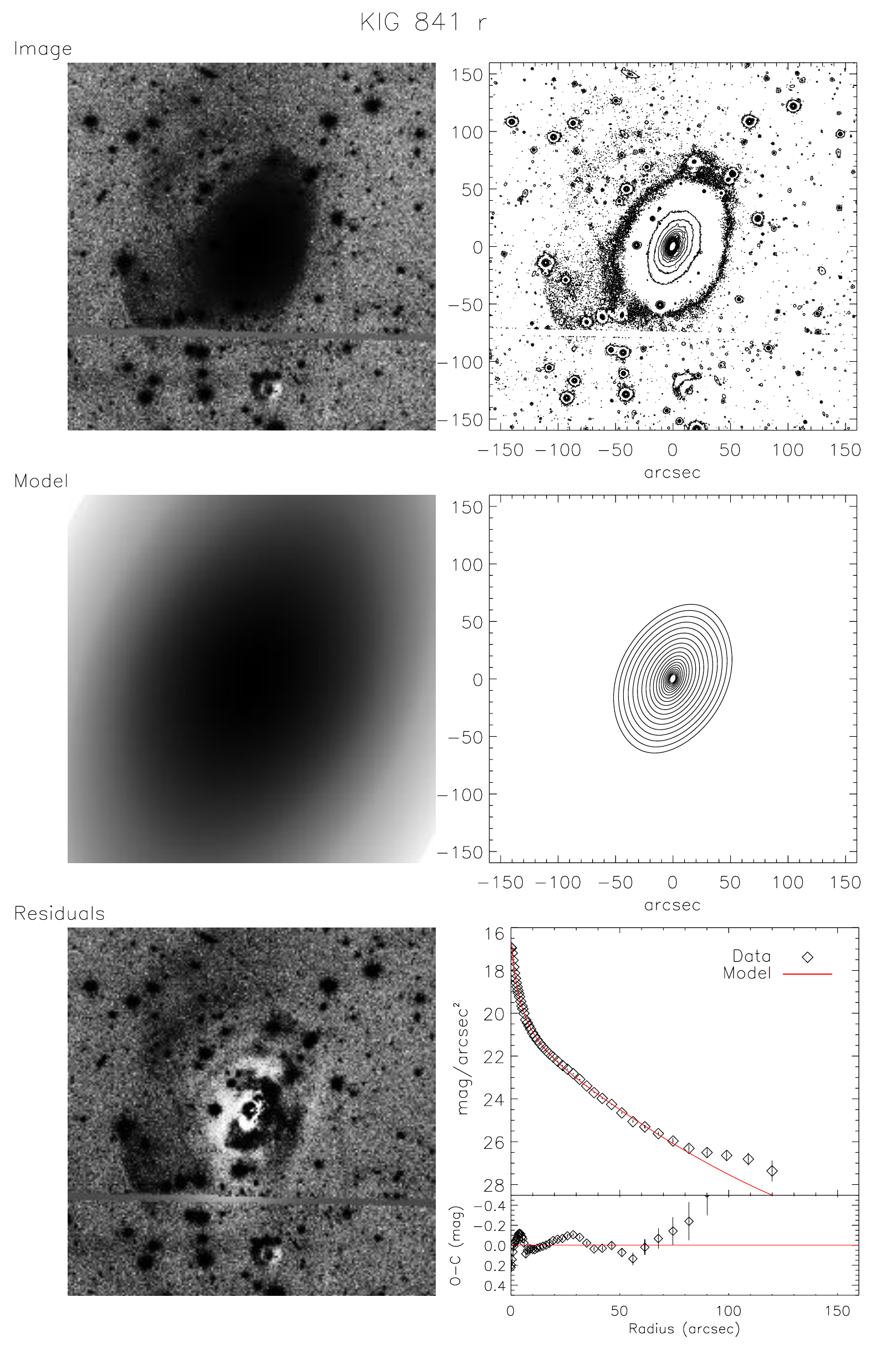}}
\caption{As in Figure~\ref{figure-5}. Summary of the surface photometric analysis of KIG 841.
 In both bands we show the B+D models. We show 20 isophote levels, between 500 and 
2 $\sigma$ of the sky level, for the original and model images. The gap in the 4K CCD and a 
dust feature on the CCD that is not removed by the flat-fielding are visible.}
\label{figure-A10}
\end{figure*}
%--------------------------end figure A10 -----------------------

\end{appendix}

\end{document}